\documentclass[hyper]{JHEP3}
\pdfoutput=1
\usepackage{graphicx}
\usepackage{epstopdf}

\input{epsf}

\usepackage{epsfig}
\usepackage{amssymb}
\usepackage{amsfonts}
\usepackage{amsbsy}
\usepackage{amsmath}

\def\Dob{{\rm \overline{D0}}}
\def\Dsb{{\rm \overline{D6}}}
\def\Dtb{{\rm \overline{D2}}}

\title{The elliptic genus from split flows and Donaldson-Thomas invariants}
\author{Andr\'es Collinucci$^1$ and Thomas Wyder$^2$\\

$^1$ Institute for Theoretical Physics, Vienna University of Technology, \\
Wiedner Hauptstr. 8-10, 1040 Vienna, Austria\\

$^2$ Instituut voor Theoretische Fysica, KU Leuven, \\
Celestijnenlaan 200D, B-3001 Leuven, Belgium \\

{\tt andres AT itf.fys.kuleuven.be,  thomas AT itf.fys.kuleuven.be}

}

\abstract{We analyze a mixed ensemble of low charge D4-D2-D0 brane states on the quintic and show that these can be successfully enumerated using attractor flow tree techniques and Donaldson-Thomas invariants. In this low charge regime one needs to take into account worldsheet instanton corrections to the central charges, which is accomplished by making use of mirror symmetry. All the charges considered can be realized as fluxed \mbox{D6-D2-D0} and $\overline{D6}$-D2-D0 pairs which we enumerate using DT invariants. Our procedure uses the low charge counterpart of the picture developed Denef and Moore. By establishing the existence of flow trees numerically and refining the index factorization scheme, we reproduce and improve some results obtained by Gaiotto, Strominger and Yin. Our results provide appealing evidence that the strong split flow tree conjecture holds and allows to compute exact results for an important sector of the theory. Our refined scheme for computing indices might shed some light on how to improve index computations for systems with larger charges.}
\preprint{KUL-TF-08/25}

\begin{document}

\section{Introduction}\label{intro}
The BPS spectrum of four dimensional black holes originating from type II string theory compactified on a Calabi-Yau manifold has been studied intensively over the last decade. Important results have been obtained by studying the microscopic D-brane description of these states, and additionally, the discovered connection between this picture and the supergravity approximation of string theory has been and still is deepening. In \cite{Moore:1998pn}, \cite{Moore:1998zu}, G. Moore studied the attractor mechanism devised to describe BPS solutions, and put forth a correspondence between spherically symmetric solutions to the black hole attractor equations and BPS states in string theory. However, this correspondence turned out not to always hold. The remedy, developed in \cite{Denef:2000nb}, is to consider not only single attractor flows, but also `split attractor flows', which are conjectured to be in one to one correspondence with multi-centered BPS solutions. Recently, however, some ambiguities have been discovered for so called `scaling solutions': these are multicentered solutions which allow different centers to approach each other to the extent that their throats `melt together'. The special feature to which they owe their name is that the distance between their centers is not fixed, but rather a `scaling' modulus. The fact that their throats melt together renders them very similar to single centered solutions in many respects, and it seems that single flows could correspond to such multi-centered solutions. However, at present, their meaning in relation to the split attractor flow conjecture is not completely clear.

The microscopic string theory meaning of split attractor flows was elucidated in \cite{Denef:2002ru}. The microscopic counterparts of multi-centered SUGRA solutions are D-branes wrapped along the internal compactification space, that decay into separate, chirally intersecting D-branes. 

In the IIB picture, black holes are made of D3-branes wrapped along special Lagrangian three-cycles of the internal Calabi-Yau manifold, whereby these D3's (and their corresponding three-cycles) can split up into intersecting D3's, by moving in the complex structure moduli space of the CY across some `line of marginal stability'.

In the IIA picture, the black holes considered in the literature are made of D4-D2-D0 systems wrapped on the even cycles of the internal CY. The physical picture corresponding to split attractor flows in this case is that of Sen's open string tachyon condensation picture. More specifically, D4-D2-D0 configurations can dynamically decay into D6/$\Dsb$ pairs with lower branes on them, when one moves across a line of marginal stability in the K\"ahler moduli space of the CY \cite{Denef:2007vg}.

In light of the existence of mirror symmetry, which relates the classical physics of D-branes wrapped on the three-cycles of a CY manifold $X$ to the quantum corrected physics of D-branes wrapped on the even cycles of a mirror CY manifold $Y$, it is clear that these two descriptions of split flows are related.
This relation was in part exploited in \cite{Denef:2001xn}, where mirror symmetry was used explicitly to scan the possible spectrum of IIA split flows.
\vskip 1cm
In this paper, we will focus on D4-D2-D0 systems with low charges on the quintic Calabi-Yau, and systematically search for their D6/$\Dsb$ descriptions by looking for split attractor flows. Given that the charges will be very low, the attractor equations will set the horizon size to be typically very small with respect to the string scale. This means that the supergravity approximation will break down in the vicinity of the horizon, and higher curvature corrections from perturbative worldsheet $\alpha'$ effects should be taken into account. The attractor equations for low D-brane charges will also typically drive the sizes of the wrapped internal cycles to the stringy scale, and the Calabi-Yau itself away from the large volume regime. At this point worldsheet instanton corrections to the central charges of the branes will become dominant. By using mirror symmetry, or more specifically, by using explicit numerical approximations for the solutions to the Picard-Fuchs equation of the mirror quintic, we will be able to fully take these corrections into account.

Having sorted out the split flow structure of the so-called `polar states' (i.e. roughly, those states with low D2-D0 charges), we will compute their BPS indices by means of Donaldson-Thomas theory, defined in \cite{DT1, DT2, DT3}. The latter is conjectured \cite{Iqbal:2003ds, MNOP1, MNOP2, Dijkgraaf:2006um} to provide an index (in some appropriate region in moduli space) for the topologically twisted gauge theory describing D6-D2-D0 systems. This will allow us to put together indices for the polar D4-D2-D0 states of the form $\sim \Omega_{DT} \overline{\Omega_{DT}}$ in the spirit of the picture developed in \cite{Denef:2007vg}. This factorization of the indices can be thought of as a microscopic building block of the well-known OSV factorization of the black hole index for large charges (\cite{Ooguri:2004zv}), as described in \cite{Denef:2007vg}. It turns out, however, that a refinement to this factorization scheme is necessary, in order to account for special, \emph{non-singular} loci in the moduli spaces of the theory on the D6/$\Dsb$ stack. Note, that these special loci are \emph{not} due to $\Dob$ or D2 branes becoming coincident\footnote{Such singular loci are automatically resolved by the DT theory.}, but are instead due to loci where the D6-$\Dsb$ tachyon degrees of freedom get enhanced.

Our results turn out to numerically match, and improve upon the geometric and CFT computations made in \cite{Gaiotto:2006wm, Gaiotto:2007cd} from an M-theory point of view, in order to compute the so-called \emph{modified elliptic genus} of the MSW CFT, \cite{Maldacena:1997de}. The goal of this paper is twofold: Our results provide a strong test to the `split attractor flow conjecture', which roughly states that the entire spectrum of string theory four-dimensional black holes (or D-particles) is encoded in the existence of (split) attractor flows. Our second main goal is to devise an exact computational scheme of BPS indices. Our calculations illustrate the sensitivity of BPS index computations to the so called \textsl{area code}. They also demonstrate their usefulness in computing exact results for elliptic genera. Furthermore, they suggest a refinement to the factorization scheme of the BPS indices that might point to a better treatment of systems with larger charges.

This paper is organized as follows: The next three sections, \ref{sec:splitattrflows} to \ref{mirror}, cover background material. In section \ref{sec:splitattrflows}, we summarize the split attractor flow picture from both the supergravity and the microscopic perspective, thereby stating the split attractor flow conjecture. In section \ref{sec:branessheavesDT}, we give an introduction to and physically motivate the use of Donaldson-Thomas theory, and concretely state our strategy for computing BPS indices. In section \ref{mirror}, we briefly explain the necessary facts about mirror symmetry and setup our conventions to compute central charges of D-brane systems in terms of the solutions to the Picard-Fuchs equations. The reader familiar with these concepts can skip parts or all of these sections and proceed directly to sections \ref{results} and \ref{fibration}, where we provide our results for the BPS indices of the D4-D2-D0 systems we are interested in, compare them to existing results, and thoroughly discuss these. In section \ref{fibration} we propose a refined procedure for putting together BPS indices. In the appendices, we provide explicit information about the structures of the split flows for two interesting cases. The purpose is to show how the computation of BPS indices is very sensitive to the region of the K\"ahler moduli space in which one computes them.

\section{Split attractor flows and BPS states in string theory} \label{sec:splitattrflows}
In this section we will discuss one of our main tools, split attractor flow trees, and their correspondence to BPS states in string theory. We will start by giving some general definitions and clarifying our notation and conventions. In this paper, we investigate a type IIA compactification of string theory on the quintic Calabi-Yau threefold $X$ as a study model. We will use $H$ as a basis element for the second cohomology $H^2(X,\mathbb{Z})$. The normalization automatically inherited is $\int_X H^3=5$. The setup we will be studying throughout consists of BPS objects on the quintic formed by a D4-brane wrapped on a hyperplane class divisor $P=H \in H_4(X)$, with additional lower dimensional D2-D0-brane charge. A general system with D6-D4-D2-D0-brane charge is represented by a polyform $\Gamma$, which is an element of the integer valued even degree cohomology $$H^{*}(X,\mathbb{Z})=H^{0}(X,\mathbb{Z})\oplus H^{2}(X,\mathbb{Z})\oplus H^{4}(X,\mathbb{Z})\oplus H^{6}(X,\mathbb{Z})\,.$$ Using our basis of the second cohomology $H$, we write the polyforms as follows:
\begin{equation}
  \Gamma =p^0+p H+\frac{q}{5} H^2+\frac{q_0}{5}H^3. \label{gammadef}
\end{equation}
Note that the volume-form is related to our basis as $\omega=\frac{1}{5}H^3$, hence factors of $\frac{1}{5}$ are included for normalization.
We define a symplectic intersection form $\langle .\, ,\,.\rangle:H^{*}\times H^{*}\rightarrow \mathbb{C}$ \label{intform} as follows:
\begin{equation}
  \langle\Gamma_1,\Gamma_2\rangle=\int_X \Gamma_1\wedge \Gamma_2^{*},
\end{equation} 
where $\Gamma_2^{*}$ is obtained from $\Gamma_2$ by inverting the sign of the two- and six-form components. This intersection product computes the net chiral spectrum emanating from strings stretched between the two brane systems $\Gamma_1$ and $\Gamma_2$. Writing the complexified K\"ahler moduli as $t^A=B^A+iJ^A$ we define the following polyform in IIA as
\begin{equation}
  \Omega =-e^{K/2}\cdot\, e^{B+iJ}=-\sqrt{\frac{3}{4J^3}}\, e^{B+iJ},
\end{equation}
where $K$ denotes the K\"ahler potential. It is normalized with respect to the symplectic product defined previously. We will also use the holomorphic version,
\begin{equation}\label{holperiod}
  \Omega_{\textrm{hol}}=e^{-K/2}\, \Omega =-e^{B+iJ}.
\end{equation}
The central charge of a brane system $\Gamma$ is now defined as
\begin{equation}
  Z=\langle\Gamma,\Omega\rangle = -\int_X e^{-(B+iJ)}\, \Gamma \,. \label{zlargev}
\end{equation}

In the next subsection, we will quickly review a few relevant points about attractor flow trees starting with their supergravity interpretation, followed by a short discussion of their microscopic D-brane/quiver interpretation and validity beyond the large radius regime.

\subsection{The split attractor flow conjecture}
The technique of split flow trees plays a central role in obtaining our results. Single flows / split flow trees are believed to be an existence criterion for BPS solutions in supergravity (and as we will later discuss more broadly, for general BPS solutions in string theory). They are graphical depictions of the flow of the K\"ahler moduli (or complex structure moduli in the mirror type IIB picture) belonging to a BPS solution of supergravity. The flow starts at the background value $t_{\infty}$ at radial infinity and moves towards one attractor point $t_{*}$ at the horizon of the black hole, for a single centered solution, or possibly several attractor points $(t_{1 *},...,t_{m *})$, one for each center, when dealing with a multicentered solution. It is governed by the famous black hole attractor equations, which were extended to multicentered black holes in \cite{Denef:2000nb}.

A split flow tree is built as follows. One follows the \textsl{incoming branch} \footnote{The term `incoming branch' refers to the part of the flow tree connecting the background point and the first split point.} of a flow tree from radial infinity towards a putative attractor point, until one hits a \textsl{wall of marginal stability} for two constituents $\Gamma_1, \Gamma_2$ such that $\Gamma=\Gamma_1+\Gamma_2$. This is defined as the hypersurface in moduli space where the phases of the central charges align, $\textrm{arg}(Z_1)=\textrm{arg}(Z_2)$. The length of a central charge measures the mass of the corresponding state, whereas the phase indicates which $\mathcal{N}=1$ supersymmetry of the original $\mathcal{N}=2$ supersymmetry is preserved by the state. If the phases of two central charges align, the two states are mutually BPS (i.e. preserve the same supersymmetry) and the binding energy of the BPS bound state vanishes,
\begin{equation}
  |Z_{1+2}|=|Z_1|+|Z_2|,
\end{equation}
or equivalently,
\begin{equation}
  \textrm{Re}(\bar{Z}_1Z_2)>0,\qquad \textrm{Im}(\bar{Z}_1Z_2)=0.
\end{equation}
We will now reserve the notion of a wall of \emph{marginal} stability as opposed to a wall of \emph{threshold} stability for the case when the two charges are non-local: $\langle\Gamma_1,\Gamma_2\rangle\neq 0$. If one reaches the wall from the side where $\langle\Gamma_1,\Gamma_2\rangle (\textrm{arg}(Z_1)-\textrm{arg}(Z_2))>0$, the decay of $\Gamma \rightarrow \Gamma_1+\Gamma_2$ is energetically favored. Microscopically, the non-zero value of $\langle\Gamma_1,\Gamma_2\rangle$ means that there are chiral strings stretched between the two constituent branes, which make the decay (or the recombination if one goes in reverse) possible. One then follows the flows of the constituents, which might decay again according to the same scheme, until every end branch flows towards an attractor point. For a given total charge $\Gamma$ one might in general find several different split flow trees, and maybe also a single flow, all contributing to the total index of BPS states with this total charge. In general, the BPS spectrum of states $\Omega(\Gamma)$ remains invariant under infinitesimal variations of the background moduli $t_{\infty}$, but it can jump when the moduli are driven through a wall of marginal stability. This is intuitively clear; a certain split might either not be possible anymore because one can only reach the appropriate wall of marginal stability from the unstable side, or, alternatively, a new type of split becomes possible, as one can now reach this wall from the stable side. In such a case one has taken the background into a different \textsl{area}, and the fact that there are different basins of attraction in moduli space has led to the name \textsl{area code} for the background. Hence, an index of BPS states should more precisely be denoted as $\Omega(\Gamma,t_{\infty})$.

Apart from the walls of marginal stability which separate regions in moduli space between which a BPS index can jump there is a second type of wall that is of importance for our work: walls of \emph{threshold} stability. The distinction between these two kinds of walls was first discussed in \cite{Denef:2001ix} and is explained in more detail in \cite{deBoer:2008fk}. The threshold conditions are the same as for marginal stability
\begin{equation}
  \textrm{Re}(\bar{Z}_1Z_2)>0,\qquad \textrm{Im}(\bar{Z}_1Z_2)=0\,,
\end{equation}
however this time, the two charges are mutually local $\langle\Gamma_1,\Gamma_2\rangle =0$. Microscopically this means that there are no tachyonic strings between the two branes that can condense to merge $\Gamma_1$ with $\Gamma_2$. Most importantly, a BPS index cannot jump when crossing the threshold stability wall with the background modulus. What can happen is that the flow tree changes topology: One can imagine, for example, a three centered solution, with one `satellite' center bound to either of the two others, which then changes sides and is bound to the other center after crossing the threshold wall with the background modulus. We will be dealing with both kinds of walls later on.

The split attractor flow conjecture from \cite{Denef:2007vg} states that, for a given background $t_{\infty}$ in K\"ahler moduli space, the existence of a split attractor flow tree starting at the background $t_{\infty}$, with a given total charge $\Gamma$ and endpoints corresponding to $\Gamma_i$, is equivalent to the existence of a multicentered BPS solution in supergravity, with centers $\Gamma_i$. Additionally, the number of split flow trees and hence the total number of states with a given charge in a fixed background is conjectured to be finite, at least when charge quantization is imposed. One important subtlety which became clear recently, is that there may be a problem with this conjecture, or another refinement might be necessary. Namely, it became apparent that there are multicentered solutions which correspond to single flows, namely the so called \textsl{scaling solutions}, \cite{Denef:2007vg}. As mentioned in the introduction, scaling solutions are multicentered black holes with two (or more) centers lying so close together in spacetime, that their throats have melted together in the supergravity description. These were at the core of the deconstruction of a D4-D0 black hole in \cite{Denef:2007yt} which might be a next step towards understanding the CFT dual of an asymptotically $AdS_2$ black hole (\cite{Gaiotto:2004ij}, \cite{Gaiotto:2004pc}).

\subsection{Microscopic D-brane picture}
The analysis of this paper is performed outside of the range of validity of the supergravity approximation to type II string theory. Strictly speaking, it is the uncorrected supergravity action that entered the derivation of the split attractor flow equations from \cite{Denef:2000nb}, \cite{Denef:2000ar}. Nevertheless, some ingredients of the derivation should be valid outside of the original regime. For example, arguments leading to stability or decay of a BPS state on one or the other side of a wall of marginal stability are still based on BPS mass and $\mathcal{N}=1$ phase encoded in the central charge. As discussed in chapter \ref{mirror}, this central charge does receive quantum corrections which have to be taken into account, and, fortunately, can be taken into account. A correspondence between split flow trees and BPS states in string theory, valid outside of the large radius regime is physically supported by the quiver picture of bound states of D-branes which can be glued together through tachyon condensation when following an inverse flow \cite{Denef:2002ru}. In a flow tree, background and attractor values of the K\"ahler modulus retain their meaning in the quiver picture. A smooth interpolation between the microscopic D-brane quiver picture and the supergravity picture of multicentered solutions was given \cite{Denef:2002ru}. The quiver picture has now been placed in the broad categorial framework as reviewed in \cite{Douglas:2000gi}. We will now briefly discuss a few central points of interest.

At low energy the $\mathcal{N}=1, d=4$ quiver gauge theory describing several D-branes can be dimensionially reduced to one dimension, yielding supersymmetric quantum mechanics. Light strings strechting between the D-branes become chiral multiplets which are represented by arrows in the quiver picture and their masses can be seen as arising from a D-term potential. Let us focus on the case with two D-branes for the moment. The symplectic intersection defined in \eqref{intform} between the two charges,
\begin{equation}
  \langle\Gamma_1 ,\Gamma_2\rangle=n_{+}-n_{-},
\end{equation}
now computes the difference of positively and negatively charged bifundamental fermionic zero modes. When one approaches a marginal stability wall in moduli space from the side where
\begin{equation}
  \langle\Gamma_1 ,\Gamma_2\rangle \cdot (\alpha_1-\alpha_2)>0,
\end{equation}
where the $\alpha$'s are the respective phases, there will be tachyonic strings present between the two branes. Tachyon condensation on this side creates a bound state of total charge $\Gamma=\Gamma_1+\Gamma_2$.

The graphical representation of a flow tree associated to a charge and its constituents itself somewhat loses its direct interpretation as the variation of the CY geometry along the radial coordinate of a BPS solution. However, microscopically one can think of splitting or glueing D-branes together through tachyon condensation when following a split flow tree in moduli space. Essentially, a split flow tree belonging to the total charge $\Gamma$ and $2n$ constituents $(\Gamma_1,...,\Gamma_{2n})$ can be reduced to a set of data
\begin{equation}
  (t_{\infty};t_{1, \textrm{\tiny{split}}},...,t_{n, \textrm{\tiny{split}}};t_{1 *},...,t_{2n *}), \label{dataset}
\end{equation}
consisting of $3n+1$ points, a background value $t_{\infty}$, a set of n split points $t_{i, \textrm{\tiny{split}}}$ ($j=(1,..,n)$), and 2n attractor points, $t_{i *}$ ($j=1,...,2n$), one for each center. Note that this is based on the assumption that a charge always splits into two at a split point, however a more general situation is also possible where a charge splits into three or more constituents at a split point. This would require a slight generalization of these statements. With this notation, a single flow would just be denoted $(t_{\infty};t_{*})$. The full correspondence between these sets of data \eqref{dataset} and BPS states in string theory is referred to as the \emph{strong version of the split attractor flow conjecture}. As will be discussed later on, our results offer some nice support for a strong version of the conjecture. However, in section \ref{fibration}, we argue that one has to take into account that certain indices corresponding to split flow trees have to be calculated with greater care, as the corresponding moduli spaces do not factorize, but rather have a non-trivial fibration structure.
\section{Branes, Sheaves and Donaldson-Thomas invariants} \label{sec:branessheavesDT}
\subsection{Two different views on DT invariants}
The goal of our program is to describe all D4-D2-D0 configurations  as bound states of D6-D2-D0 and ${\rm \overline{D6}}$-D2-D0 with U(1) fluxes on them, whereby the D4 charge is induced by the latter fluxes. This will allow us to factorize the indices of D4 systems as products of D6 and $\rm{\overline{D6}}$ indices. In order to compute the BPS indices of the D6 and $\rm{\overline{D6}}$ systems, we will make use of the Donaldson-Thomas invariants. These can be understood in two complementary ways: As objects counting curves and points, which correspond to D2 and $\Dob$-branes on the D6 (or $\rm{\overline{D6}}$), or as the Witten indices of the worldvolume gauge theory on the latter. 
 
More precisely, the invariant $N_{\textrm{DT}}(\beta, n)$ computes the Witten index of a system with a D2 brane wrapping a curve of homology class $\beta$, and a collection of $\rm{\overline{D0}}$'s, such that the total D0 charge equals $n$. Although the U(1) flux on the D6 interacts with these lower branes, it does not alter the Witten index. In mathematical terms, the DT invariants compute the dimensions of the moduli spaces of the ideal sheaves corresponding to curves and points on the Calabi-Yau. They are indeed conjectured \cite{MNOP1, MNOP2} to contain equivalent information as the Gopakumar-Vafa invariants \cite{Gopakumar:1998ii, Gopakumar:1998jq}, which count the states of M2 branes with momentum, where the M2's are wrapped on holomorphic curves. 
 
The other way to view these invariants, is by treating the worldvolume gauge theory on the D6-brane as follows. It is known, that the Born-Infeld theory on a D6-brane is simply the reduction of $\mathcal{N}=1$, $d=10$ SYM on $\mathbb{R}^{4,1} \times X_{CY}$ down to $X_{CY}$. This yields a maximally supersymmetric ($\mathcal{N}=2$) $d=6$ SYM theory on the Calabi-Yau. However, because the quintic Calabi-Yau (or any \emph{proper}) Calabi-Yau manifold only admits one covariantly constant spinor, the theory is automatically topologically twisted. Nekrasov et al. \cite{Iqbal:2003ds} devised a trick to compute the path integral of this Euclidean $6$-d topological theory. By introducing a non-commutative deformation, this gauge theory now supports `small' instantons with vanishing first Chern class, but non-vanishing second Chern character, which cannot exist in an ordinary Abelian gauge theory. The size of these instantons is a modulus, just like that of non-Abelian instantons in $4$-d. One can then show that the path integral of the non-commutative theory localizes on  instantons of zero `thickness', i.e. on instantons localized on holomorphic curves and points. The Donaldson-Thomas partition function is conjectured to be the Witten index of this theory \cite{Iqbal:2003ds}. One also expects this partition function to remain unchanged after turning the non-commutative deformation off, bringing us back to the original theory we set out to describe. 

These ideal sheaves, just as ordinary Born-Infeld flux, will induce lower brane charges. Alternatively, one can think of these sheaves literally as lower dimensional brane charges. Depending on whether one is using an ideal sheaf or a dual ideal sheaf, the induced D2 charge will be dual to some curve class $\mp \beta$ (i.e. it is a $\rm \overline{D2}$ or a D2), where $\beta \in H^4(X, \mathbb{Z})$, and the D0 charge will be dual to $n\,\omega$, where
\begin{equation}
n = - \tfrac{1}{2} \, \chi(C_{\beta})+N\,,
\end{equation}
whereby $\chi(C_{\beta})$ is the ordinary Euler number of the curve in the homology class $\beta$, $N$ is the number of point-like $\Dob$'s, and $\omega$ is the volume-form of the CY.

\subsection{Main formulae for D-brane charges and indices}
Let us set our conventions for the total charge vectors for D6 and $\Dsb$-branes. The general formula for the induced charges on a D-brane (due to the WZ term in the Born-Infeld action) wrapped on a (sub)-manifold $W$ is the following:
\begin{equation}
 S^{\rm Dbrane}_{W,C} = 2 \pi \int_W C \wedge e^{-B} \, {\rm Tr} \, e^F
 \sqrt{\frac{\widehat{A}(TW)}{\widehat{A}(NW)}},
\end{equation}
where $\widehat A$ is the A-roof characteristic class, $TW$ the tangent bundle of the brane, and $NW$ its normal bundle.

In the case that will be of interest to us, that of a D6-brane carrying U(1)-flux with field-strength $F_1$, a $\Dtb$ of class $-\beta_1$ and $N_1$ $\Dob$'s, the above formula yields the following polyform:
\begin{equation}
\Gamma_{D6} = e^{F_1}\,\Big(1-\beta_1-(\tfrac{1}{2}\,\chi(C_{\beta_1})+N_1)\,\omega \Big)\,\Big(1+\frac{c_2(X)}{24}\Big)\,,
\end{equation}
where $\beta \in H^4(X, \mathbb{Z})$, and $c_2(X)$ is the second Chern class of the tangent bundle of the CY threefold $X$. Similarly, a $\Dsb$ with flux $F_2$ will bind to a D2 of class $\beta_2$ and $N_2$ $\Dob$'s to give the following total charge vector:
\begin{equation}
\Gamma_{\Dsb} = -e^{F_2}\,\Big(1-\beta_2+(\tfrac{1}{2}\,\chi(C_{\beta_2})+N_2)\,\omega \Big)\,\Big(1+\frac{c_2(X)}{24}\Big)\,,
\end{equation}
The modification with respect to the general formula is the addition of D2 and D0 charge in the form of sheaves, which can be thought of as generalizations of bundles (U(1) fluxes). Notice that the D6 will bind with a $\Dtb$, the $\Dsb$ with a D2, but both will bind to $\Dob$'s.\\

Our goal is to enumerate numbers of D4-D2-D0 BPS states using the D6 / $\Dsb$ tachyon condensation picture and split flow trees. To derive our results presented in section \ref{results} we will use the index for D4-D2-D0 BPS states of total charge $\Gamma$ from \cite{Denef:2007vg}:
\begin{equation}
  \Omega(\Gamma)=\sum_{\Gamma\rightarrow\Gamma_1+\Gamma_2}(-1)^{|\langle\Gamma_1,\Gamma_2\rangle |-1}|\langle\Gamma_1,\Gamma_2\rangle |\,\Omega (\Gamma_1)\,\Omega (\Gamma_2),
\end{equation}
with the sum running over all possible first splits $\Gamma\rightarrow\Gamma_1+\Gamma_2$, belonging to a full split flow tree, and $\langle \Gamma_1,\Gamma_2\rangle$ is the symplectic intersection of the two charges, as defined previously. The microscopic logic behind this formula is that all degrees of freedom in a D6/$\Dsb$ can be factorized as the degrees of freedom on the gauge theories of the D6 and $\Dsb$ plus the degrees of freedom of the tachyon field, which are counted by the intersection product. As we will later see, this formula will need some refinement. To evaluate the number of states for the two building blocks after the first split we will make use of the Donaldson-Thomas invariants for the quintic, which may naively be seen as counting the number of D6-D4-D2-D0 BPS states. A split will contribute a term to the index of the D4 system as follows:
\begin{equation}
  \Delta \Omega(\Gamma_{D4})=(-1)^{|\langle\Gamma_{D6},\Gamma_{\Dsb}\,\rangle |-1}\,|\langle\Gamma_{D6},\Gamma_{\Dsb}\rangle | \, N_{\textrm{DT}}(\beta_1, n_1) \, N_{\textrm{DT}}(\beta_2, n_2)\,,
\end{equation}
where 
\begin{equation}
n = \tfrac{1}{2}\,\chi(C_{\beta})+N\,.
\end{equation}
The full index for the D4 will then be constructed by adding up all possible contributions of this form. We will see, however, that this will require some care as it is not always trivial whether one is enumerating the D6-D2-D0 BPS states in the background where indeed $N_{\textrm{DT}}$ yields the correct number of states. This will be discussed in more detail, later on.

For convenience, let us state some DT invariants for the quintic in the following table, including all the ones used to derive our results.
\begin{center}
\begin{tabular}{|c||c|c|c|c|}
    \hline
    \textbf{DT invariants: quintic} & $\mathbf{n=0}$ & $\mathbf{n=1}$ & $\mathbf{n=2}$ & $\mathbf{n=3}$\\
    \hline
    \hline
    $\mathbf{\beta=0}$ & 1 & 200 & 19'500 & 1'234'000\\
    \hline
    $\mathbf{\beta=1}$ & 0 & 2875 & 569'250 & 54'921'125\\
    \hline
    $\mathbf{\beta=2}$ & 0 & 609'250 & 124'762'875 & 12'448'246'500\\
    \hline
    $\mathbf{\beta=3}$ & 0 & 439'056'375 & 76'438'831'000 & 7'158'676'736'750\\
    \hline
\end{tabular}
\end{center}
By the conjectured identity between the generating functional for  GV invariants and DT invariants \cite{Dijkgraaf:2006um}, one can easily obtain these from \cite{Klemm:2004km}, \cite{Huang:2006hq}, where the GV invariants for the quintic were calculated up to high order.

\section{Mirror symmetry and instanton corrected central charges}\label{mirror}
Since we are interested in configurations with low D-brane charges, the attractor flow equations drive the horizon size of solutions to be very small in string scale units. This brings us automatically far outside the supergravity regime, requiring higher curvature corrections. However, as will be confirmed in this paper, the main tool of analysis, namely split attractor flow tree techniques, will retain their meaningfulness as predicted by the strong split attractor flow conjecture.

Another important consequence of the low D-brane charges, is that the attractor equations will typically drive the cycles and the CY itself to stringy sizes. Worldsheet instanton corrections become important in this regime. As a consequence, the central charges of brane systems receive important worldsheet instanton corrections, which need to be taken into account. In type IIA string theory this would be an impossible task. Luckily, mirror symmetry solves the problem as the central charges are exactly determined classically by the periods of the holomorphic three-form in type IIB string theory on the mirror CY manifold. Note that the mass of a BPS saturated brane wrapping an even cycle corresponds exactly to the notion of quantum volume of that cycle from \cite{Greene:2000ci}. The general scheme set up is to identify an integral basis of three-cycles, calculate the periods of these cycles, find the explicit mirror map and in this way define the quantum volume of any even dimensional cycle of the mirror. We will handle this task by means of numerical approximations with Mathematica\footnote{We thank F. Denef for sharing his Mathematica code with us \cite{Denef:2001xn}.}. We use the mirror symmetry induced monomial-divisor map, to convert K\"ahler to complex structure modulus, and map $(D6,D4,D2,D0)$ brane systems $\Gamma_A$ into their $(D3,D3,D3,D3)$ brane mirrors $\Gamma_B$, $L:H^{2*}(X,\mathbb{Z})\rightarrow H^3(Y,\mathbb{Z})$ and analyze the attractor flows of the exact central charges in complex structure moduli space (or more precisely in the five-fold cover w-plane). An asymptotic comparison of the IIA and IIB periods near the LCS point leads to 
\begin{equation}
  L= \left( \begin{array}{cccc} -1 & 0 & 0 & 0\\0 & 1 & 0 & 0\\\frac{25}{12} & \frac{11}{2} & 1 & 0\\ 0 & -\frac{25}{12} & 0 & -1 \end{array}\right) .
\end{equation}

We proceed with some illustrative remarks on the aspects of mirror symmetry we use, and the reader interested in details is advised to consult \cite{Greene:2000ci}.

Let us start by noting some well known facts about our example. The mirror $Y$ \cite{Candelas:1990rm} of the quintic $X$ is described by an algebraic quotient of a hypersurface in $C\mathbb{P}^4$, again given by a degree five polynomial of Fermat type:
\begin{equation}
  \sum_{i=1}^5x_i^5-5\psi x_1x_2x_3x_4x_5=0,
\end{equation}
where $\psi$ denotes the complex structure modulus (mirror to the K\"ahler modulus). $Y$ is now defined under the identifications $x_i \equiv \omega^{k_i}x_i$ with $\omega =e^{2\pi i/5}$ and $\sum_i k_i=0, k_i\in \mathbb{Z}$.

Mirror symmetry relates the even cycles of real dimension $2j$ on the CY $X_A$ to the three-cycles $\gamma_{i}$ of the mirror CY manifold $X_B$. The periods $\Pi_i = \int_{\gamma_i} \Omega$, of the holomorphic three-form on the $\gamma_{i}$ have leading logarithmic behavior $\textrm{log}^j(z)$ near $z=0$ (LCS point), using the coordinate $z=\psi^{-5}$. These periods are solutions to the generalized hypergeometric equation, known as the Picard-Fuchs equation,
\begin{equation}
  \Big{[}z\,\partial_z \prod_{i=1...q}(z\,\partial_z +\beta_i-1)-z\prod_{j=1...p}(z\,\partial_z+\alpha_j)\Big{]}u=0,
\end{equation}
where the $\alpha_j$ and the $\beta_i$ are model dependent constants, which in the case of the quintic are  $\alpha_j =\frac{j}{5}, \beta_i=1$, $j=1,...,4,  i=1, 2, 3$. 
A class of solutions is given in terms of Meijer-functions (G-functions) $U_j(z)$, which can be expressed as
\begin{equation}
  U_j(z)=\frac{1}{(2\pi i)^j} \oint \frac{\Gamma (-s)^{j+1}\prod_{i=1}^4\Gamma(s+\alpha_i)((-1)^{j+1}z)^s}{\Gamma (s+1)^{3-j}} ds,
\end{equation}
for the quintic. This particular basis of periods is related to three branching points (LCS point, conifold point and Gepner point) which are connected by appropriately chosen branch cuts. This is important, as we will deal with attractor flows crossing these branch cuts in moduli space throughout our analysis.  Associated to these branch cuts are three types of monodromies, which can be expressed as matrices acting on the periods. The monodromy $T(0)$ around the LCS point ($z=0 ,\psi=\infty$) and the monodromy $T(\infty )$ around the Gepner point ($z=\infty ,\psi =0$) act on the period vector ${\bf U}(z)=(U_j(z))_{j=1...4}$ as follows,
\begin{eqnarray}
  {\bf U}(e^{2\pi i}z) &=&\, T(0)\, {\bf U}(z),\qquad |z| << 1 \nonumber\\
  {\bf U}(e^{2\pi i}z) &=&T(\infty )\, {\bf U}(z),\qquad  |z|>> 1 .
\end{eqnarray}
The third monodromy matrix of course follows directly from the other two, as a monodromy can always be seen either as `around one of the branching points' or, equivalently as a monodromy `around the two other branching points' in the appropriate directions. For the monodromy around the conifold point one has
\begin{eqnarray}
  T(1)&=&T(\infty )T(0)^{-1}\qquad \textrm{Im}(z)<0,\nonumber\\
  T(1)&=&T(0)^{-1}T(\infty )\qquad \textrm{Im}(z)>0 .
\end{eqnarray}
Adhering to the conventions in \cite{Denef:2001xn}, we will work with a period basis defined by the vector ${\bf \Pi} = L {\bf U}$, where $L$ is the following matrix:
\begin{equation}
  L=\frac{8\,i\,\pi^2}{125}\,\left( \begin{array}{cccc} 0 &5 & 0 & 5\\0 & 1 & -5 & 0\\0 & -1 & 0 & 0\\1 & 0 & 0 & 0\end{array}\right) .
\end{equation}
Knowing the exact periods one can calculate the K\"ahler potential
\begin{equation}
  e^{-K} = i\int_Y\Omega_B\wedge\overline{\Omega_B} = i\,{\bf \Pi}^{\dagger}(z)\cdot I^{-1} \cdot {\bf \Pi}(z)\,,
\end{equation}
where $\Omega_B$ is the holomorphic $(3,0)$-form of type a IIB CY compactification $\Omega_B$, and $I$ is the symplectic intersection matrix given by
\begin{equation}
  I=\left( \begin{array}{cccc} 0 & 0 & 0 & -1\\0 & 0 & 1 & 0\\0 & -1 & 0 & 0\\1 & 0 & 0 & 0\end{array}\right) \,.
\end{equation}
The fully quantum corrected central charges of a IIB charge vector ${\bf q} = (p^0, p, q, q_0) $ can now be computed as follows:
\begin{equation}
Z(\Gamma) = e^{K/2}\, {\bf q} \cdot {\bf \Pi}\,.
\end{equation}

\section{Enumeration of D4-D2-D0 BPS states on the quintic}\label{results}
In this section we present our results on split flow trees and BPS indices of a type IIA string theory compactification on the quintic. Our main results offer strong evidence that the split attractor flow conjecture holds accurately for the BPS states under investigation, suggesting a `strong' correspondence between flow tree data and BPS solutions of the worldsheet instanton corrected theory. We will see, however, that one has to take care of some subtleties in order to establish a correspondence for all charge systems considered. We will also compare our results to the calculations of the modified elliptic genus done by means of geometric and CFT techniques, and by the use of modular invariance.

Let us first get acquainted with the relevant information about the modified elliptic genus. By growing the M-theory circle, the D4-D2-D0 configurations oxidize to M5-M2-KK-mode configurations. In the limit where the M-theory circle is large with respect to the CY threefold, the system is effectively described by a string. The corresponding CFT on it is referred to as the MSW CFT \cite{Maldacena:1997de}. The supersymmetric index of this theory, the modified elliptic genus, is constrained to be a \emph{weak Jacobi form}, as explained in \cite{Dijkgraaf:2000fq}. Roughly speaking, this mathematical object is a function $Z(q, \bar q, z)$ of two complex variables, $q$ and $z$, that behaves as a modular form with respect to $q$ and as an elliptic function with respect to $z$. This weak Jacobi form can be decomposed in a basis of theta-functions as follows:
\begin{equation}
Z(q, \bar q, z) = \sum_{k=0}^{4} Z_k(q)\, \Theta_k (\bar q, z)\,,
\end{equation}
where the theta-function is given by
\begin{equation}
\Theta_k (\bar q, z) = \sum_n (-1)^{n+k}\,\bar q^{\frac{5}{2}\,(n+\frac{k}{5}+\frac{1}{2})^2}\, z^{5\,n+k+\frac{5}{2}}\,,
\end{equation}
and the $Z_{k}(q)$ are meromorphic functions of the variable $q$. The sum over the variable $k$ corresponds physically to a sum over possible $U(1)$ fluxes on the worldvolume of the D4-brane that cannot be written as the pullback of some two-form on the ambient space. If $k$ reaches five, then the flux can be represented as the pullback $\imath_*(H)$ of the hyperplane class on the CY space. Hence, the sum terminates at four. The $Z_k$'s are organized as follows: $Z_0$ corresponds to a sum over states with no added D2-charge, and increasing D0-charge as the powers of $q$ increase. Each coefficient in the $q$-expansion corresponds to the index of a state with fixed D0-charge. Similarly, the $Z_k$'s correspond to states with $k$ units of added D2-charge. Schematically, the first few terms of $Z_k$ will look as follows
\begin{equation}
Z_k(q) = q^{-\alpha} (\#+ \# \, q+ \# \, q^2 + \ldots)\,,
\end{equation}
where $\hat{q}_0 \equiv q_0-\frac{1}{2}D^{AB}q_Aq_B$, and $\alpha$ is the higher possible value of $\hat{q}_0$ for a given $k$.
We will only concern ourselves with the $Z_k(q)$ functions, as they contain all the relevant information. A stringent mathematical property of weak Jacobi modular forms is the fact that they are entirely determined by their \emph{polar part}, i.e. terms with negative powers of $q$. These terms correspond to charge configurations that satisfy $\hat{q}_0>0$, where the variables are defined in \eqref{gammadef}. We refer to such configurations as \emph{polar states}.

Whether a charge is polar or non-polar has important implications for the existence of attractor flows in the large volume regime. If the central charge $Z(\Gamma, t)$ vanishes on a regular point in the moduli space, then no single-centered solution exists. In the large radius approximation, the central charge of system with $(D4,D2,D0)$ charge $(p,q,q_0)$ reads $Z=\langle \Omega,\Gamma \rangle = -\frac{5p}{2} t^2+q t-q_0$. Under a shift of the $B$-field, $B\rightarrow B-\frac{1}{5}q$, this becomes $Z=-\frac{5p}{2}t^2-\hat{q}_0$. Of course, for this shift in the B-field not to affect the BPS spectrum one has to assume that the background K\"ahler modulus does not cross a wall of marginal stability. Writing the K\"ahler modulus as $t=B+i J$ and setting $Z=0$ leads to $B \cdot J=0$ and $\frac{1}{2}(J^2-B^2)=\hat{q}_0$, from which one can deduce that one needs $\hat{q}_0>0$ to have a solution\footnote{Recall that $J$ is in the self-fual part of $H^2(P,\mathbb{Z})$.}. Thus, it turns out that in the large radius approximation, $\hat{q}_0>0$ is the condition for a single flow to crash at a regular point in moduli space, whereas for $\hat{q}_0<0$ the flow will terminate at a regular attractor point. We will see shortly that this is no longer true when working in the regime where the central charge receives dominant instanton corrections.

Clearly, there is no algorithm to find all attractor flows for the charge systems under investigation. However, we did follow a certain set of rules when gathering all attractor flows contributing to an index. One might very loosely say that we chose the most convenient and least complicated topological background sector in moduli space, but let us be a bit more elaborate on that. First of all, we searched for a value of the background complexified K\"ahler modulus where no single flow exists for a given system. We then searched for splits into a single D6-brane (potentially with U(1) flux, D2 ($\Dtb$) and $\Dob$-branes) and a single $\Dsb$ (again potentially carrying lower-dimensional brane charge) and we restricted our attention to D2's (or $\Dtb$'s) wrapped on holomorphic curves of up to degree three, either rational or elliptic. A next criterion was to select a topological sector of the background where there is no single flow, but only split flows are possible. We focused our search on split flows into single D6/$\Dsb$ pairs. However, we also looked for splits into higher rank stacks, whenever it seemed reasonable to expect these, but did not find any.

Note that when using a DT-invariant to count a number of D6-D2-D0 states, one must be careful not to overcount the states involved in a flow tree. Such issues are discussed in more detail on a case by case basis in what follows and in the appendix.
 
\subsection{Polar states: $\hat{q}_0>0$}
  
We now analyze the polar states, moving from more polar to less polar. The following table, \ref{polarcharges}, summarizes the polar charge systems with one $p=1$ magnetic D4-brane wrapped on a hyperplane class divisor of the quintic. We label charge systems by their deviation in D2-brane charge $\Delta q$ and D0-brane charge $\Delta q_0$ as measured from the most polar state with charge vector
\begin{equation}
  \Gamma=H+\frac{1}{2}H^2+\frac{7}{12}H^3,
\end{equation}
which means D2-brane charge $q=\frac{5}{2}$ and D0-brane charge $q_0=\frac{35}{12}$. Thus, $\hat{q}_0=q_0-\frac{1}{10}q^2=\frac{55}{24}$. In the `charge shift' notation it is denoted as $\Delta q=0, \Delta q_0=0$.
\begin{center}\label{polarcharges}
\begin{tabular}{|c|c|c||c|c|}
    \hline
    \multicolumn{5}{|c|} {\textbf{Polar brane systems}}\\
    \hline
    \multicolumn{3}{|c||}{D2/D0 shifts} & Total charge & \footnotesize{Reduced D0 brane charge $\hat{q}_0$}\\
    \hline
    \hline
    1. & $\mathbf{\Delta q=0}$ & \!\!\!\!\!$\mathbf{\Delta q_0=0}$ & $H+\frac{1}{2}H^2+\frac{7}{12}H^3$ & \,\,$55/24\approx 2.29$\\
    \hline
    2. & $\mathbf{\Delta q=0}$ & $\mathbf{\Delta q_0=-1}$ & $H+\frac{1}{2}H^2+\frac{23}{60}H^3$ & \,\,$31/24\approx 1.29$\\
    \hline
    3. & $\mathbf{\Delta q= 1}$ & $\mathbf{\Delta q_0=-1}$ & $H+\frac{7}{10}H^2+\frac{23}{60}H^3$ & $83/120\approx 0.69$\\
    \hline
    4. & \;\, $\mathbf{\Delta q=-1}$ & $\mathbf{\Delta q_0=-2}$ & $H+\frac{3}{10}H^2+\frac{11}{60}H^3$ & $83/120\approx 0.69$\\
    \hline
    5. & $\mathbf{\Delta q=0}$ & $\mathbf{\Delta q_0=-2}$ & $H+\frac{1}{2}H^2+\frac{11}{60}H^3$ & \quad $7/24\approx 0.29$\\
    \hline
\end{tabular}
\end{center}
\vskip 2mm
In the following, we will use small pictograms to denote our flow trees. Indices corresponding to a flow tree will be calculated right after the first split point, as in \cite{Denef:2007vg}. The various branches of a pictogram will be denoted by the corresponding brane charge, where we denote D6- or D4-branes with a subscript indicating possible worldvolume flux and bound (anti-) D2- or D0-branes. When stating a flux dual to a curve on a D4-brane or stating the type of curve on which one wraps a (anti-) D2-brane, the curves of degree $d$ will be denoted by $C_d$ when they are rational (i.e. genus zero), otherwise the genus will be explicitly stated.

\begin{enumerate}
\item $\mathbf{\Delta q=0, \Delta q_0=0}$:\\
 We start with the most polar state: a pure D4-brane ($\hat{q}_0\approx 2.29$). One needs to turn on a half-integer flux for anomaly cancellation as the hyperplane class divisor does not support a spin structure, \cite{Minasian:1997mm}, \cite{Freed:1999vc}. This pure fluxed D4-brane has charge $\Gamma = H+\frac{H^2}{2}+(\frac{\chi (P)}{24}+\frac{1}{2}F^2)\,\omega =H+\frac{1}{2}H^2+\frac{7}{12}H^3 $ where $\omega \in H^6(X,\mathbb{Z})$ denotes the volume form on $X$. One finds a split flow tree with two endpoints, corresponding to charges
\begin{eqnarray*}   
\Gamma_1&=&1+H+\frac{H^2}{2}+\frac{c_2(X)}{24}+\frac{H^3}{6}+\frac{c_2(X)\cdot H}{24},\\   \Gamma_2&=&-1-\frac{c_2(X)}{24}.
\end{eqnarray*}
It looks and is enumerated as follows:
\begin{figure}[h]
\setlength{\unitlength}{1cm}
\centering 
\begin{picture}(3,2.3)
%\put(0,0) {\tfbox{\includegraphics[width=0.25\textwidth,angle=0]{00}}}
\put(0,0) {\includegraphics[width=0.25\textwidth,angle=0]{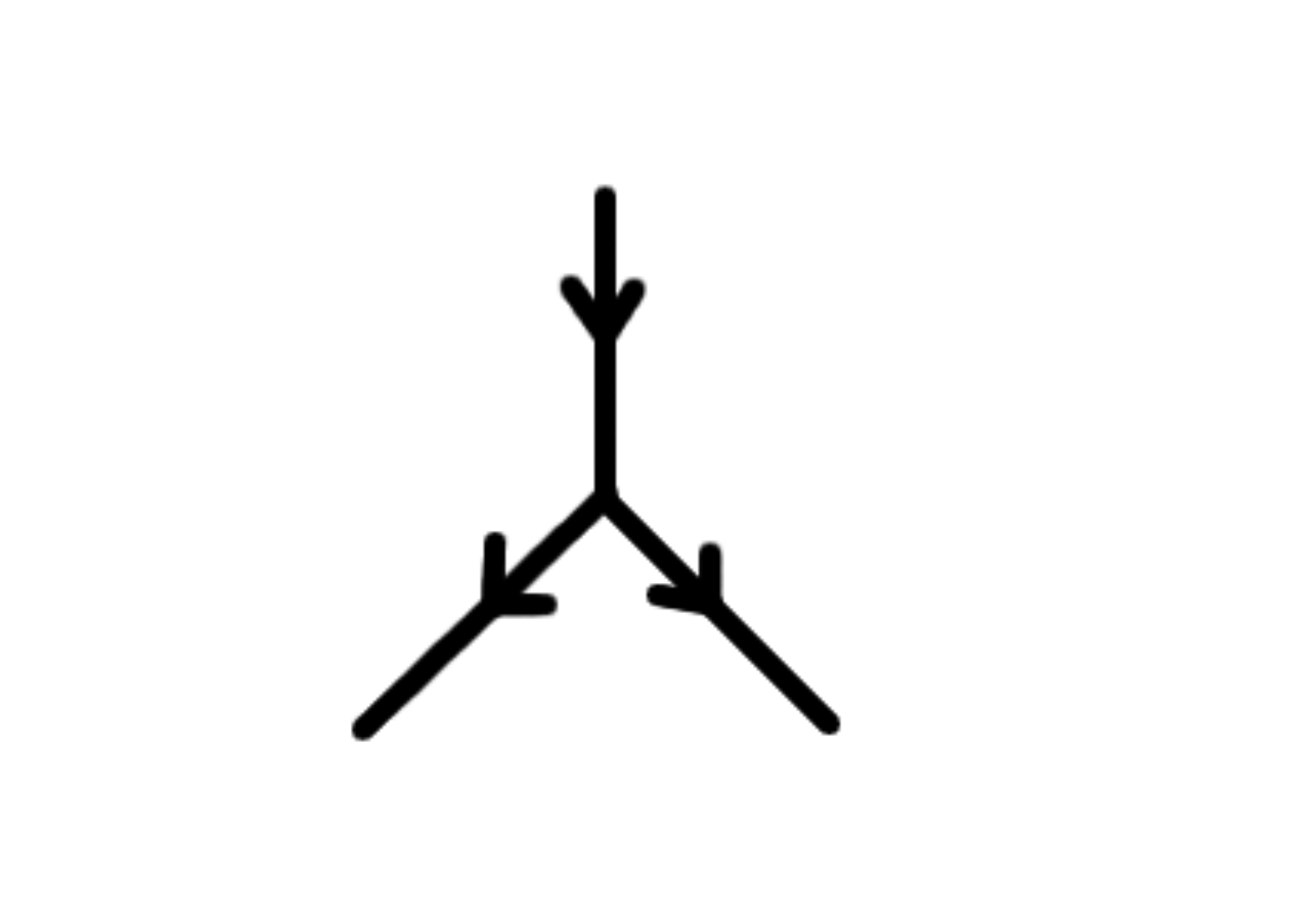}}
\put(1.5,2.25) {$D4_{\frac{H}{2}}$}
\put(0.6,0.15) {$D6_H$}
\put(2.25,0.15) {$\overline{D6}$}
\put(4,1.3) {$=$}
\put(5,1.3) {$5$.}
\end{picture}
%\caption{$\Omega =5$.}
\end{figure}

The first center is a D6-brane with one unit of worldvolume flux turned on, the second center is a pure $\Dsb$. Note that the correction from the second Chern class of the quintic $c_2(X)$ on D2- and D0-brane charges has been taken into account for both centers. As discussed in \cite{Denef:2001xn}, the attractor points for these centers lie on copies of the conifold singularity. They might be viewed as some `microscopic building blocks' of an empty hole. The BPS index reads
\begin{equation}
  \Omega=(-1)^{|\langle\Gamma_1,\Gamma_2\rangle|-1}|\langle\Gamma_1,\Gamma_2\rangle|N_{\textrm{DT}}(0,0)\cdot N_{\textrm{DT}}(0,0)=(-1)^4\cdot 5\cdot 1\cdot 1=5.
\end{equation}

\item $\mathbf{\Delta q=0, \Delta q_0=-1}$:\\
The next polar state is obtained by binding one $\Dob$-brane $N=1$ to the D4-brane. This system has total charge $\Gamma = H+\frac{H^2}{2}+(\frac{\chi (P)}{24}+\frac{1}{2}F^2-N)\,\omega =H+\frac{H^2}{2}+\frac{23}{60}H^3 $. We find a split flow tree with three endpoints corresponding to a D6-brane with one unit of worldvolume flux, a pure $\Dsb$ and an $\Dob$. Depending on the background value of the K\"ahler modulus, the $\Dob$-brane is on the side of the D6 or of the $\Dsb$. There is a threshold wall for the $\Dob$ running through the fundamental wedge. We find what is expected. The $\Dob$ binds to the D6 on one side of the TH wall and it binds to the $\Dsb$ on the other side of the wall. One possibility is the flow tree of the type
\vskip 2mm
\begin{figure}[h!]
\setlength{\unitlength}{1cm}
\centering 
\begin{picture}(3,2.3)
%\put(0,0) {\tfbox{\includegraphics[width=0.25\textwidth,angle=0]{00}}}
\put(0,0) {\includegraphics[width=0.25\textwidth,angle=0]{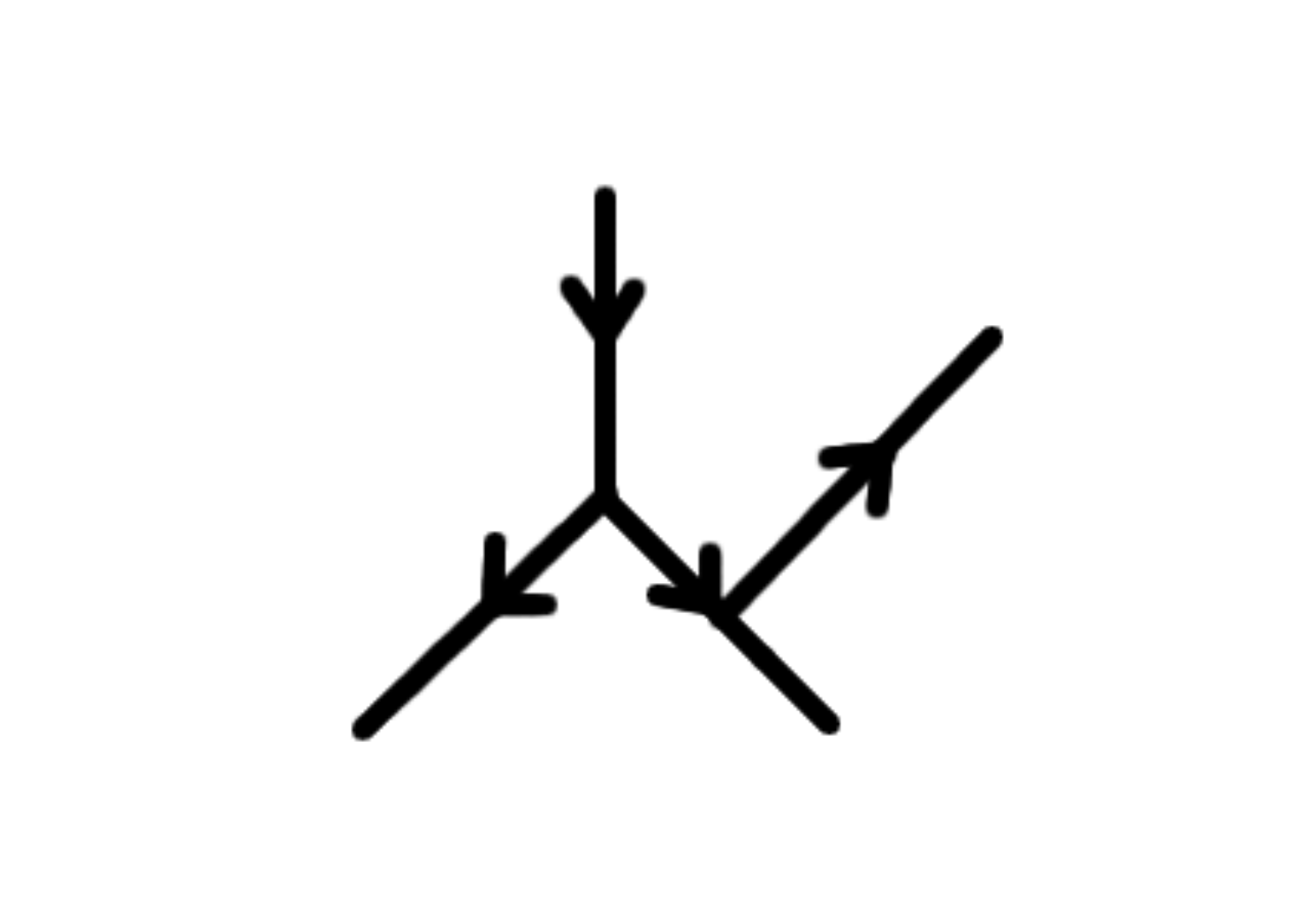}}
\put(1,2.35) {$D4_{\frac{H}{2},\overline{D0}}$}
\put(0.6,0.15) {$D6_H$}
\put(2.25,0.15) {$\overline{D6}$}
\put(2.65,1.8) {$\overline{D0}$}
\put(4,1.3) {$=$}
\put(5,1.3) {$-800$,}
\end{picture}
\end{figure}

leading to the BPS index
\begin{equation*}
  \Omega=(-1)^{|\langle\Gamma_1,\Gamma_2\rangle|-1}|\langle\Gamma_{1},\Gamma_2\rangle|N_{\textrm{DT}}(0,1)\cdot N_{\textrm{DT}}(0,0)=(-1)^3\cdot 4\cdot 200\cdot 1=-800.
\end{equation*}
Note, that in calculating this index we have now in some sense approximated the degeneracy (index) by treating the whole flow tree as though it corresponded to a two-centered solution (instead of a three-centered one). The physical justification for this, which is based on the supergravity picture (which is a priory not valid at low charge), is that, during the first split, the $\Dsb$ (or the D6-brane, depending on the side of the threshold wall) with a bound $\Dob$ behaves approximately as a single particle in spacetime, since the distance between the $\Dsb$ and the $\Dob$ is negligible compared to the distance between the D6 and the $\Dsb$. Of course, this does not immediately tell us how one would go about computing the exact index for a three-centered solution. We will follow this logic throughout our computations. In this case here, the DT invariant correctly counts the number of BPS states. It can and will happen, though, that one has to work harder to get the correct index for a whole flow tree. In that case a more detailled area code analysis becomes relevant.\\
\item $\mathbf{\Delta q=1, \Delta q_0=-1}$:\\
The next state ($\hat{q}_0\approx 0.69$) is obtained in the D4-picture by turning on a flux dual to a degree one rational curve. This leads to the total charge  $\Gamma = H+\frac{7}{10}H^2+(\frac{\chi (P)}{24}+\frac{1}{2}F^2)\omega =H+\frac{7}{10}H^2+\frac{23}{60}H^3$. We find a split flow tree ending with a D6 with one unit of worldvolume flux and a $\Dsb$ with a sheaf corresponding to a D2-brane wrapped on a degree one rational curve. The holomorphic Euler character for this curve is $n_2=\chi_h(C_1)=1$, so we set $(\beta_2,n_2)=(1,1)$.  The split flow tree looks like
\vskip 2mm
\begin{figure}[h!]
\setlength{\unitlength}{1cm}
\centering 
\begin{picture}(3,2.3)
%\put(0,0) {\tfbox{\includegraphics[width=0.25\textwidth,angle=0]{00}}}
\put(0,0) {\includegraphics[width=0.25\textwidth,angle=0]{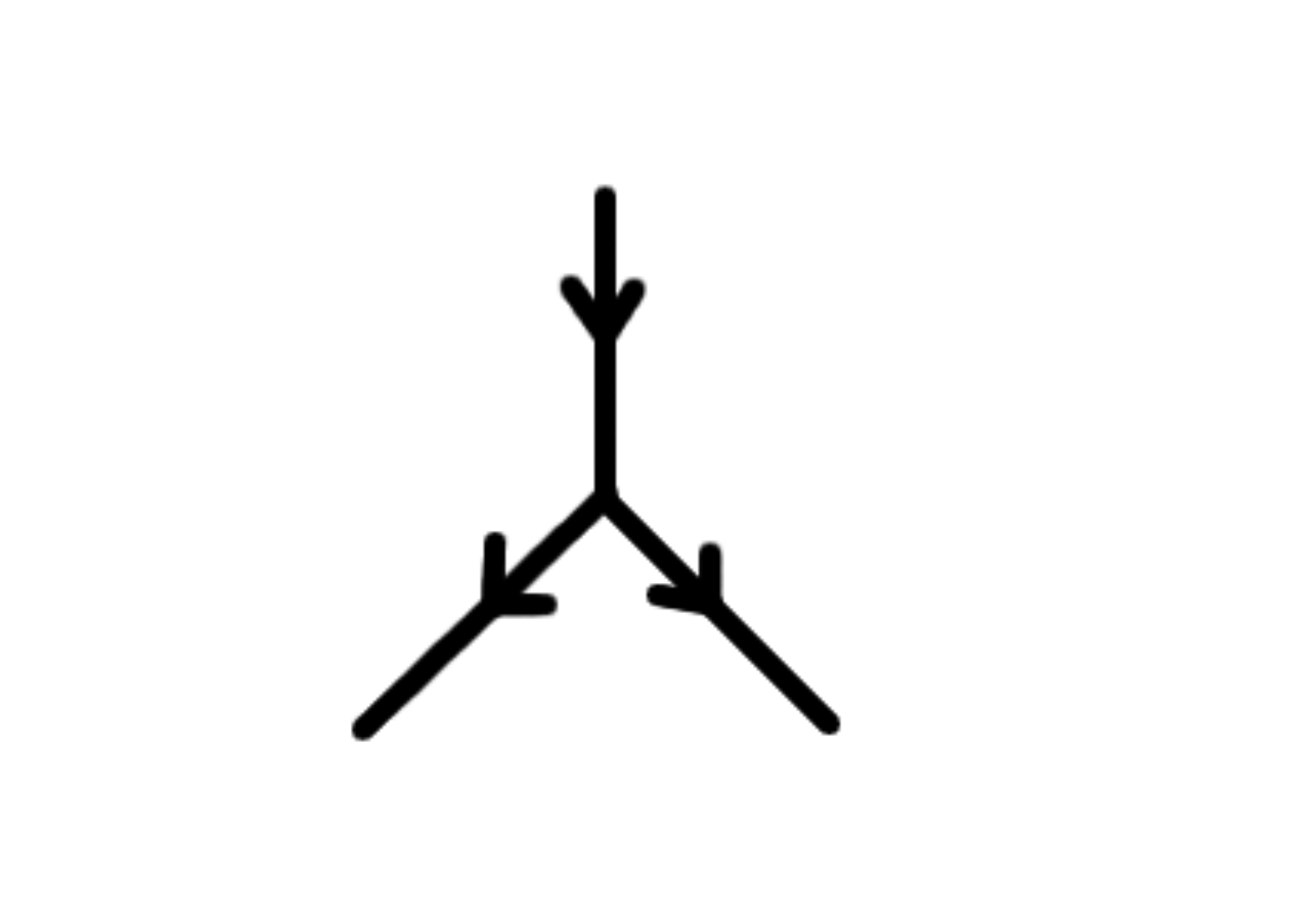}}
\put(1,2.35) {$D4_{\frac{H}{2}+C_1}$}
\put(0.6,0.15) {$D6_H$}
\put(2.25,0.15) {$\overline{D6_{D2(C_1)}}$}
\put(4,1.3) {$=$}
\put(5,1.3) {$8'625$,}
\end{picture}
\end{figure}

which leads to the BPS index
\begin{equation*}
  \Omega=(-1)^{|\langle\Gamma_1,\Gamma_2\rangle|-1}|\langle\Gamma_{1},\Gamma_2\rangle|N_{\textrm{DT}}(0,0)\cdot N_{\textrm{DT}}(1,1)=(-1)^2\cdot 3\cdot 1\cdot 2'875=8'625.
\end{equation*}

\item $\mathbf{\Delta q=-1, \Delta q_0=-2}$:\\
Similarly there is the state ($\hat{q}_0\approx 0.69$) with total charge $\Gamma = H+\frac{3}{10}H^2+(\frac{\chi (P)}{24}+\frac{1}{2}F^2)\omega =H+\frac{3}{10}H^2+\frac{11}{60}H^3$. We find the split flow tree with a D6 with one unit of worldvolume flux and a sheaf corresponding to an $\Dtb$ wrapped on a degree one rational curve on one side, and an $\Dsb$ on the other side, as centers. We therefore set $(\beta_1,n_1)=(1,-1)$. The split flow tree looks like
\pagebreak
\vskip 2mm
\begin{figure}[h!]
\setlength{\unitlength}{1cm}
\centering 
\begin{picture}(3,2.3)
%\put(0,0) {\tfbox{\includegraphics[width=0.25\textwidth,angle=0]{00}}}
\put(0,0) {\includegraphics[width=0.25\textwidth,angle=0]{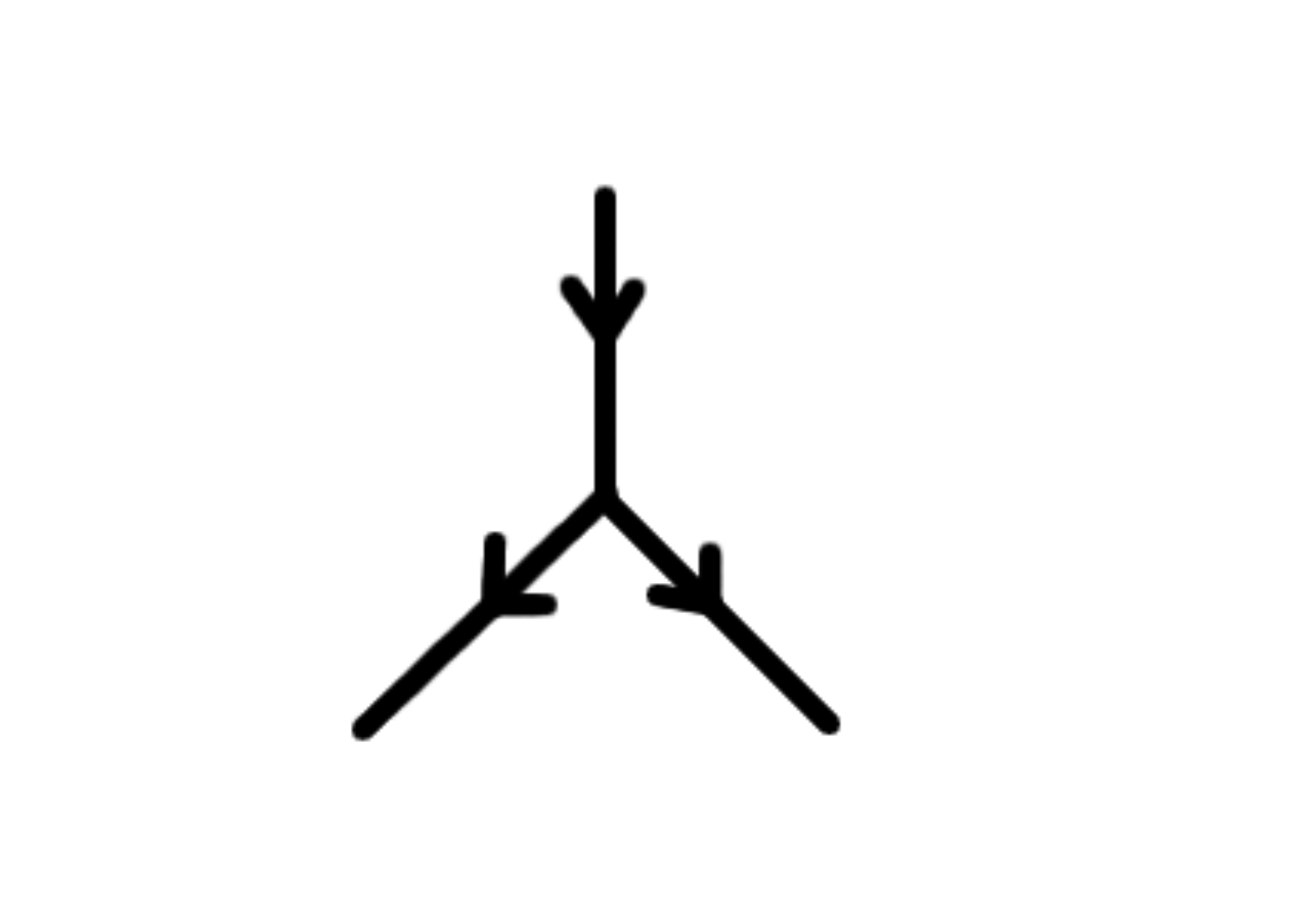}}
\put(1,2.35) {$D4_{\frac{H}{2}+\overline{C_1}}$}
\put(0.2,0.18) {$D6_{H,\overline{D2}(C_1)}$}
\put(2.25,0.15) {$\overline{D6}$}
\put(4,1.3) {$=$}
\put(5,1.3) {$8'625$,}
\end{picture}
\end{figure}

 and we obtain the BPS index
\begin{equation*}
  \Omega=(-1)^{|\langle\Gamma_1,\Gamma_2\rangle|-1}|\langle\Gamma_{1},\Gamma_2\rangle|N_{\textrm{DT}}(1,1)\cdot N_{\textrm{DT}}(0,0)=(-1)^2\cdot 3\cdot 1\cdot 2'875=8'625.
\end{equation*}

\item $\mathbf{\Delta q=0, \Delta q_0=-2}$:\\
Next, one can bind two $N=2$ $\Dob$'s to the D4. This yields a total charge $\Gamma = H+\frac{H^2}{2}+(\frac{\chi (P)}{24}+\frac{1}{2}F^2-N)\omega =H+\frac{H^2}{2}+\frac{11}{60}H^3 $. One finds one split flow tree of the schematic form
\begin{figure}[h]
\setlength{\unitlength}{1cm}
\centering 
\begin{picture}(3,2.3)
%\put(0,0) {\tfbox{\includegraphics[width=0.25\textwidth,angle=0]{00}}}
\put(0,0) {\includegraphics[width=0.25\textwidth,angle=0]{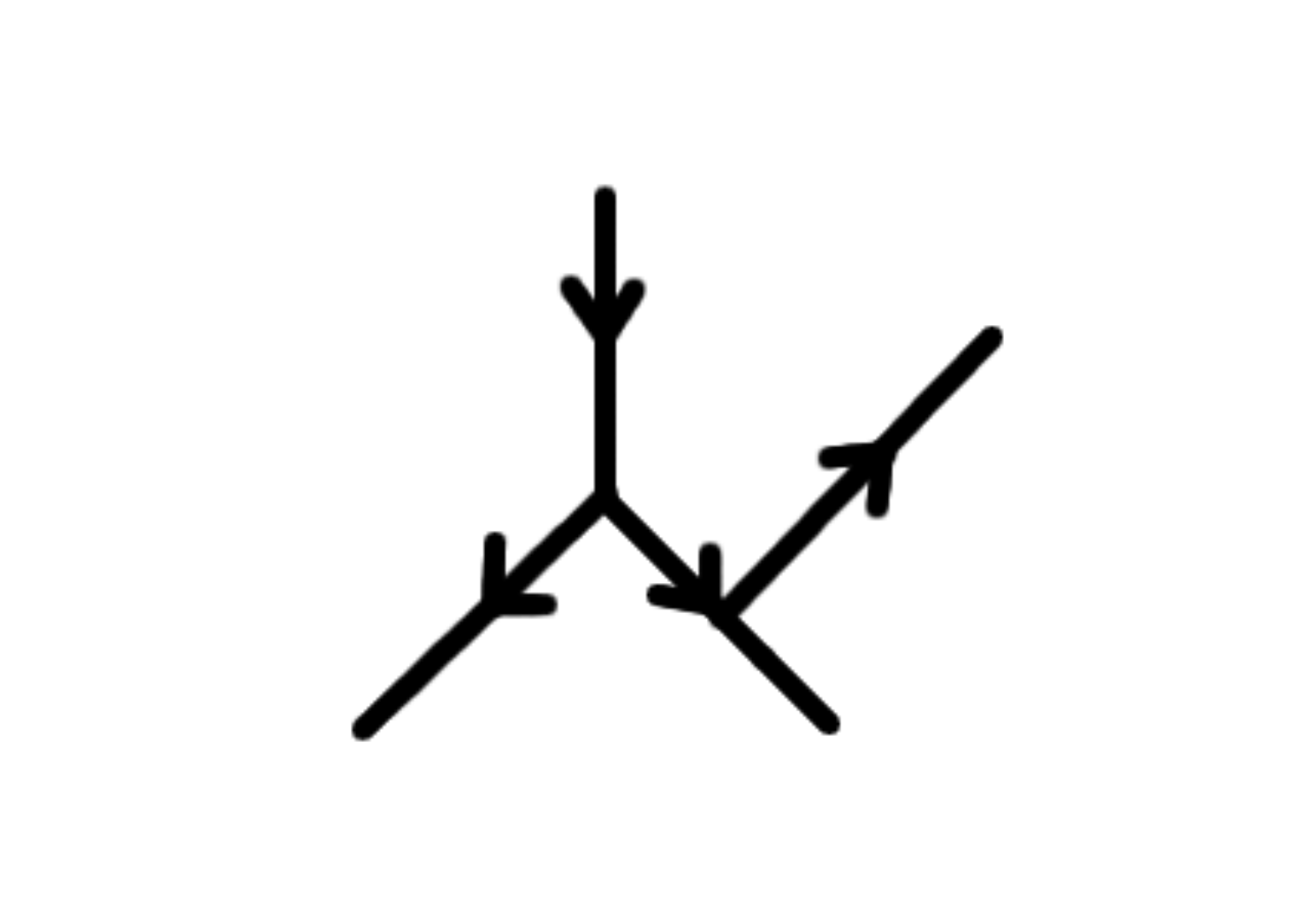}}
\put(1,2.35) {$D4_{\frac{H}{2},2 \overline{D0}}$}
\put(0.6,0.15) {$D6_H$}
\put(2.25,0.15) {$\overline{D6}$}
\put(2.65,1.8) {2 $\overline{D0}$}
\put(5,1.3) {$=58'500$,}
\end{picture}
\end{figure}

and one is led to the BPS index 
\begin{equation*}
  \Omega=(-1)^{|\langle\Gamma_1,\Gamma_2\rangle|-1}|\langle\Gamma_{1},\Gamma_2\rangle|N_{\textrm{DT}}(0,2)\cdot N_{\textrm{DT}}(0,0)=(-1)^2\cdot 3\cdot 19'500\cdot 1=58'500.
\end{equation*}
\end{enumerate}
One can check that all the results for the polar states exactly match the corresponding numbers from the elliptic genus obtained in \cite{Gaiotto:2006wm}.
\subsection{Non-polar states: $\hat{q}_0<0$}
Here is a summary of the non-polar charges we will consider.
\begin{center}
\begin{tabular}{|c|c|c||c|c|}
    \hline
    \multicolumn{5}{|c|} {\textbf{Non-polar brane systems}}\\
    \hline
    \multicolumn{3}{|c||}{D2/D0 shifts} & Total charge & \footnotesize{Reduced D0 brane charge $\hat{q}_0$}\\
    \hline
    \hline
    6. & $\mathbf{\Delta q=2}$ & $\mathbf{\Delta q_0=-1}$ & $H+\frac{9}{10}H^2+\frac{23}{60}H^3$ & $-13/120\approx -0.11$\\
    \hline
    7. & $\mathbf{\Delta q=1}$ & $\mathbf{\Delta q_0=-2}$ & $H+\frac{7}{10}H^2+\frac{11}{60}H^3$ & $-37/120\approx -0.31$\\
    \hline
    8. & $\mathbf{\Delta q=0}$ & $\mathbf{\Delta q_0=-3}$ & $H+\frac{1}{2}H^2+\frac{-1}{60}H^3$ & \quad $17/24\approx -0.71$\\
    \hline
    9. & $\mathbf{\Delta q=2}$ & $\mathbf{\Delta q_0=-2}$ & $H+\frac{9}{10}H^2+\frac{11}{60}H^3$ & \;\, $83/120\approx -1.11$\\
    \hline
\end{tabular}
\end{center}

\begin{enumerate}
\setcounter{enumi}{5}

\item $\mathbf{\Delta q=2, \Delta q_0=-1}$:\\
The first non-polar state ($\hat{q}_0\approx -0.11$) can be obtained in the D4-picture by turning on flux dual to a degree two rational curve. This system has charge $\Gamma = H+\frac{9}{10}H^2+(\frac{\chi (P)}{24}+\frac{1}{2}F^2)\omega =H+\frac{9}{10}H^2+\frac{23}{60}H^3 $. One finds the split flow tree
\pagebreak
\begin{figure}[h!]
\setlength{\unitlength}{1cm}
\centering 
\begin{picture}(3,2.3)
%\put(0,0) {\tfbox{\includegraphics[width=0.25\textwidth,angle=0]{00}}}
\put(0,0) {\includegraphics[width=0.25\textwidth,angle=0]{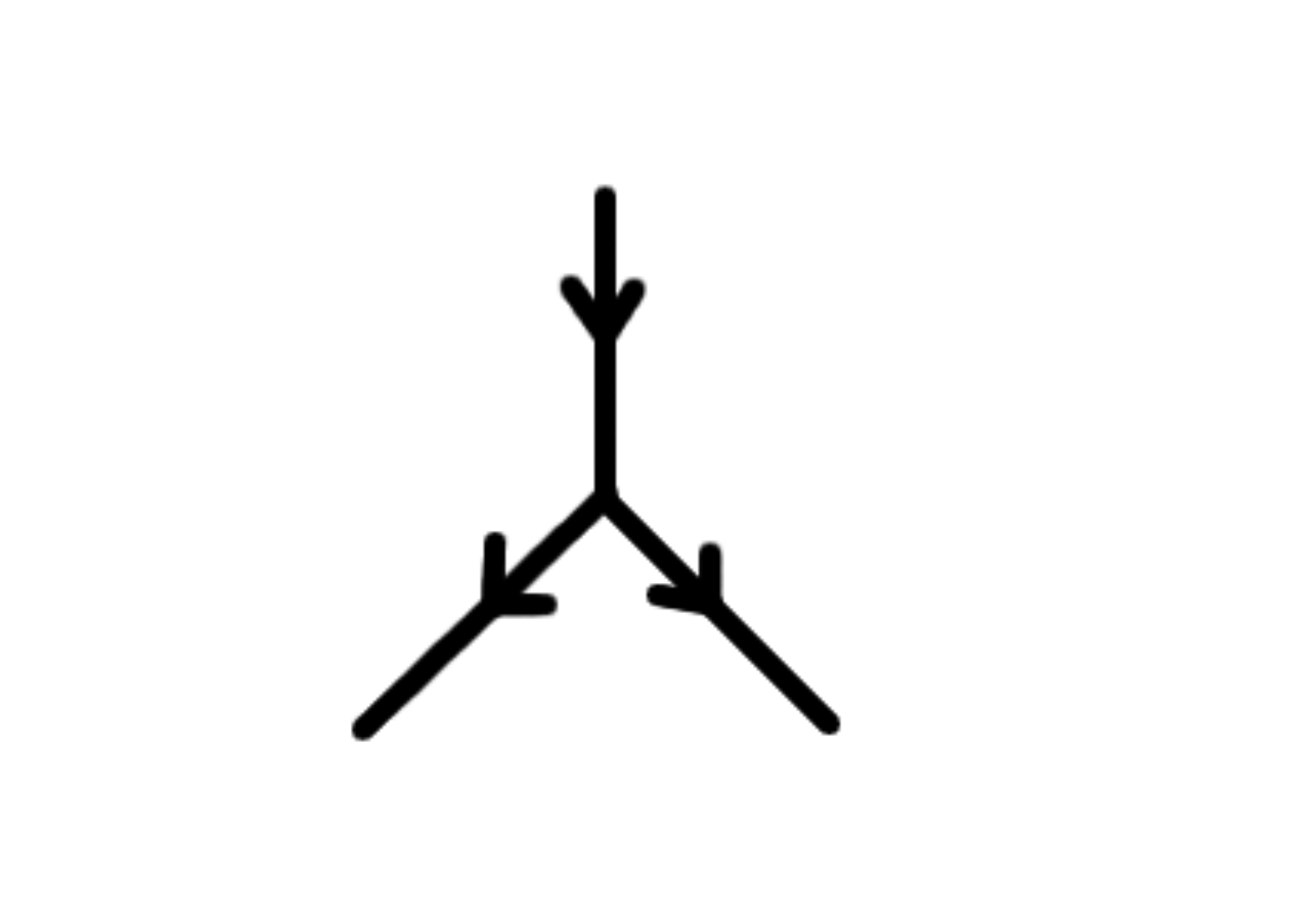}}
\put(1,2.35) {$D4_{\frac{H}{2}+C_2}$}
\put(0.6,0.15) {$D6_H$}
\put(2.25,0.15) {$\overline{D6_{D2(C_2)}}$}
\put(4,1.3) {$=$}
\put(5,1.3) {$-1'218'500$,}
\end{picture}
\end{figure}

which leads to the index
\begin{equation*}
  \Omega=(-1)^{|\langle\Gamma_1,\Gamma_2\rangle|-1}|\langle\Gamma_{1},\Gamma_2\rangle|N_{\textrm{DT}}(0,0)\cdot N_{\textrm{DT}}(2,1)=(-1)^1\cdot 2\cdot 1\cdot 609'250=-1'218'500.
\end{equation*}

\item $\mathbf{\Delta q=1, \Delta q_0=-2}$:\\
The next state ($\hat{q}_0\approx -0.31$) can be viewed as a D4-brane with worldvolume flux dual to a degree one rational curve as well as ($N=1$) one $\Dob$, corresponding to the total charge $\Gamma = H+\frac{7}{10}H^2+(\frac{\chi (P)}{24}+\frac{1}{2}F^2-N)\omega =H+\frac{7}{10}H^2+\frac{11}{60}H^3 $. This sytem is the first with a non-trivial area code. The correct index receives different contributions in different background regimes. The reader interested in the details of the area code is referred to the appendix \ref{areacode12}. Tuning the background to the simplest case, two split flow trees contribute. First, there is the split into a D6-brane with one unit of worldvolume flux, a pure $\Dsb$ and a D2-D0 halo particle. The halo is on the anti-D6 side after the first split. Second, there is the split into the D6-brane with one unit of worldvolume flux, a $\Dsb$ with sheaf corresponding to a D2 on a degree one rational curve and one $\Dob$, where the $\Dob$ is on the $\Dsb$ side after the first split. This corresponds schematically to
\begin{figure}[h]
\setlength{\unitlength}{1cm}
\centering 
\begin{picture}(3,2.3)
\put(-5,0) {\includegraphics[width=0.25\textwidth,angle=0]{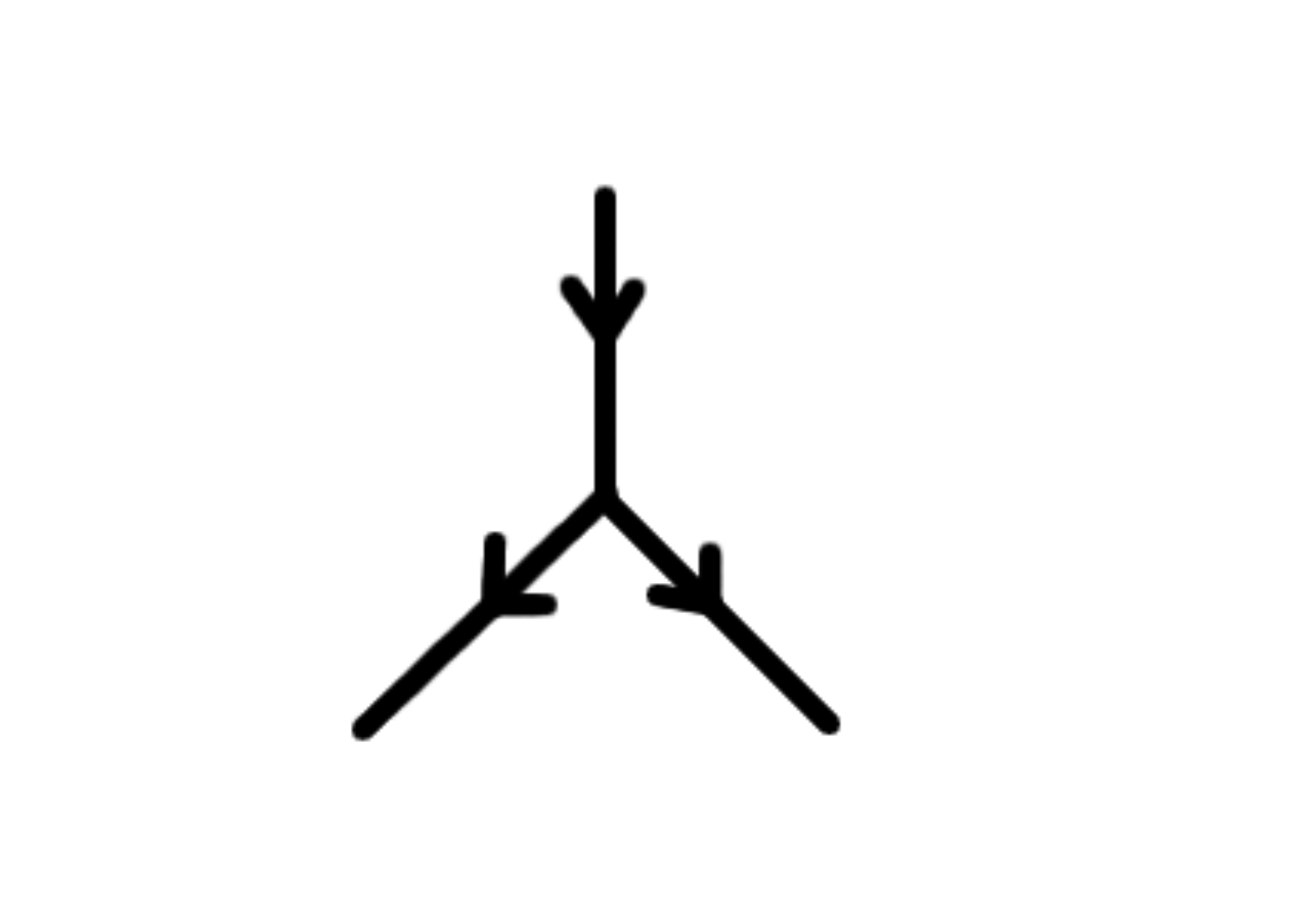}} 
\put(0,0) {\includegraphics[width=0.25\textwidth,angle=0]{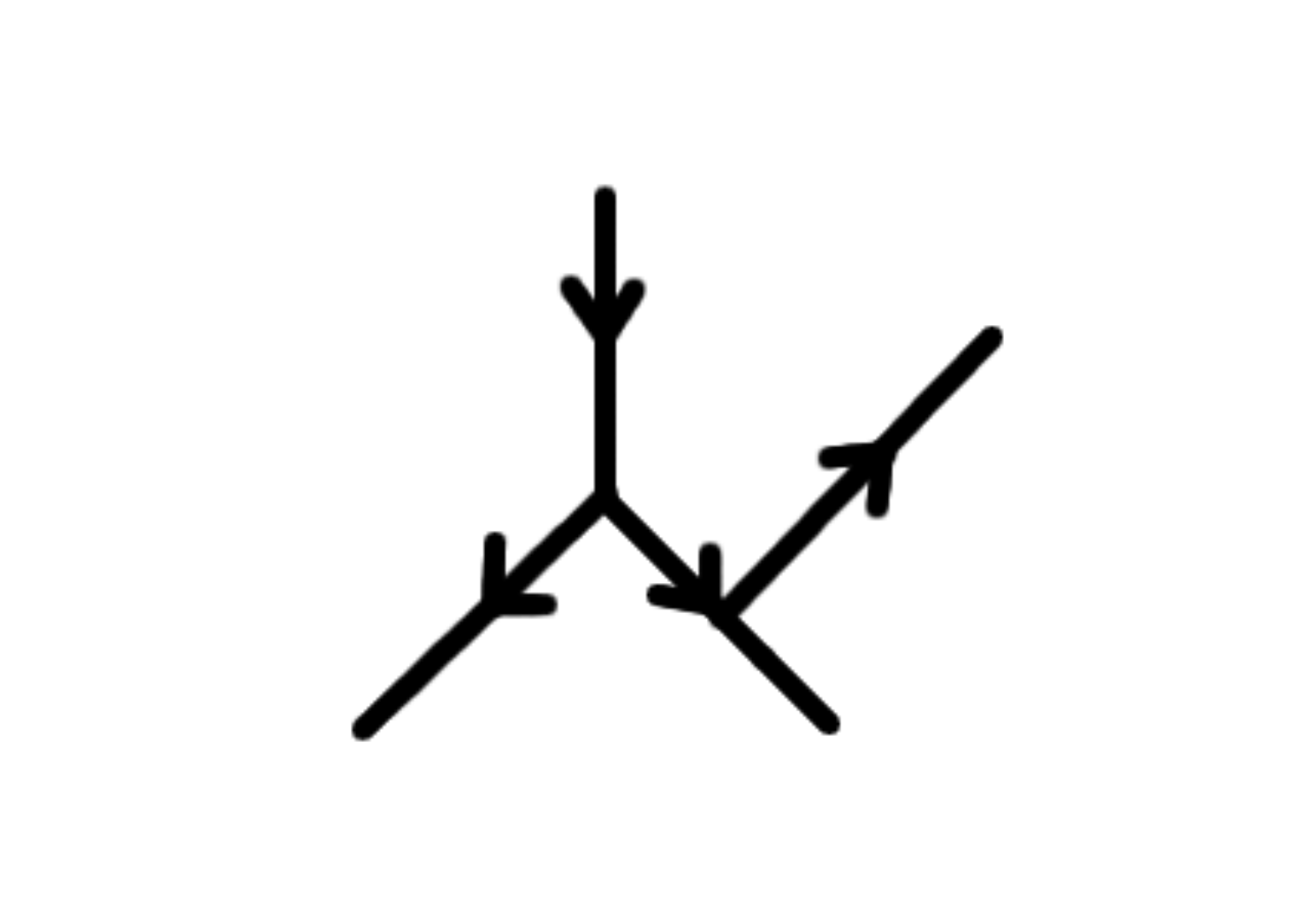}}
\put(-4.2,2.35) {$D4_{\frac{H}{2}+C_1,\overline{D0}}$}
\put(-4.4,0.15) {$D6_H$}
\put(-2.75,0.15) {$\overline{D6}_{D2(C_1),\overline{D0}}$}
\put(0,1.3) {$+$}
\put(1,2.35) {$D4_{\frac{H}{2}+C_1,\overline{D0}}$}
\put(0.6,0.15) {$D6_H$}
\put(2.25,0.15) {$\overline{D6}$}
\put(2.55,1.8) {$D2_{\overline{D0}}$}
\put(4.5,1.3) {$=$}
\put(5.3,1.3) {$-1'138'500.$}
\end{picture}
\end{figure}

These two flow trees sum up according to the following:
\begin{equation*}
  \Omega=(-1)^{|\langle\Gamma_1,\Gamma_2\rangle|-1}|\langle\Gamma_{1},\Gamma_2\rangle|N_{\textrm{DT}}(0,0)\cdot N_{\textrm{DT}}(1,2)=(-1)^1\cdot 2\cdot 1\cdot 569'250=-1'138'500.
\end{equation*}

\item $\mathbf{\Delta q=0, \Delta q_0=-3}$:\\
One then considers a state ($\hat{q}_0\approx -0.71$) which can be seen as a D4-brane with three $\Dob$'s ($N=3$). The total charge is $\Gamma = H+\frac{1}{2}H^2+(\frac{\chi (P)}{24}+\frac{1}{2}F^2-N)\omega =H+\frac{1}{2}H^2+\frac{-1}{60}H^3$. One finds two contributions to the total index. The first split flow tree is
\begin{figure}[h]
\setlength{\unitlength}{1cm}
\centering 
\begin{picture}(3,2.3)
%\put(0,0) {\tfbox{\includegraphics[width=0.25\textwidth,angle=0]{00}}}
\put(0,0) {\includegraphics[width=0.25\textwidth,angle=0]{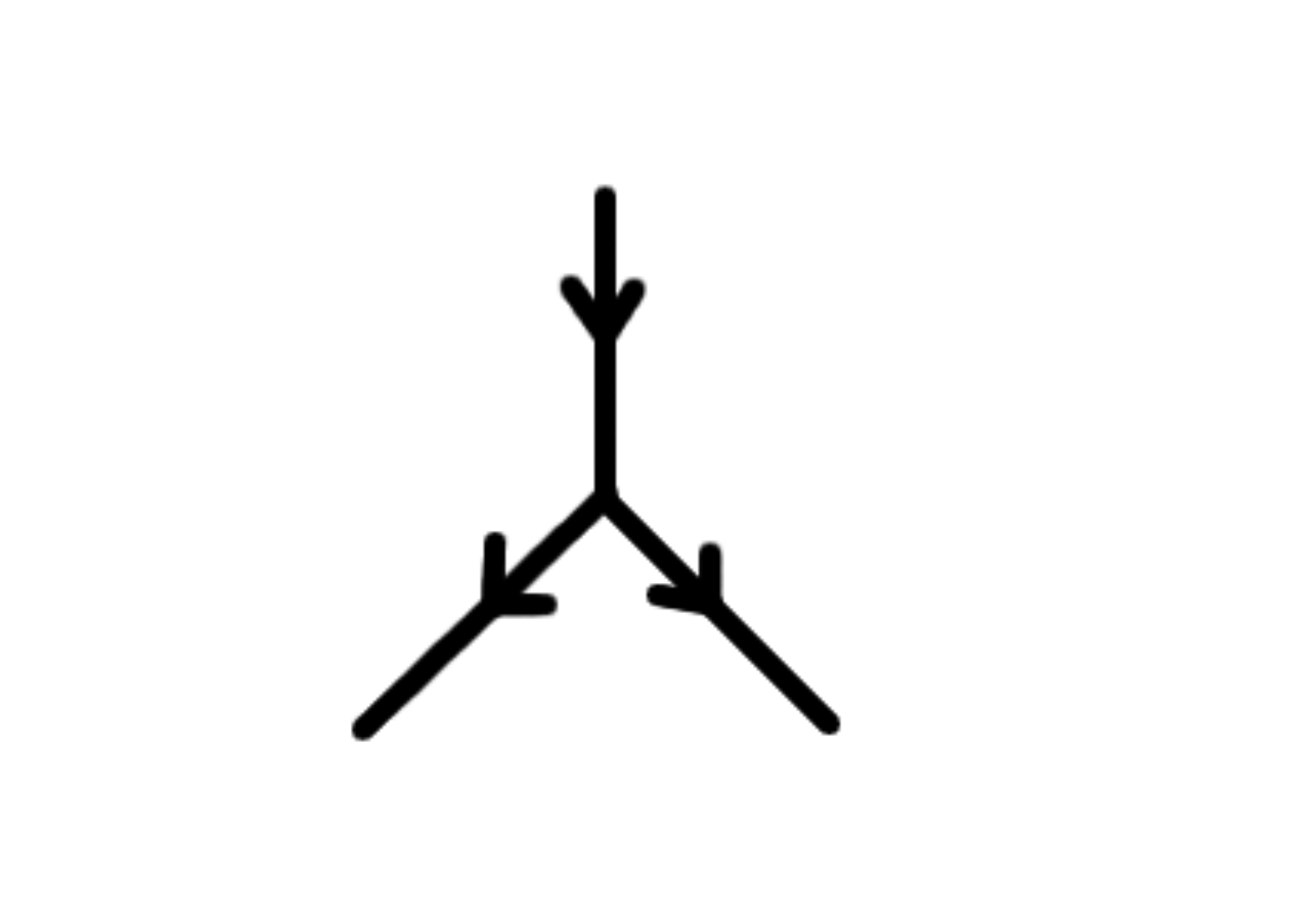}}
\put(1,2.35) {$D4_{\frac{H}{2}+C_1 +\overline{C_1}}$}
\put(0.2,0.18) {$D6_{H,\overline{D2}(C_1)}$}
\put(2.25,0.15) {$\overline{D6_{D2(C_1)}}$}
\put(4,1.3) {$=$}
\put(5,1.3) {$8'265'625$,}
\end{picture}
\end{figure}

and contributes by
\begin{equation*}
  \Omega_A=(-1)^{|\langle\Gamma_1,\Gamma_2\rangle|-1}|\langle\Gamma_{1},\Gamma_2\rangle|N_{\textrm{DT}}(1,1)\cdot N_{\textrm{DT}}(1,1)=(-1)^0\cdot 1\cdot 2'875\cdot 2'875=8'265'625.
\end{equation*}
The second type of split flow tree is one with three centers, a D6-brane with one unit of worldvolume flux, a pure $\Dsb$ and three $\Dob$'s. Again, a threshold wall interpolates between areas where the $\Dob$'s are on the D6 side after the first split, or on the $\Dsb$ side, respectively. The flow tree looks like
\vskip 2mm
\begin{figure}[h!]
\setlength{\unitlength}{1cm}
\centering 
\begin{picture}(3,2.3)
%\put(0,0) {\tfbox{\includegraphics[width=0.25\textwidth,angle=0]{00}}}
\put(0,0) {\includegraphics[width=0.25\textwidth,angle=0]{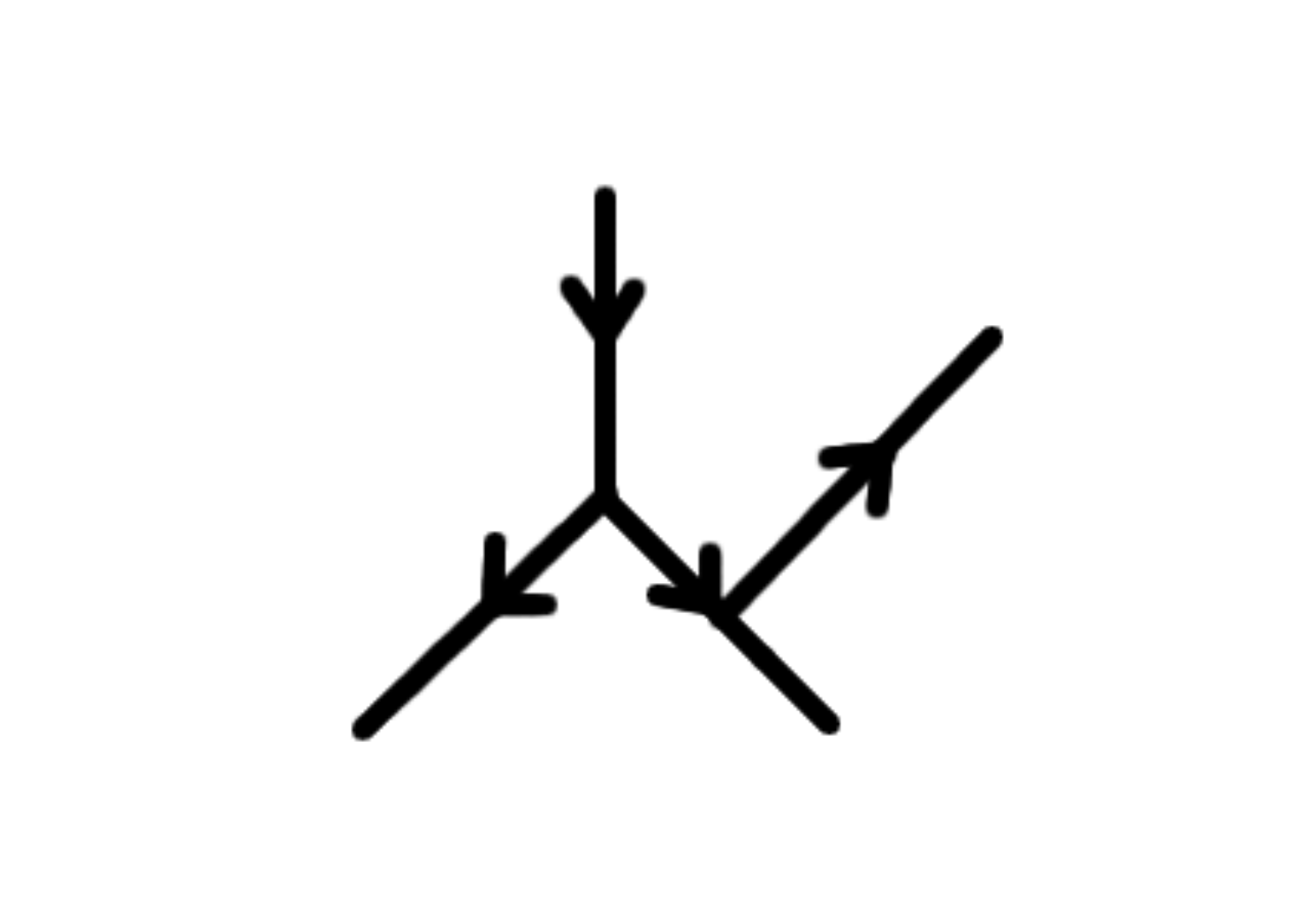}}
\put(1,2.35) {$D4_{\frac{H}{2},3 \overline{D0}}$}
\put(0.6,0.15) {$D6_H$}
\put(2.25,0.15) {$\overline{D6}$}
\put(2.65,1.8) {3 $\overline{D0}$}
\put(4,1.3) {$=$}
\put(5,1.3) {$-2'468'000$,}
\end{picture}
\end{figure}

and contributes
\begin{eqnarray*}
  \Omega_B&=&(-1)^{|\langle\Gamma_1,\Gamma_2\rangle|-1}|\langle\Gamma_{1},\Gamma_2\rangle|N_{\textrm{DT}}(0,3)\cdot N_{\textrm{DT}}(0,0)\nonumber\\
  &=&(-1)^1\cdot 2\cdot 1'234'000\cdot 1=-2'468'000,
\end{eqnarray*}
Altogether this leads to
\vskip 2mm
\begin{figure}[h]
\setlength{\unitlength}{1cm}
\centering 
\begin{picture}(3,2.3)
\put(-5,0) {\includegraphics[width=0.25\textwidth,angle=0]{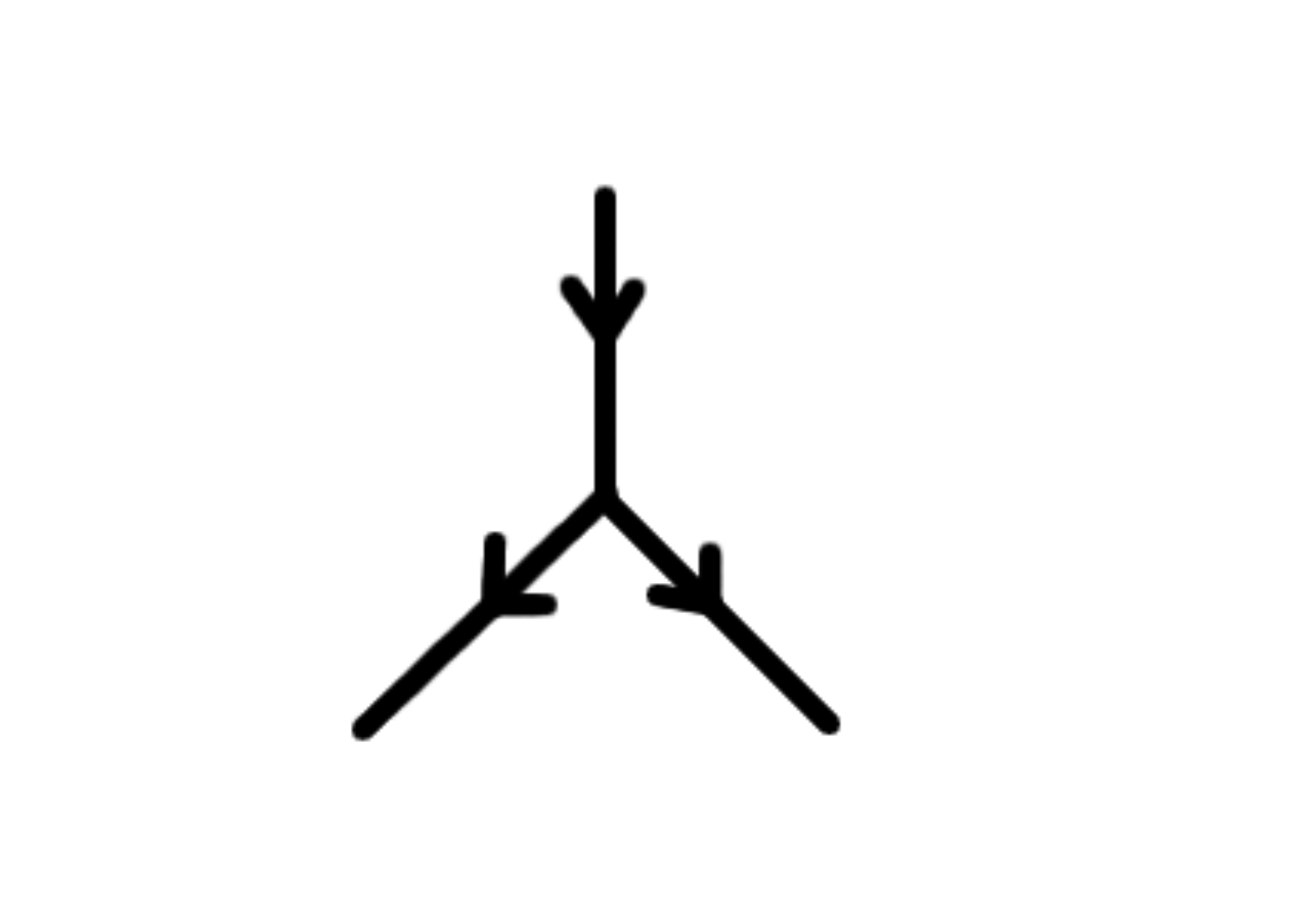}} 
\put(0,0) {\includegraphics[width=0.25\textwidth,angle=0]{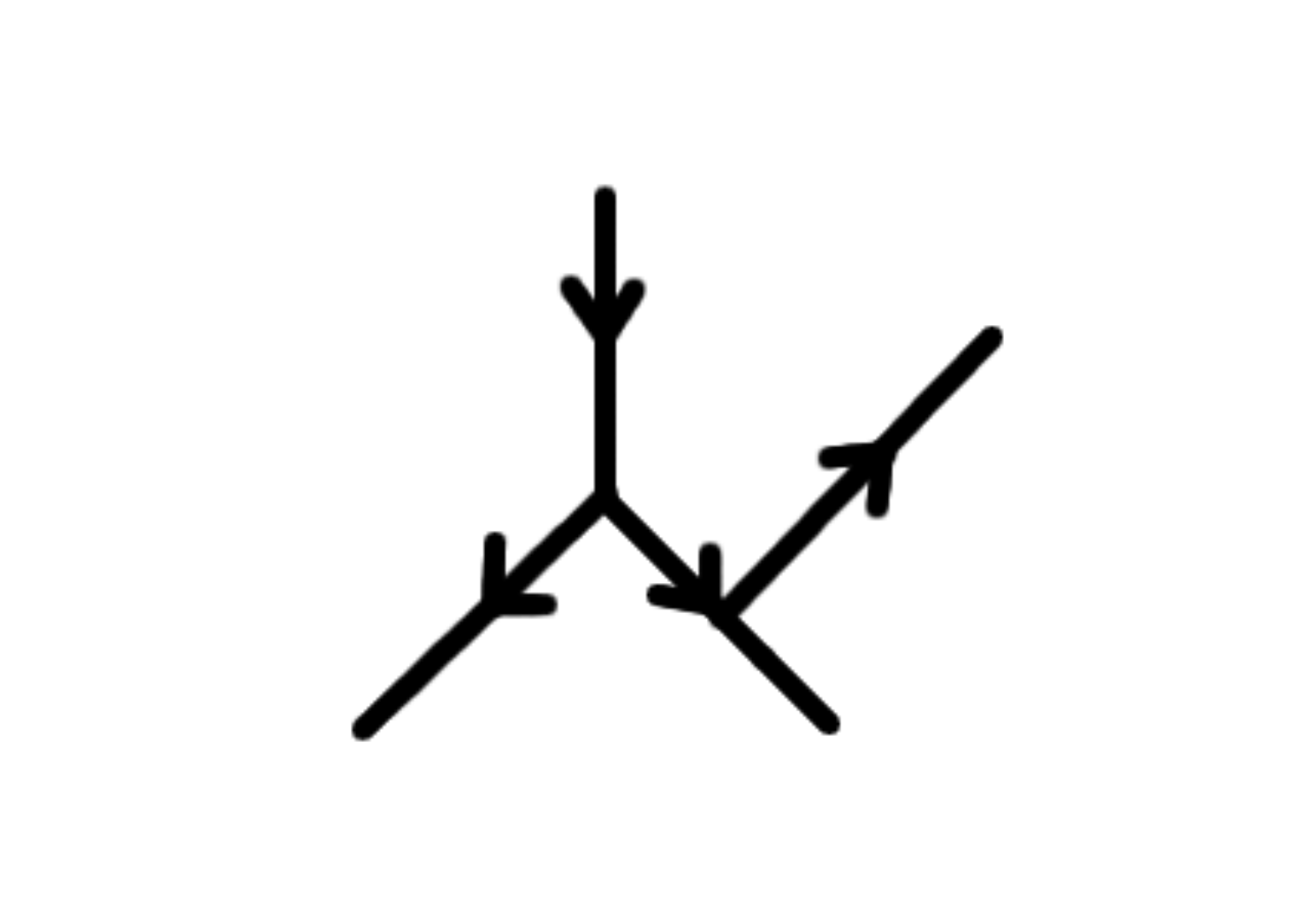}}
\put(-4.2,2.38) {$D4_{\frac{H}{2}+C_1+\overline{C_1}}$}
\put(-4.8,0.18) {$D6_{H,\overline{D2}(C_1)}$}
\put(-2.75,0.15) {$\overline{D6}_{D2(C_1)}$}
\put(0,1.3) {$+$}
\put(1,2.35) {$D4_{\frac{H}{2},3\overline{D0}}$}
\put(0.6,0.15) {$D6_H$}
\put(2.25,0.15) {$\overline{D6}$}
\put(2.55,1.8) {$3 \overline{D0}$}
\put(4.5,1.3) {$=$}
\put(5.3,1.3) {$5'797'625.$}
\end{picture}
\end{figure}

\item $\mathbf{\Delta q=2, \Delta q_0=-2}$:\\
The last state ($\hat{q}_0\approx -1.11$) we consider has the most complicated area code. In the D4 picture, one obtains it by turning on flux dual to a degree two rational curve as well as binding one anti-D0-brane ($N=1$). The total charge reads $\Gamma = H+\frac{9}{10}H^2+(\frac{\chi (P)}{24}+\frac{1}{2}F^2-N)\omega =H+\frac{9}{10}H^2+\frac{11}{60}H^3 $. Again, the reader can find a detailled discussion in the appendix \ref{areacode22}. Choosing a convenient background, one finds three split flow trees.\\
The first two contribute as follows:
\vskip 5mm
\begin{figure}[h]
\setlength{\unitlength}{1cm}
\centering 
\begin{picture}(3,2.3)
\put(-5,0) {\includegraphics[width=0.25\textwidth,angle=0]{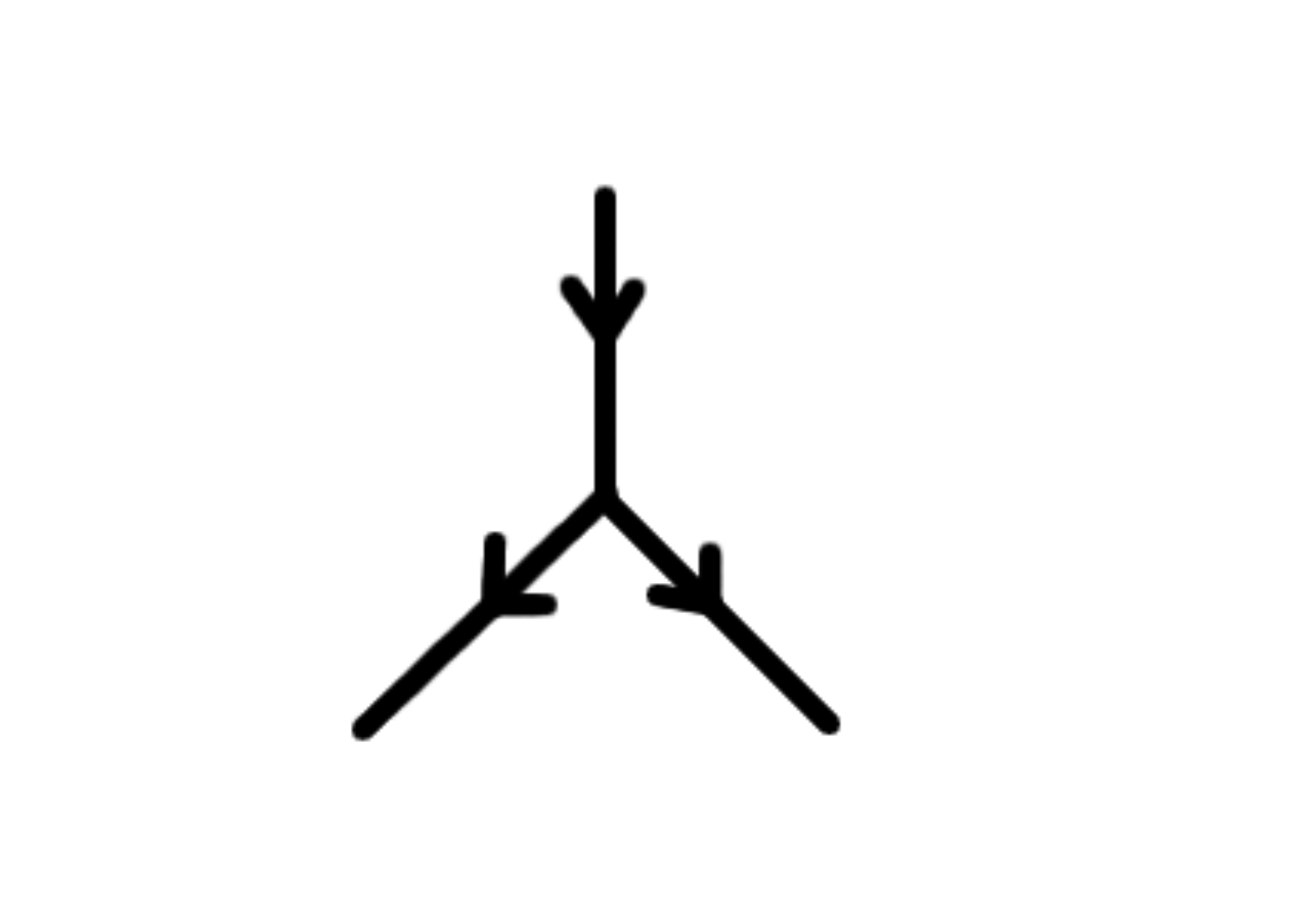}} 
\put(0,0) {\includegraphics[width=0.25\textwidth,angle=0]{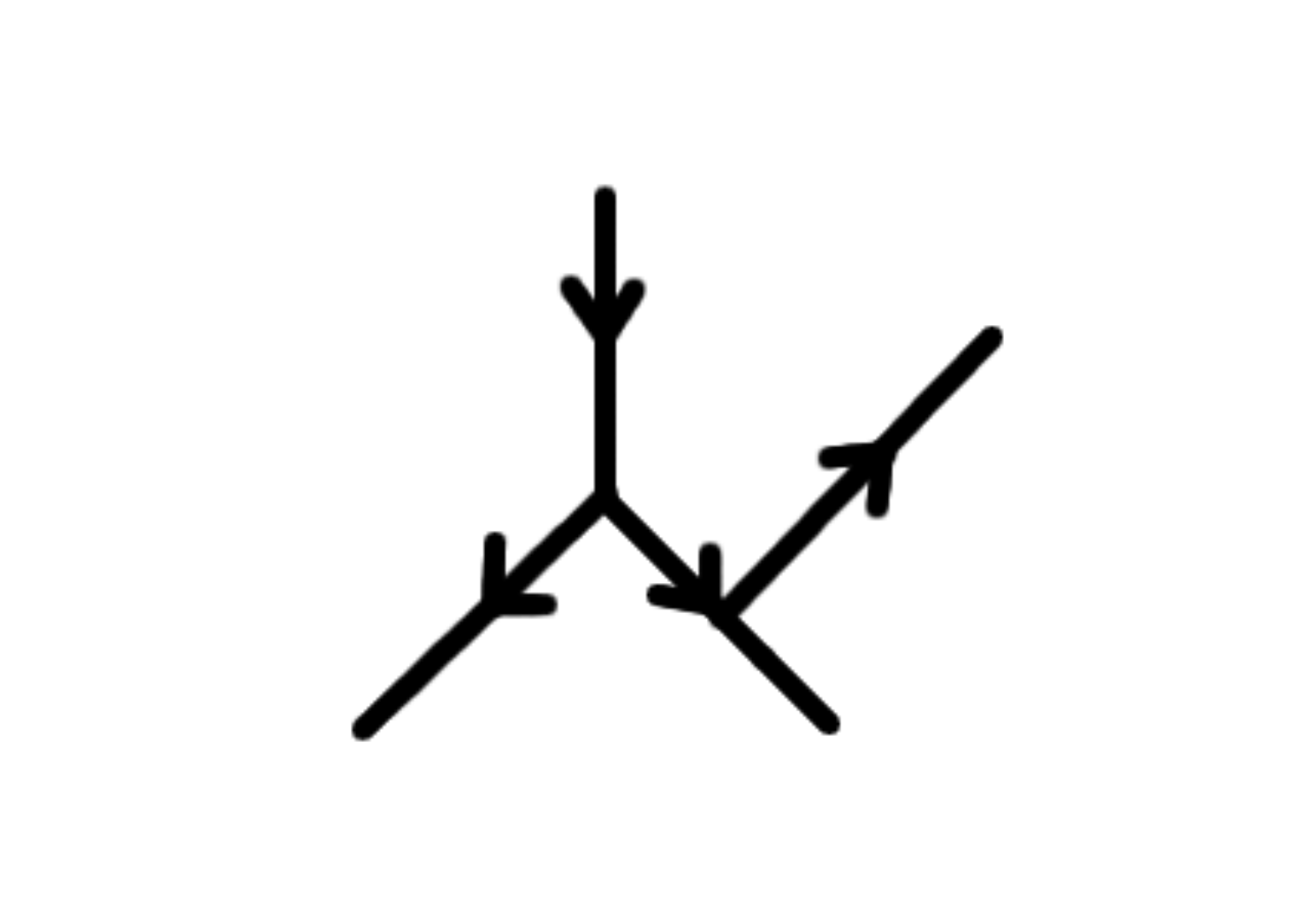}}
\put(-4.2,2.35) {$D4_{\frac{H}{2}+C_2,\overline{D0}}$}
\put(-4.4,0.15) {$D6_H$}
\put(-2.75,0.15) {$\overline{D6}_{D2(C_2),\overline{D0}}$}
\put(0,1.3) {$+$}
\put(1,2.35) {$D4_{\frac{H}{2}+C_2,\overline{D0}}$}
\put(0.6,0.15) {$D6_H$}
\put(2.25,0.15) {$\overline{D6}$}
\put(2.55,1.8) {$D2_{\overline{D0}}$}
\put(4.5,1.3) {$=$}
\put(5.3,1.3) {$-124'762'875,$}
\end{picture}
\end{figure}

which is calculated according to
\begin{eqnarray*}
  \Omega_B&=&(-1)^{|\langle\Gamma_1,\Gamma_2\rangle|-1}|\langle\Gamma_1,\Gamma_2\rangle|N_{\textrm{DT}}(0,0)\cdot N_{\textrm{DT}}(2,2)\nonumber\\
  &=&(-1)^0\cdot 1\cdot 1\cdot 124'762'875=124'762'875.
\end{eqnarray*}
Note that the subscript of the index will become clear in the appendix, where the area code for this charge system is discussed in detail. The next contribution arises from a split flow tree with a D6 with two units of flux and additional sheaf corresponding to a $\Dtb$ wrapped on a degree three rational curve $(\beta_1,n_1)=(3,1)$ as well as a $\Dsb$ with one unit of worldvolume flux. This flow tree looks like
\vskip 2mm
\begin{figure}[h!]
\setlength{\unitlength}{1cm}
\centering 
\begin{picture}(3,2.3)
%\put(0,0) {\tfbox{\includegraphics[width=0.25\textwidth,angle=0]{00}}}
\put(0,0) {\includegraphics[width=0.25\textwidth,angle=0]{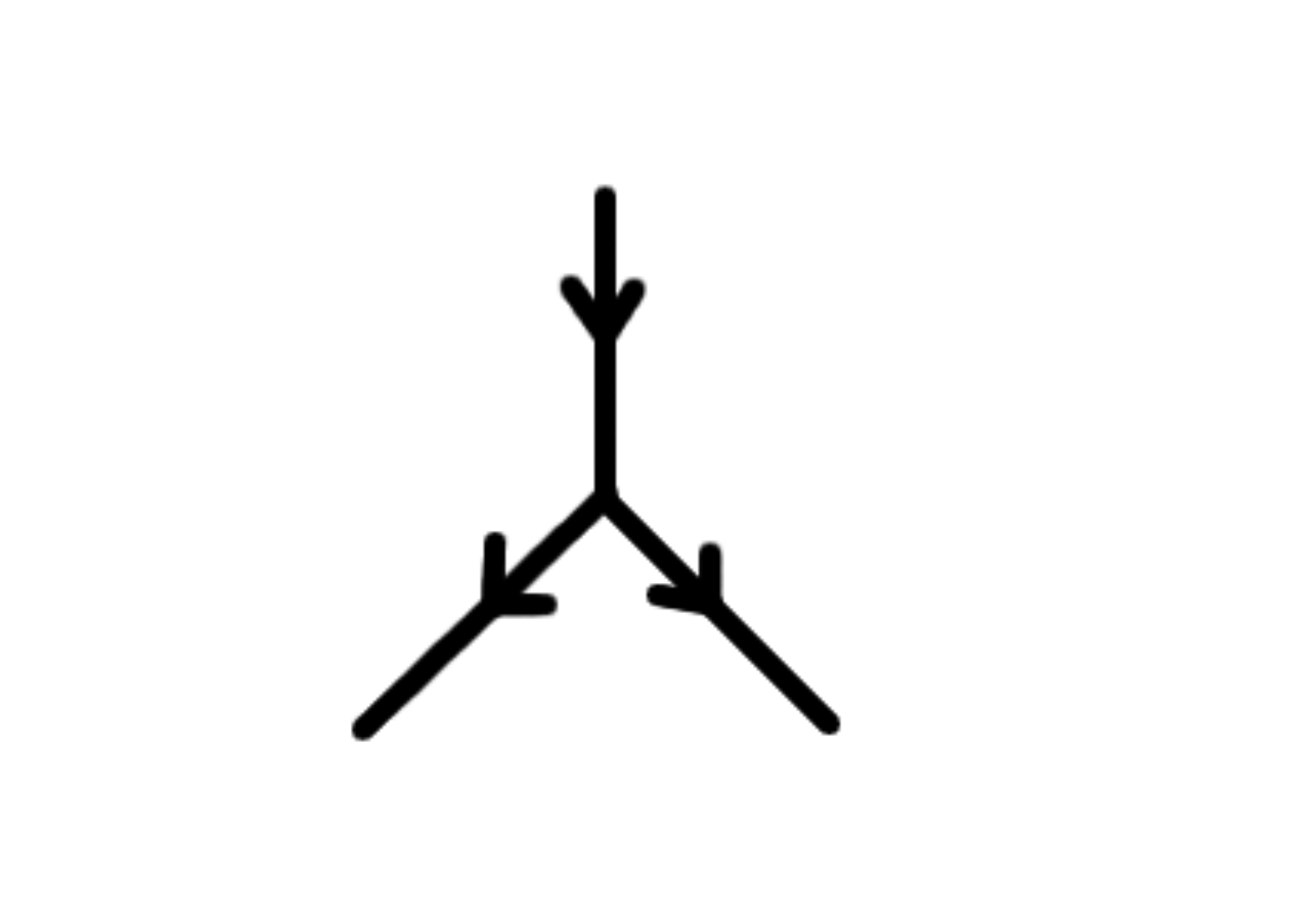}}
\put(1.5,2.25) {$D4_{\frac{H}{2},\overline{C_3}}$}
\put(0.1,0.19) {$D6_{2H,\overline{D2}(C_3)}$}
\put(2.25,0.15) {$\overline{D6}_H$}
\put(4,1.3) {$=$}
\put(5,1.3) {$317'206'375$.}
\end{picture}
\end{figure}

 One might at first have thought that this flow tree would yield the index
\begin{equation*}
  \Omega_C=(-1)^{|\langle\Gamma_1,\Gamma_2\rangle|-1}|\langle\Gamma_1,\Gamma_2\rangle|N_{\textrm{DT}}(0,0)\cdot N_{\textrm{DT}}(3,1)=439'056'375\, ,
\end{equation*}
but closer inspection shows that one would overcount the number of BPS states corresponding to this flow tree. Namely, one would also enumerate the number of BPS states corresponding to the case, when the D6-brane would have a $\Dtb$ on an elliptic degree three curve, $C_3{}^{g=1}$, and an extra $\Dob$. However, one does not find this type of split flow tree. A simple trick allows us to substract the right number of states from $439'056'375$: Imagine that this type of split existed. In that case one might also find a threshold wall for the $\Dob$. If one would take the background into a region, where the $\Dob$ flips side after the first split, one would then enumerate this split as follows:
\begin{equation*}
  \Omega_{C2}=(-1)^0\cdot 1\cdot N_{\textrm{DT}}(0,1)\cdot N_{\textrm{DT}}(3,0)=200\cdot 609'250=121'850'000.
\end{equation*}
Again, the subscript `$C2$' of the index will become clear in the appendix, where the area code is discussed in detail. As the index cannot jump at a threshold wall, this is presumably also the index which we have to substract. One then obtains
\begin{equation*}
  \Omega_{C1}=\Omega_C-\Omega_{C2}=439'056'375-121'850'000=317'206'375.
\end{equation*}
This can also be graphically depicted by
\vskip 2mm
\begin{figure}[h!]
\setlength{\unitlength}{1cm}
\centering 
\begin{picture}(3,2.3)
\put(-5,0) {\includegraphics[width=0.25\textwidth,angle=0]{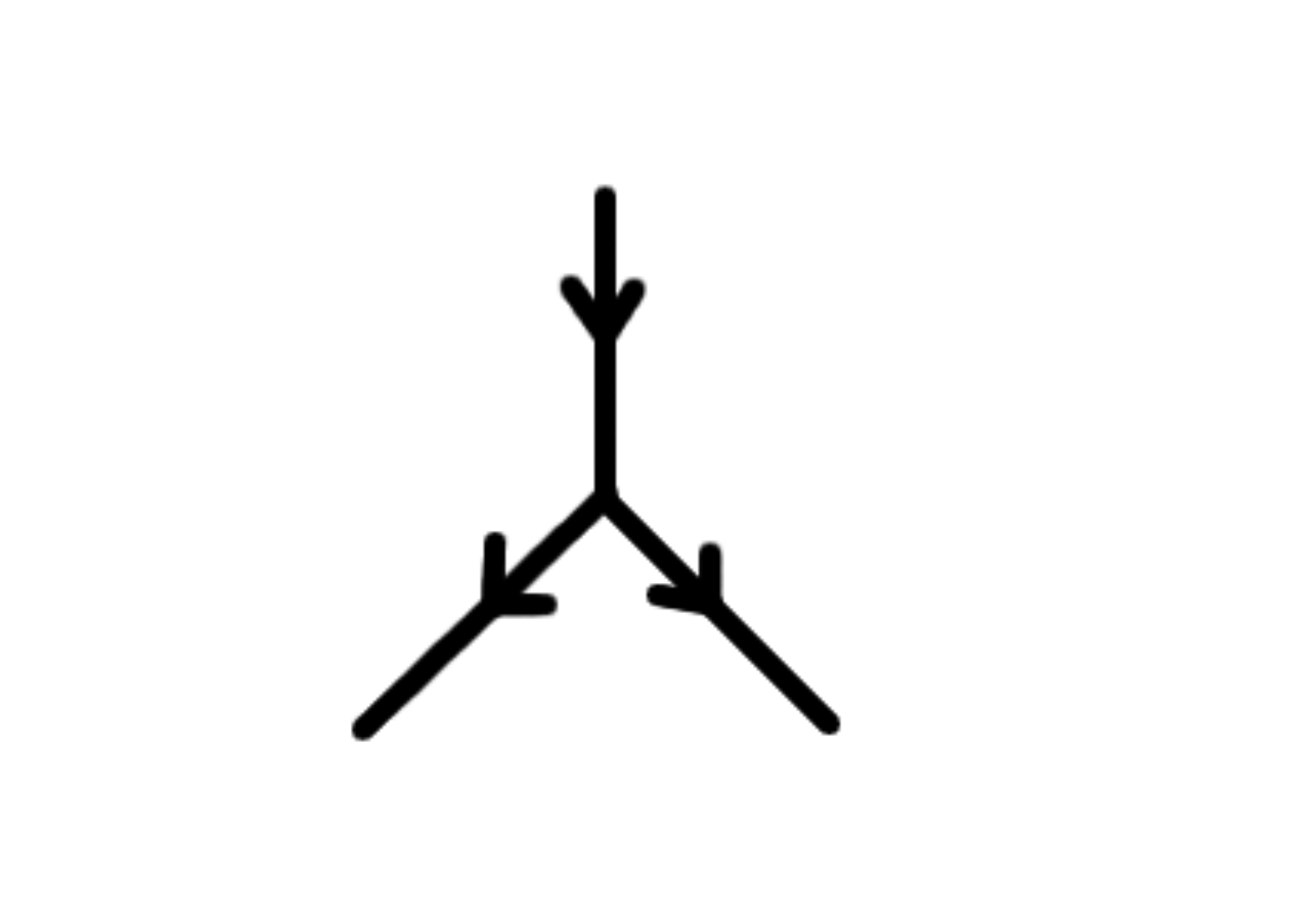}} 
\put(0,0) {\includegraphics[width=0.25\textwidth,angle=0]{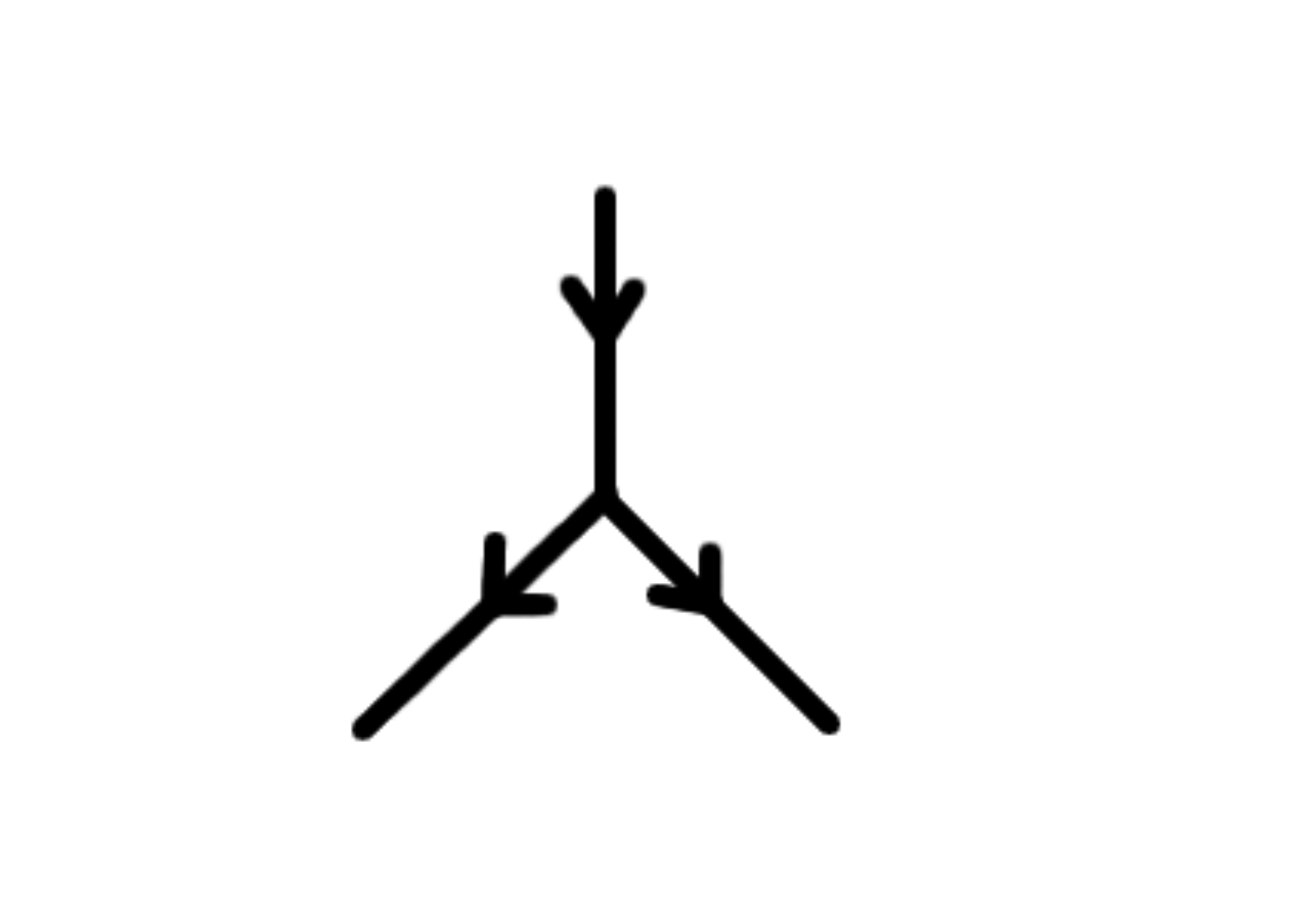}}
\put(-4.2,2.35) {$D4_{\frac{H}{2}+\overline{C_3}}$}
\put(-4.8,0.15) {$D6_{2H,\overline{D2}(C_3)}$}
\put(-2.75,0.15) {$\overline{D6}_H$}
\put(0,1.3) {$+$}
\put(1,2.35) {$D4_{\frac{H}{2}+\overline{C_3{}^{g=1}},\overline{D0}}$}
\put(-0.3,0.15) {$D6_{2H,\overline{D2}(C_3{}^{g=1})}$}
\put(2.25,0.15) {$\overline{D6}_H$}
\put(4.5,1.3) {$=$}
\put(5.3,1.3) {$439'056'375,$}
\end{picture}
\end{figure}

and
\pagebreak
\begin{figure}[h!]
\setlength{\unitlength}{1cm}
\centering 
\begin{picture}(3,2.3)
%\put(-5,0) {\includegraphics[width=0.25\textwidth,angle=0]{00}} 
\put(-4,1.3) {$439'056'375$}
\put(0,0) {\includegraphics[width=0.25\textwidth, angle=0]{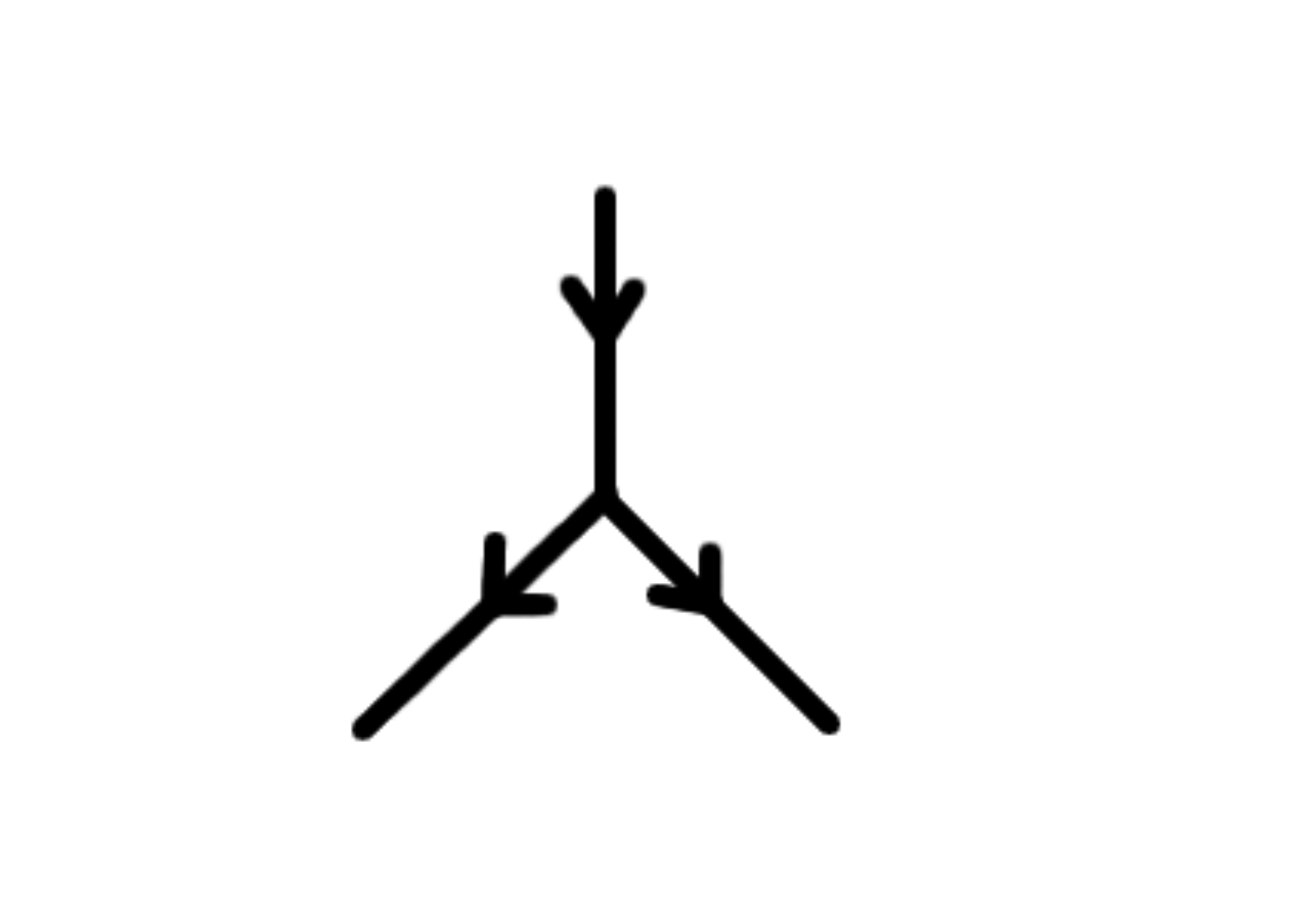}}
\put(0,1.3) {$-$}
\put(1,2.35) {$D4_{\frac{H}{2}+\overline{C_3{}^{g=1}},\overline{D0}}$}
\put(-0.3, 0.15) {$D6_{2H,\overline{D2}(C_3{}^{g=1})}$}
\put(2.25,0.15) {$\overline{D6}_H$}
\put(4.5,1.3) {$=$}
\put(5.3,1.3) {$317'206'375.$}
\end{picture}
\end{figure}

Finally, one can sum up to obtain the total BPS index:
\vskip 2mm
\begin{figure}[h!]
\setlength{\unitlength}{1cm}
\centering 
\begin{picture}(3,2.3)
\put(-5.5,0) {\includegraphics[width=0.25\textwidth,angle=0]{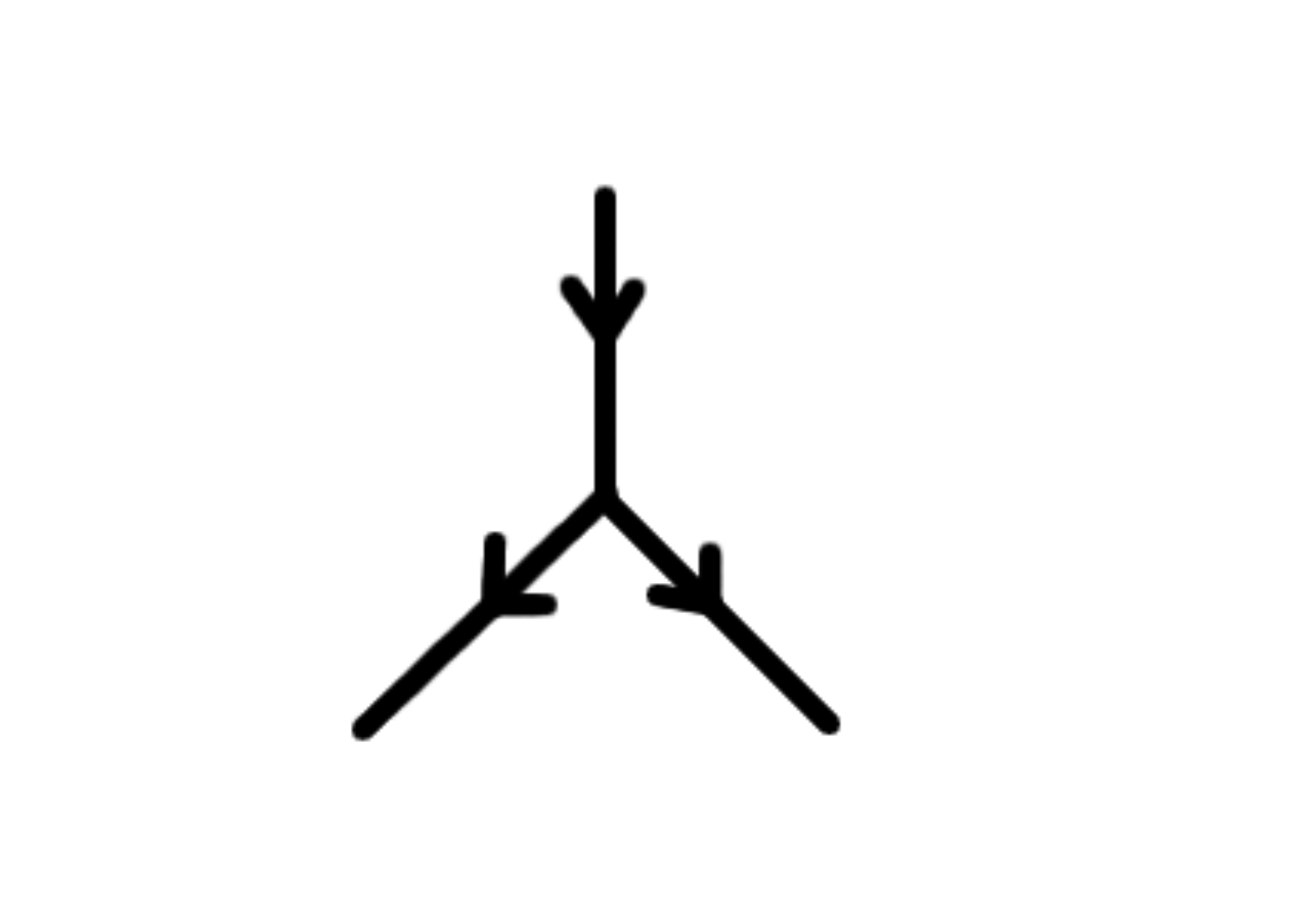}} 
\put(-1.8,0) {\includegraphics[width=0.25\textwidth,angle=0]{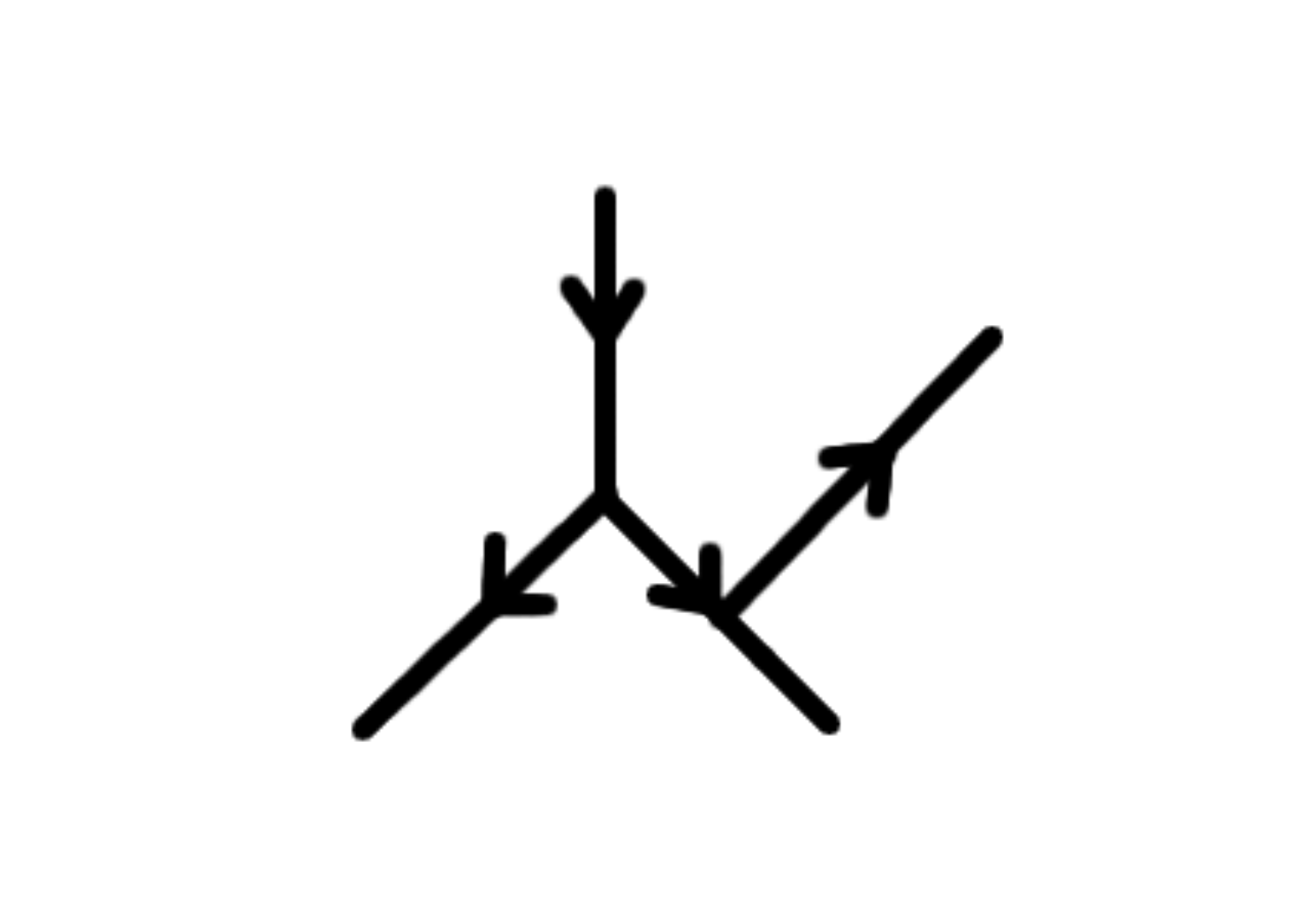}}
\put(2,0) {\includegraphics[width=0.25\textwidth,angle=0]{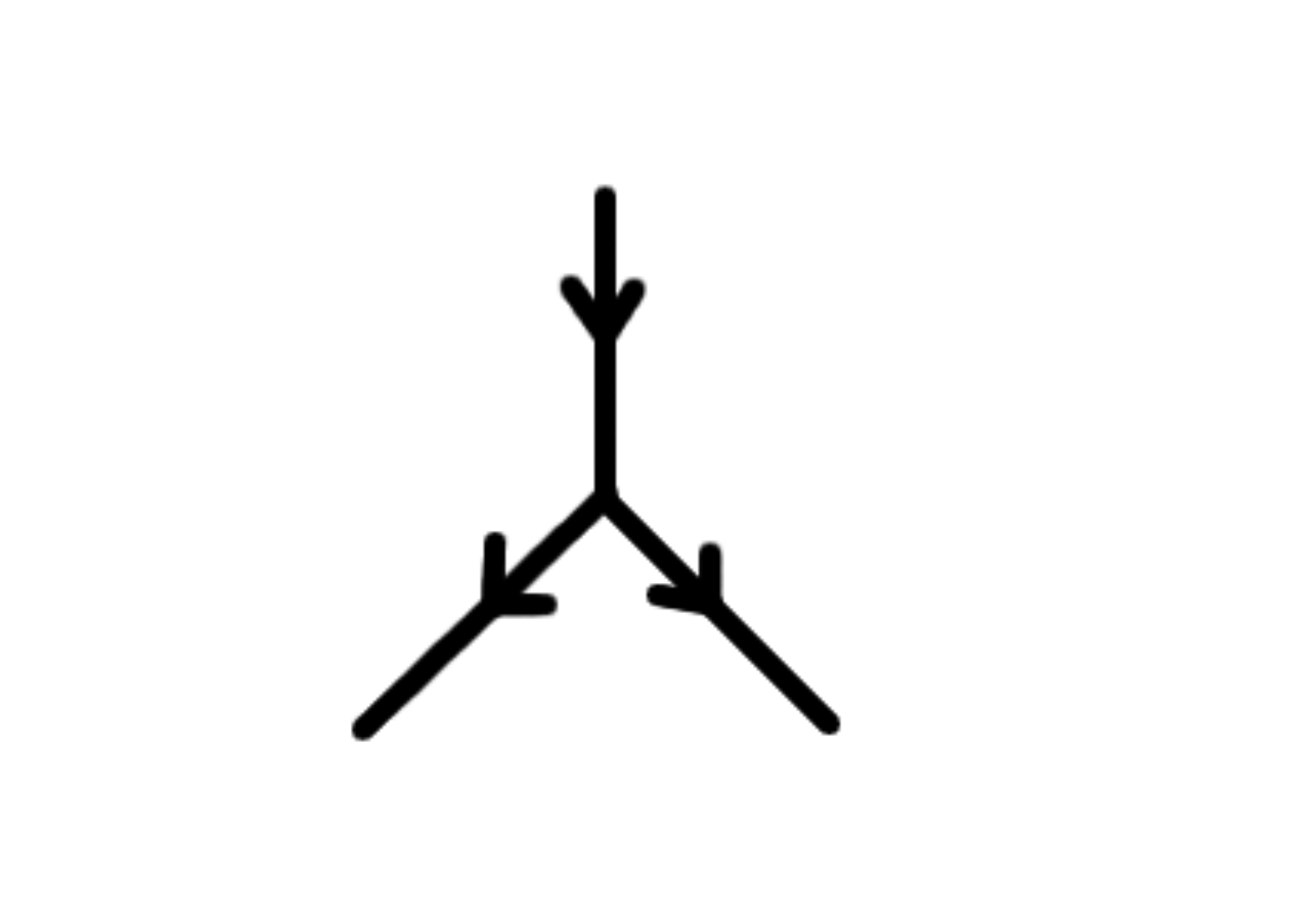}}
\put(-4.7,2.35) {$D4_{\frac{H}{2}+C_2,\overline{D0}}$}
\put(-5.2,0.15) {$D6_H$}
\put(-3.5,0.15) {$\overline{D6}_{D2(C_2),\overline{D0}}$}
\put(-2,1.3) {$+$}
\put(-0.7,2.35) {$D4_{\frac{H}{2}+C_2,\overline{D0}}$}
\put(-1.1,0.15) {$D6_H$}
\put(0.55,0.15) {$\overline{D6}$}
\put(0.85,1.8) {$D2_{\overline{D0}}$}
\put(1.9,1.3) {$+$}
\put(3.2,2.35) {$D4_{\frac{H}{2},\overline{C_3}}$}
\put(2.2,0.15) {$D6_{2H,\overline{D2}(C_3)}$}
\put(4.2,0.15) {$\overline{D6}_H$}
\put(5.5,1.3) {$=$}
\put(6.5,1.3) {$441'969'250.$}
\end{picture}
\end{figure}
\end{enumerate}
Before continuing to analyze and extend our results, let us add one more important remark at this point. Overviewing all of our polar and non-polar states and their split flow trees obtained numerically, and taking into account that a BPS index cannot possibly jump when crossing a wall of threshold stability\footnote{As explained earlier in this paper, unlike a wall of marginal stability, a wall of threshold stability means that there are no tachyon fields between the would-be products of a decay process. Therefore, the decay is \emph{kinematically} impossible.} , we may safely conclude that there is indeed a background region where all of our indices are valid simultaneously, making our enumeration valid as a whole.
\subsection{Non polar states without any single flows}
In the large radius regime the notion of a polar state automatically coincided with a state that does not support a single attractor flow. On the other hand, a non-polar state allowed to write down an attractor point for a single flow. As mentioned previously and as we have now seen, this property is altered when using our instanton corrected central charges, exhibiting an interesting difference with the large radius regime. As discussed, three of the non-polar states do not allow for any single-centered attractor flows\footnote{Note that also the fourth non-polar state does not allow single flows in certain background regimes.}. This might have been expected, as the criterion in \cite{Denef:2007vg} for the existence of a crash in the flow was based on the central charge, which gets strong instanton corrections in the regime we are working in. The criterion for having a crash in the single flow is therefore no longer $\hat{q}_0>0$. Put differently, non-polarity no longer guarantees the existence of single centered solutions.

\subsection{Comparison to the elliptic genus}
The following table summarizes our results and compares them with the predictions from the modular form (for non-polar states, obtained from the knowledge of the polar states) as well as their results obtained by performing a geometric counting for the D4-D2-D0 moduli space, and then improved by analyzing the degeneracies of the $(0,4)$-MSW-CFT dual states from \cite{Gaiotto:2006wm} in dilute gas approximation. The latter is just denoted with the short term `CFT results' in the following table.
\begin{center}
\begin{tabular}{|c||c|c|c|}
    \hline
    \textbf{Reduced D0 charge $\hat{q}_0$} & \footnotesize{Modular form prediction} & \footnotesize{CFT results} & \footnotesize{\textbf{Split flows and DT}}\\
    \hline
    \hline
    $2.29$ & +5 & +5 & +5\\
    \hline
    $1.29$ & -800 & -800 & -800\\
    \hline
    $0.69$ & +8'625 & -8'625 & +8'625\\
    \hline
    $0.69$ & +8'625 & -8'625 & +8'625\\
    \hline
    $0.29$ & +58'500 & +58'500 & +58'500\\
    \hline
    $-0.11$ & -1'218'500 & -1'218'500 & -1'218'500\\
    \hline
    $-0.31$ & -1'138'500 & +1'138'500 & -1'138'500\\
    \hline
    $-0.71$ & +5'817'125 & +5'797'625 & +5'797'625\\
    \hline
    $-1.11$ & +441'969'250 & +441'969'250 & +441'969'250\\
    \hline
\end{tabular}
\end{center}
It is intriguing to see that our results match the numbers from \cite{Gaiotto:2006wm}, but do improve some signs. At the same time our and their result differs from the exact prediction from the modular form on one of the states. We now move on to discuss our resolution to this puzzle in the next section.

\section{Flow tree index refinement: non-trivially fibered moduli spaces}\label{fibration}
Our results obtained in the last section are almost fully satisfying. The ${\Delta q=0, \Delta q_0=-3}$ state is the only case where we were not directly able to reproduce the result expected from modularity. Our index yields $+5'797'625$, whereas the expected value would be $+5'817'125$. Interestingly enough, this is off by $19'500$, which exactly equals $N_{\textrm{DT}}(0,2)$ for the quintic. Luckily, careful consideration of the geometry from the D4 perspective does shed some light on this discrepancy. In \cite{Gaiotto:2007cd}, the authors calculated this index by refining the geometric counting of the D4-moduli space. They found that the index for this state had two contributions: a D4-brane with three $\Dob$'s bound to it, and a D4-brane with two degree one rational curves on it. This is analogous to our findings, where we have a case with a D6/$\Dsb$ split with three $\Dob$'s bound to one of these two branes, and a case where there is one degree one rational curve on each of the two constituent branes. For the contribution arising from a D4 with three $\Dob$'s, careful consideration is taken of the fact that the three $\Dob$'s can be aligned, which enhances the moduli space of the D4-brane.

This raises one very interesting question. Does this special \emph{collinear locus} of the three $\Dob$-branes leave a footprint on the structure of the split flow trees? In other words, does something special happen in the supergravity picture when these branes align?
One possible answer to this question might be, that the state with collinear $\Dob$-branes corresponds to a \emph{scaling solution}. As mentioned previously, these are multi-centered supergravity solutions that are parametrically connected to single-centered solutions. In other words, the distance between the centers is no longer fixed by the BPS condition, but it becomes a modulus. States corresponding to scaling solutions make the index computation less obvious, as one has to attribute a factorized index to a single attractor flow. Properly accounting for the index of a scaling solution might in principle be at the origin of the missing part in our result.

However, we were not able to find any single-centered solutions, or single flows for that matter, corresponding to the $(0,-3)$-state. This may be due to finiteness in the numerical precision of our approach, as every single-centered flow seems to want to go straight through the Gepner point in the CY moduli space. Should a scaling solution exist (and this be the only additional contribution), we could predict its BPS index to be $19'500$. This might for the first time allow a check on a prediction of a BPS index for a scaling solution from quantizing the moduli space, \cite{deBoer:2008zn}. However, we do not believe that a scaling solution is at the heart of this problem. 

In our language, the moduli space of the D4-brane translates into the degrees of freedom of the tachyonic strings stretching between the D6 and the $\Dsb$. The actual number of tachyonic degrees of freedom is computed by the DSZ intersection product. Our index factorization scheme does not `see' this possibility for the three $\Dob$'s to align. Let us make this more precise.

Suppose we just had a D6 with flux $F_1=H$, and a $\Dsb$ flux $F_2=0$. Then the DSZ intersection product gives $5$. This DSZ product is actually computing an index via the Riemann-Roch theorem
\begin{equation}
\int_X {\rm ch}(F_2^*) \, {\rm ch}(F_1) \, {\rm Td}(X) = \int_X {\rm ch}(H) \, {\rm Td}(X)\,.
\end{equation}
The tachyon field between the two branes is a section of $F_2^* \otimes F_1$, and this index is just counting the basis elements in the space of sections of this bundle. In this case, the general form of the tachyon will be a polynomial of degree one:
\begin{equation}
T = a_1\,x_1 + \ldots + a_5\,x_5\,.
\end{equation}
Hence, we see where the `$5$' comes from. After tachyon condensation, a D4-brane will emanate at the locus $T=0$\footnote{From the point of view of the D4 divisor, the moduli space is a projectivization of the moduli space of sections of the tachyon bundle.}.

Suppose now, that we also turn on an ideal sheaf on the D6, localized at a point $p$ on $X$. If we now compute the DSZ product, which now becomes a Grothendieck-Riemann-Roch type of index, it will actually compute the number of sections of the bundle we had before, tensored with the ideal sheaf. In other words, the index computes the number of independent sections of the bundle $F_2^* \times F_1$ that vanish on the point $p$. Geometrically, this means that the D4 brane resulting from the condensation will be at the locus $T=0$, and that this hypersurface will actually pass through $p$. In this case, the DSZ index gives $4$. This reflects the fact that the tachyon field loses one degree of freedom due to the restriction of having to vanish on $p$. 

If we now put an ideal sheaf localized on three generic points on $X$, then this will impose three linear constraints on the tachyon field. The index in this case gives $2$. However, if these three points happen to be collinear, then the three linear constraints on $T$ are not linearly independent. The DSZ index, analogously to any index, relies on the `genericity' of the choice of the three points. It only computes a `virtual' or `expected' `dimension' of the tachyon moduli space. In the collinear case, however, the true `dimension' is $3$. That is, the system with three collinear $\Dob$'s behaves as a system with only two $\Dob$'s.

When we compute our index in the straightforward fashion, we are not properly taking this special \emph{collinear locus} in moduli space into account. Note, that this collinear locus is \emph{not} a singularity of the moduli space. It has nothing to do with having coincident points, which the Donaldson-Thomas invariants automatically take into account. 

We therefore propose a refinement of the prescription\begin{equation}
Z(A+B) = (-1)^{\langle A,\, B\rangle-1} |\langle A,\, B\rangle|\, N_{\textrm{DT}}(A)\,N_{\textrm{DT}}(B)\,,
\end{equation}
where $A$ and $B$ are D$6$ and $\Dsb$ states, respectively. In general, the $A$ and $B$ configurations have moduli spaces, $\mathcal{M}_A$, $\mathcal{M}_B$, corresponding to the displacements of the $\Dob$'s on them, plus the moduli of the Riemann surfaces on which the D$2$'s are wrapped. Alternatively, these are the moduli spaces of the ideal sheaves on the D$6$ and $\Dsb$. These spaces have Euler characteristics given by the Donaldson-Thomas invariants. The other component of the configuration is the tachyon field $T$ connecting $A$ and $B$. It has its own moduli space $\mathcal{M}_T$, whose Euler number is computed by means of the DSZ intersection number $|\langle A,\, B\rangle|$, which is nothing other than an Atiyah-Singer topological index. Na\"ively, the full moduli space should be a product
\begin{equation}
\mathcal{M}_{total} = \mathcal{M}_A \times \mathcal{M}_B \times \mathcal{M}_T\,.
\end{equation}
However, in general, the total space will be a non-trivial fibration of $\mathcal{M}_T$ over the other two factors, with the fiber dimension jumping at special loci. This is also explained from the D$4$-brane point of view in \cite{Denef:2007vg}.
In the $(0,-3)$ case, the tachyon moduli space is typically a $\mathbb{CP}^1$, and the index is $-2$. However, whenever the three $\Dob$'s align, the system behaves effectively like a two-particle state. In this case, the $\mathcal{M}_T$ gets enhanced to a $\mathbb{CP}^2$, and the index should really be $+3$. We therefore propose to heuristically define what we will refer to as \emph{DT-densities} $\mathcal{N}_{\textrm{DT}}$ such that WLOG
\begin{equation}
\int_{\mathcal{M}_A} d z^i \, \mathcal{N}_{\textrm{DT}}(A, z) = N_{\textrm{DT}}(A)\,,
\end{equation}
where the $z^i$ are coordinates of the moduli space. Actually, the virtual dimension of these moduli spaces is zero \cite{DT1, DT3}, so this integral can be written as a sum:
\begin{equation}
\sum_i \mathcal{N}_{\textrm{DT}}(A, i) = N_{\textrm{DT}}(A)\,,
\end{equation}
where the index $i$ labels a sector in the moduli space of the sheaves.
These `densities' can be thought of as the DT analogs of the top Chern class or the Euler class of a tangent bundle, which integrates over a manifold to give an Euler number. We then submit, that the proper procedure to compute the BPS index is the following refined index:
\begin{eqnarray}
Z(\mathcal{M}_{total}) &=&\int_{\mathcal{M}_A \times \mathcal{M}_B} dz^i\,d{z'}^{j} (-1)^{\langle A,\, B\rangle(z, z')-1} |\langle A,\, B\rangle (z, z')|\, \mathcal{N}_{\textrm{DT}}(A,z')\,\mathcal{N}_{\textrm{DT}}(B,z') \nonumber\\
 &=&\sum_{i,i'} (-1)^{\langle A,\, B\rangle_{i, i'}-1} |\langle A,\, B\rangle_{i, i'}|\, \mathcal{N}_{\textrm{DT}}(A,i)\,\mathcal{N}_{\textrm{DT}}(B,i')\,,
\end{eqnarray}
where $\langle A,\, B\rangle_{i,i'}$ is a function of the moduli.

In the $(0,-3)$ state, the moduli space of the D$6$ is trivial, and the moduli space $\mathcal{M}_B$ of the $\Dsb$ must be divided into two parts, $\mathcal{M}_{B_1}$ and $\mathcal{M}_{B_2}$, where the first part corresponds to the generic configuration of three particles, and the second part corresponds to the subspace where the three particles are aligned, effectively behaving like two particles. This first part will yield a tachyon index $-2$, whereas the effective two-particle state will yield a tachyon index $+3$. 

The Witten index for the supersymmetric quantum mechanics of a three particle state on $X$ has the form
\begin{equation}
\chi(X)^3 + {\rm corrections \ for \ coincidence \ loci} = -1,234,000\,,
\end{equation}
where $\chi(X) = -200$. The case of a two-particle state has
\begin{equation}
\chi(X)^2 + {\rm corrections } = +19,500\,.
\end{equation}
Hence, the two have a relative sign difference.
The total refined index $Z'$ (excluding the contribution from the curves) is then given by
\begin{eqnarray}
Z'(0,3) &=& N_{\textrm{DT}}(0,0) \times \int_{\mathcal{M}_{(0,-3)}} (-1)^{\langle (0,0),\, B \rangle-1} |\langle (0,0),\, B \rangle|\, \mathcal{N}_{DT}((0,-3))\\
&=& \int_{\mathcal{M}_{B_1}} \mathcal{N}_{DT}(0,3)\,(-1)^{\langle (0,0),\, B_1 \rangle-1} |\langle (0,0),\, B_1 \rangle| \\ 
&+& \int_{\mathcal{M}_{B_2}} \mathcal{N}_{DT}(0,3)\,(-1)^{\langle (0,0),\, B_2 \rangle-1} |\langle (0,0),\, B_2 \rangle| \\
&=& -2\, (1,234,000+19,500) + 3\, (+19,500) = -2,448,500\,.
\end{eqnarray}
We must therefore equate one flow tree anew:
\vskip 2mm
\begin{figure}[h]
\setlength{\unitlength}{1cm}
\centering 
\begin{picture}(3,2.3)
%\put(0,0) {\tfbox{\includegraphics[width=0.25\textwidth,angle=0]{00}}}
\put(0,0) {\includegraphics[width=0.25\textwidth,angle=0]{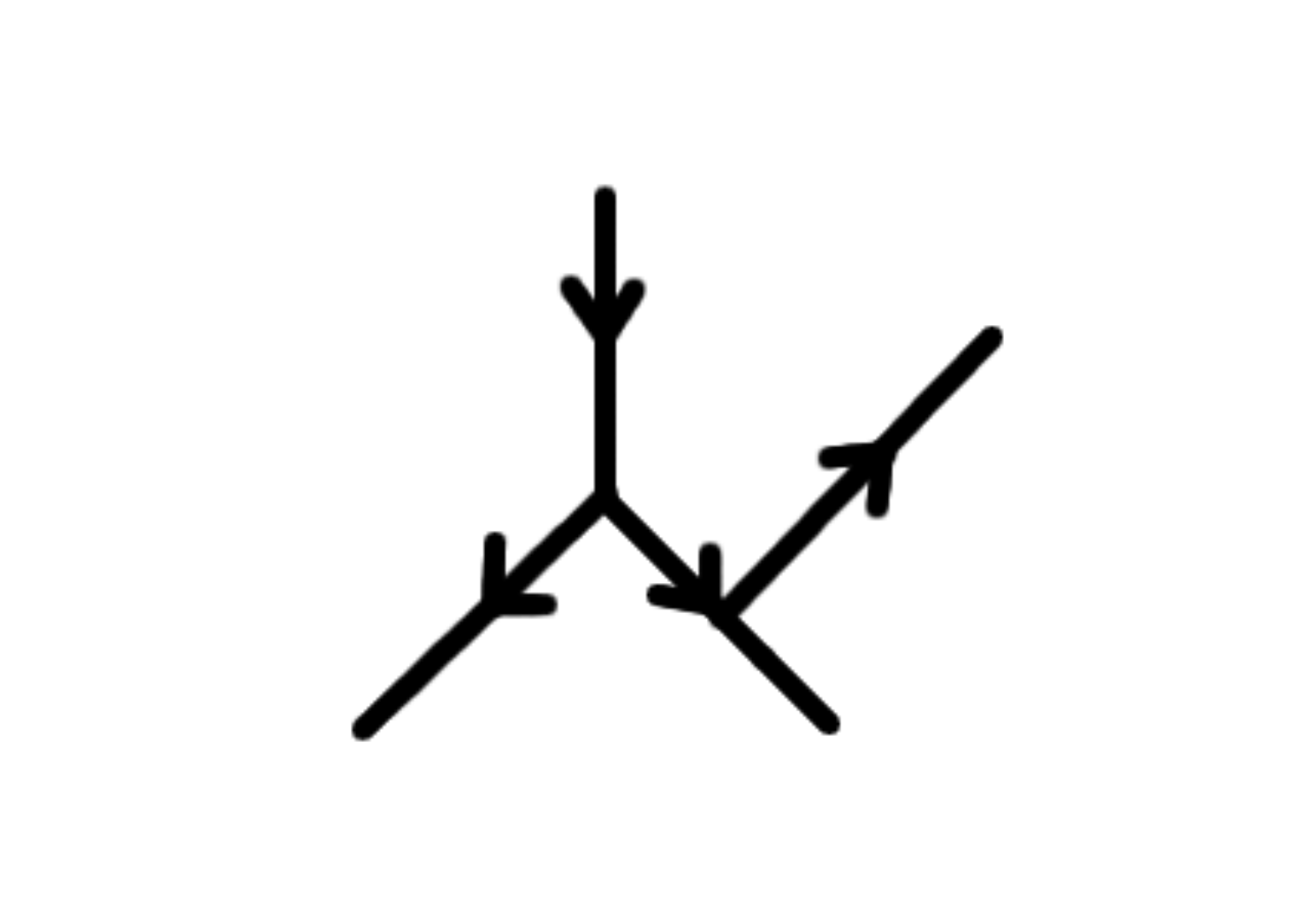}}
\put(1,2.35) {$D4_{\frac{H}{2},3 \overline{D0}}$}
\put(0.6,0.15) {$D6_H$}
\put(2.25,0.15) {$\overline{D6}$}
\put(2.65,1.8) {3 $\overline{D0}$}
\put(4,1.3) {$=$}
\put(5,1.3) {$-2'448'500$.}
\end{picture}
\end{figure}
This means
\vskip 2mm
\begin{figure}[h]
\setlength{\unitlength}{1cm}
\centering 
\begin{picture}(3,2.3)
\put(-5,0) {\includegraphics[width=0.25\textwidth,angle=0]{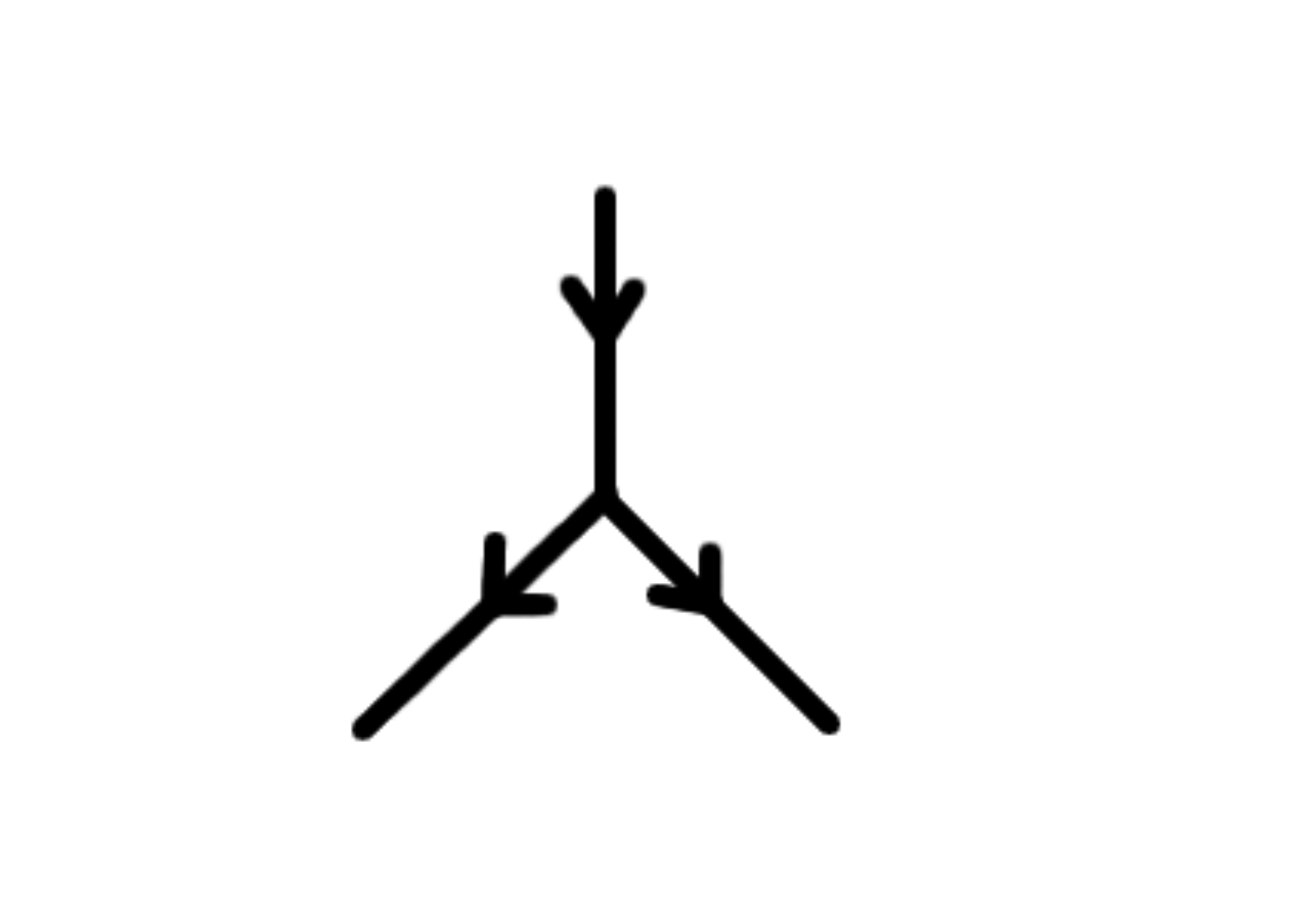}} 
\put(0,0) {\includegraphics[width=0.25\textwidth,angle=0]{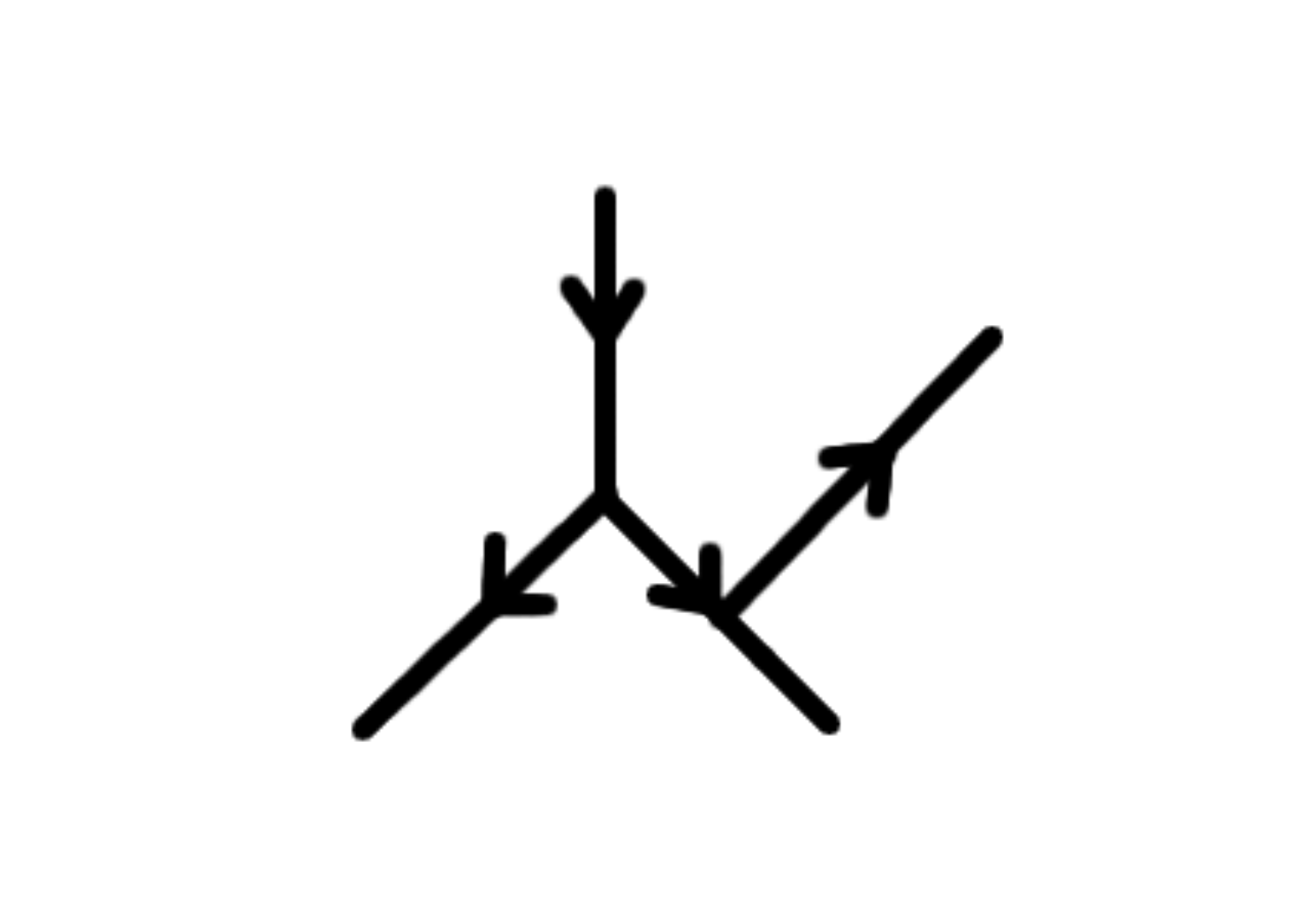}}
\put(-4.2,2.38) {$D4_{\frac{H}{2}+C_1+\overline{C_1}}$}
\put(-4.8,0.18) {$D6_{H,\overline{D2}(C_1)}$}
\put(-2.75,0.15) {$\overline{D6}_{D2(C_1)}$}
\put(0,1.3) {$+$}
\put(1,2.35) {$D4_{\frac{H}{2},3\overline{D0}}$}
\put(0.6,0.15) {$D6_H$}
\put(2.25,0.15) {$\overline{D6}$}
\put(2.55,1.8) {$3 \overline{D0}$}
\put(4.5,1.3) {$=$}
\put(5.3,1.3) {$5'817'125,$}
\end{picture}
\end{figure}
agreeing with the modular prediction! Note that this disagrees with the result in \cite{Gaiotto:2007cd}. Whereas the authors of that paper conclude the same index for the D4-picture version of the state with three $\Dob$'s, they obtain a different index for the state with the curves, because they applied a Pauli exclusion principle for these two curves. The fact that we do not encounter their problem has a nice geometrical interpretation: we consider these two curves to lie on different branes (i.e. the D6 and $\Dsb$ as opposed to the D4), so in our case, the Pauli exclusion principle does not apply.

We therefore refine the split flow conjecture as follows. Every supergravity tree flow corresponds to a family of microscopic states of fixed charges. The moduli space of the latter need not be a product of the moduli spaces of the constituents times the moduli space of the tachyon field, but can be a non-trivial fibration of the latter over the former. Therefore, the index of such states need not be a product of three indices, but will instead be an integral of \emph{index densities} over the moduli spaces of the constituent branes. If the tachyon index is a constant over these moduli spaces, then the integral factorizes, however, it will typically be a \emph{locally constant} function with discontinuous jumps.

\section{Conclusions}
In this paper, we set out to describe polar D4-D2-D0 systems (and some non-polar ones) as bound states of D6/$\Dsb$ pairs with fluxes and sheaves on them, in order to compute BPS indices for these states by factorizing them into the indices of the constituent branes times the index of the tachyon degrees of freedom between the constituent branes. 

We proceeded by looking for all relevant split attractor flows corresponding to the D6/$\Dsb$ pairs into which a given D4-D2-D0 might split. We were able to find all necessary split flows to account for the desired descriptions in a highly non-trivial way. Namely, we found that the existence of such split flows is highly sensitive the the \emph{area code}, i.e. the sector in the K\"ahler moduli space in which one starts the attractor flow. Hence, the results in this paper are a clean demonstration that split attractor flow trees are thus applicable also in this low charge regime, as had been previously expected.

We then set out to compute the BPS indices of these states by attributing indices to each possible split flow contributing to the state of a given D4-D2-D0 charge. We found that the background dependence of split flows due to marginal stability walls and threshold walls create highly intricate structures in the flow trees. Nevertheless, by making consistent choices and subtracting `doubly counted' quantities, we were able to compute the indices in a straightforward way for all but one charge system, the $(0,-3)$-state.

By carefully studying the moduli space of the $(0,-3)$-state, we realized that the latter is not a product of three moduli spaces, $\mathcal{M}_{D6} \times \mathcal{M}_{\overline{D6}} \times \mathcal{M}_{{\rm Tachyon}}$, but a non-trivial fibration of the latter over a product of the two former. This allowed us to devise a scheme to obtain the correct index for this system. One can interpret this as a refinement to the split flow tree conjecture, in the sense that an index associated to a split flow tree might have to be calculated with greater care if one is dealing with a non-trivially fibered moduli space.

Accepting our refinement to the enumeration of the $(0,-3)$-charge system, our results numerically improve upon those of \cite{Gaiotto:2006wm} and \cite{Gaiotto:2007cd}, where the M5-brane elliptic genus was computed considering the dual CFT operators and, alternatively, from the moduli space geometry from a D4-D2-D0 point of view. Our indices match up with those predicted by the modularity of the elliptic genus. 

Our results can be summarized in the following:
\begin{center}
\begin{tabular}{|c||c|c|}
    \hline
    \textbf{Reduced D0 charge $\hat{q}_0$} & \footnotesize{Modular form prediction} & \footnotesize{\textbf{Split flows and DT densities}}\\
    \hline
    \hline
    $2.29$ & +5 & +5\\
    \hline
    $1.29$ & -800 & -800\\
    \hline
    $0.69$ & +8'625 & +8'625\\
    \hline
    $0.69$ & +8'625 & +8'625\\
    \hline
    $0.29$ & +58'500 & +58'500\\
    \hline
    $-0.11$ & -1'218'500 & -1'218'500\\
    \hline
    $-0.31$ & -1'138'500 & -1'138'500\\
    \hline
    $-0.71$ & +5'817'125 & +5'817'125\\
    \hline
    $-1.11$ & +441'969'250 & +441'969'250\\
    \hline
\end{tabular}
\end{center}
This table indeed suggests a strong split flow conjecture!

Finally, our results shed light on the IIA picture factorization of D4-D2-D0 states into D6/-$\Dsb$ pairs, each accounted for by invariants (conjecturally) calculated by topological string theory. This can be seen as a first step in studying a microscopic, low charge counterpart of the large charge OSV conjecture (\cite{Ooguri:2004zv}). One may also see our results as the microscopic building blocks of the desired large charge factorized partition functions arising in the OSV conjecture. 

One can use our techniques to study some polar states on higher class divisors on the quintic, but also to study the spectrum with divisors on other Calabi-Yau's. It might be interesting to pursue this in a more exhaustive manner, and collect the new information which can be obtained.

\acknowledgments{We would like to thank Lotte Hollands, Joris Raeymaekers, Emanuel Scheidegger and Dieter Van den Bleeken for useful discussions. We would also like thank Frederik Denef for initial collaboration, useful discussions and thoroughly reading the initial draft for this paper. This work  is supported in part by the European Community's Human Potential Programme under contract MRTN-CT-2004-005104 â`Constituents, fundamental forces and symmetries of the universe', in part by the FWO-Vlaanderen, project G.0235.05, in part by the Federal Office for Scientific, Technical and Cultural Affairs through the Interuniversity Attraction Poles Programme Belgian Science Policy P6/11-P, and in part by the Austrian Research Funds FWF under grant number P19051-N16}

\begin{appendix}

\section{The area code: two interesting case studies}
This part of the appendix is a short discussion of the area code of two charge systems. As in \cite{Denef:2001xn}, we will use the w-plane to graphically represent the region of moduli space of interest for our graphical illustrations. The w-coordinate is related to the complex structure modulus $\psi$ by
\begin{equation}
  w=\frac{\textrm{ln}(|\psi |+1)}{\textrm{ln}(2)}\frac{\psi}{|\psi |}
\end{equation}
The advantage of this coordinate is that it is proportional to $\psi$ near the Gepner point and grows logarithmically in $\psi$ in the large complex structure limit. The w-plane is a fivefold cover of our moduli space and the normalization is chosen such that the five conifold point copies lie at $w=e^{2 n i\pi /5}$ for $(n=0,...,4)$.
In the illustrations in the appendices, beginning with figure \ref{typeA}, threshold walls are printed in black, walls of marginal stability for the first split in blue, and flow branches in green, the five copies of the conifold point in red. A wall of marginal stability between a D6-core and a D2-D0-halo will be printed as a dotted red line.
\subsection{D4-brane, degree one rational curve and $\overline{D0}$:  $\mathbf{\Delta q=1, \, \Delta q_0=-2}$ }{\label{areacode12}}
Here we discuss the area code of the non-polar system  with $\hat{q}_0\approx -0.31$ in more detail. Recall that this charge system can be described as a D4-brane with half a unit of worldvolume of flux turned on for anomaly cancellation as well as additional worldvolume flux dual to a degree one rational curve and one $\Dob$. This corresponds to the total charge
\begin{equation}
  \Gamma=H+\frac{7}{10}H^2+\frac{11}{60}H^3.
\end{equation}
We find two different basic types of split flow trees for this charge configuration.

\begin{figure}[h]
\begin{tabular}{cc}
    \fbox{\includegraphics[width=0.45\textwidth,angle=0]{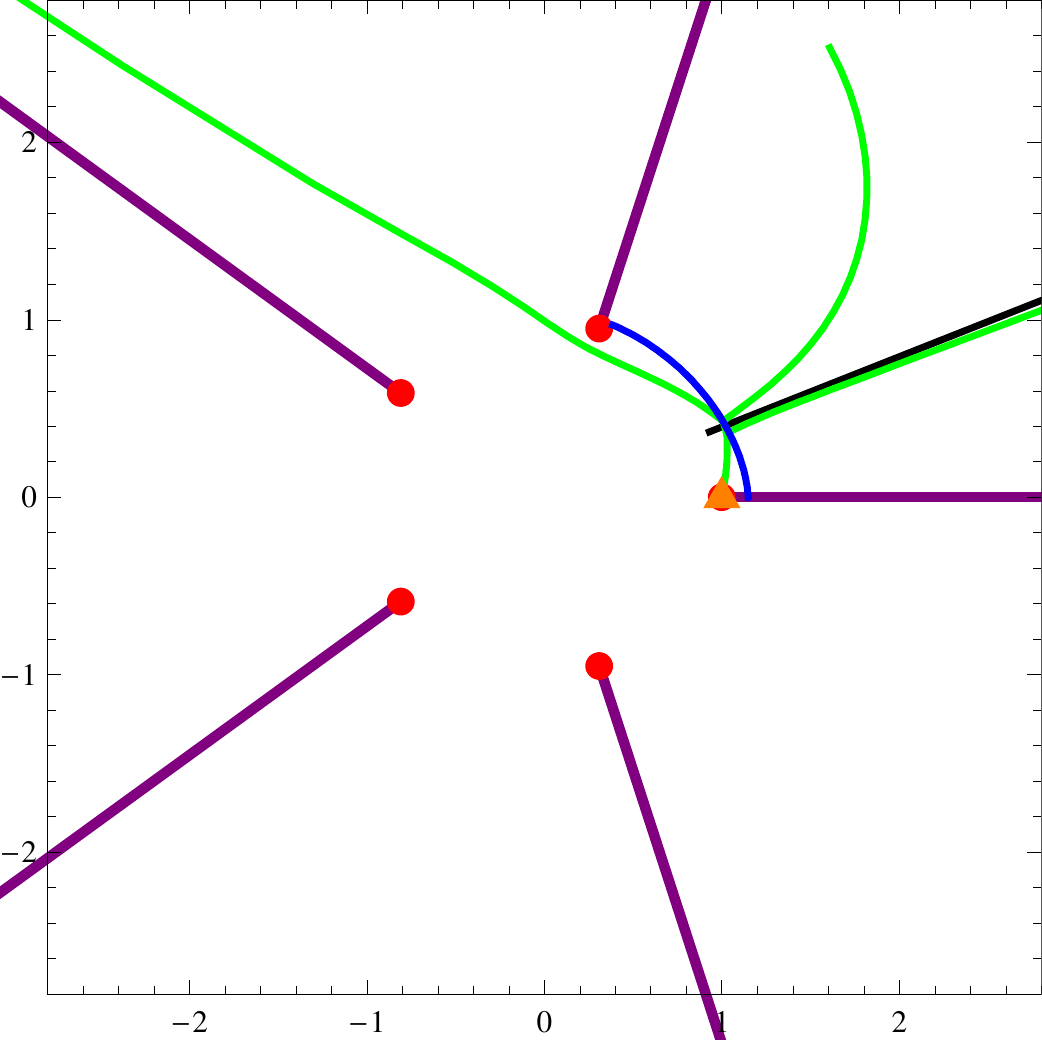}}&
    \fbox{\includegraphics[width=0.45\textwidth,angle=0]{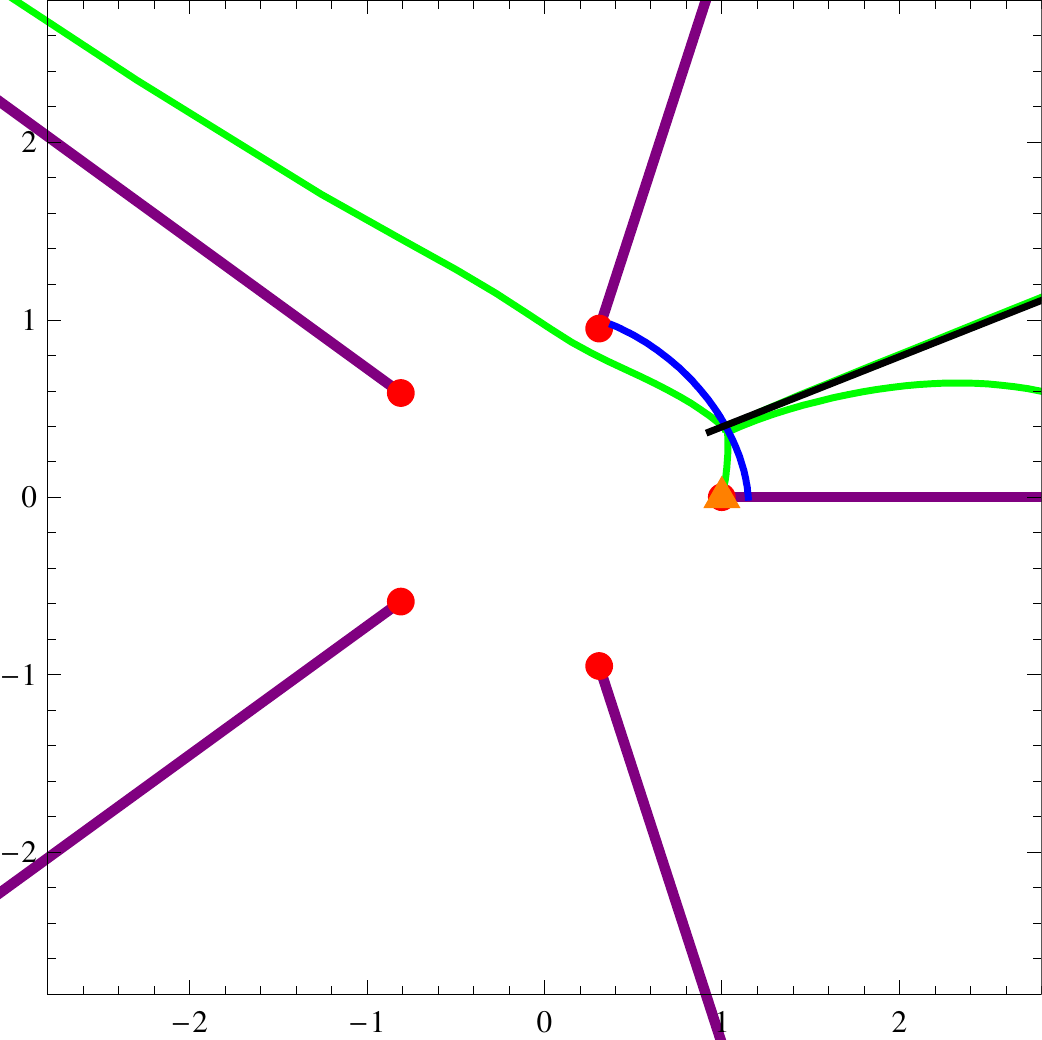}}
\end{tabular}
\caption{\textbf{Left: Type A1 split.} A split flow tree with the background above the threshold wall. The incoming branch reaches a wall of marginal stability and splits into D6-$\Dob$ and $\Dsb$-D2. The $\Dsb$-D2-branch then flows off to an attractor point outside of the fundamental wedge, not shown in the figure. The D6 and the $\Dob$-brane split just slightly below the wall after which the $\Dob$ flows off to the LCS point, and the D6 (with one unit of worldvolume-flux) flows to a copy of the conifold point. \textbf{Right: Type A2 split.} A split flow tree with the background below the threshold wall. The incoming branch reaches a wall of marginal stability and splits into D6 and $\Dsb$-D2-$\Dob$, after which the D6 flows to a copy of the conifold point. The $\Dob$ split off and heads towards the LCS point, while the $\Dsb$-D2 runs off towards an attractor point outside of the fundamental wedge, again not shown in the figure.}
\label{typeA}
\end{figure}

\begin{figure}[h]
\begin{center}
   \fbox{\includegraphics[width=0.42\textwidth,angle=0]{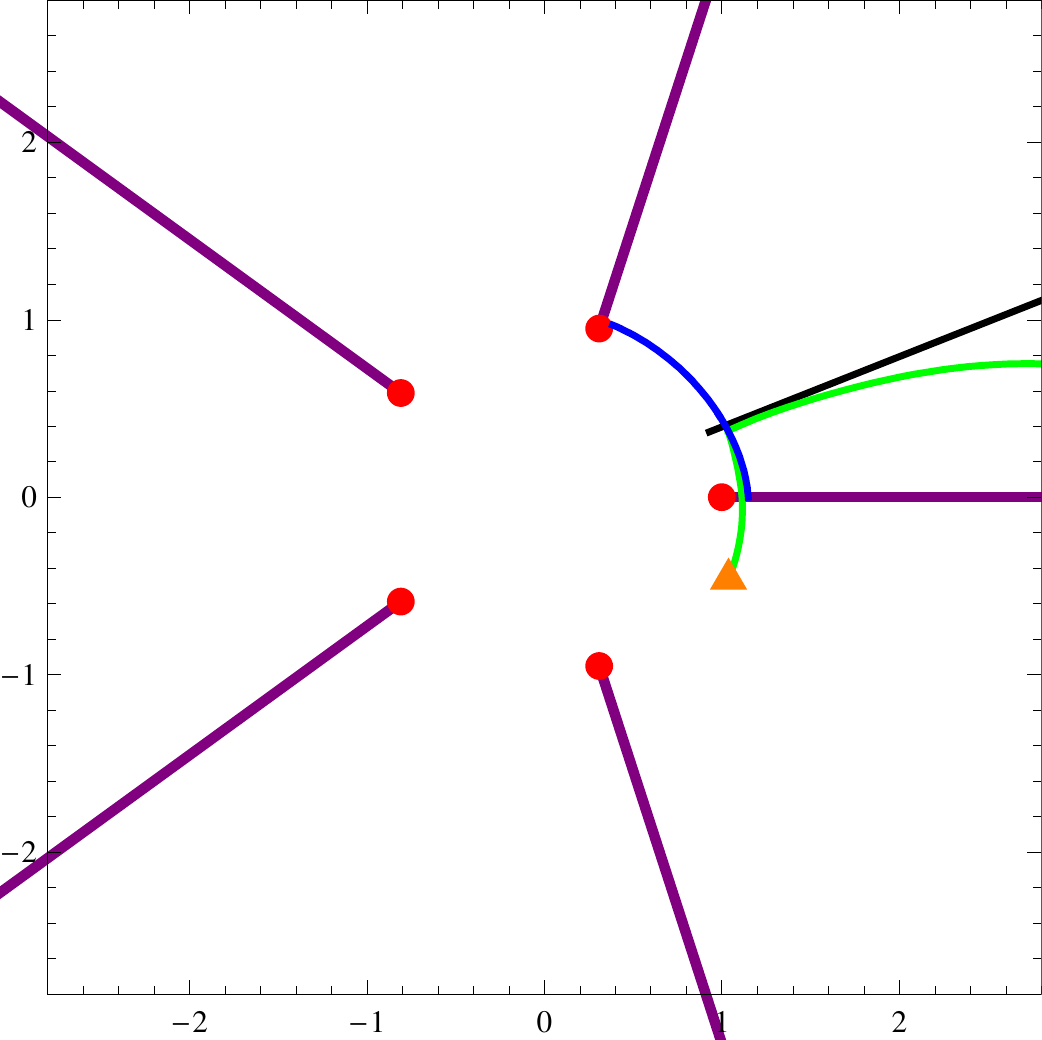}}
\end{center} 
\caption{\textbf{Type A1 split not found.} The type A1 split cannot be found when the background is chosen below the $\Dob$-threshold wall: After the first split into D6-$\Dob$ and $\Dsb$-D2 on a degree one rational curve, the D6-$\Dob$ branch crashes before either reaching a split or an attractor point.}
\label{nottypeA1}
\end{figure}

The first we will refer to as \textbf{type A} which comes in two variants. These are the ones with the following three endpoint centers: a D6 with one unit of worldvolume flux, a $\Dsb$ with a sheaf corresponding to a D2 on a degree one rational curve, and a $\Dob$. The two variants correspond to the two sides of the $\Dob$-threshold wall where one can place the background. Note that one finds that the `incoming branch' of the flow tree does not cross the $\Dob$-TH wall.
\begin{itemize}
\item If one starts above the TH wall, the $\Dob$ remains on the D6-side after the first split (and then splits off to run towards the LCS point): this is denoted the \textsl{type A1} variant. An example can be found on the left-hand side in figure \ref{typeA}. This split contributes in the background region above the TH wall, but it does not exist when starting below the wall: this is exemplified in figure \ref{nottypeA1}. Schematically: \begin{eqnarray} (D4_{C_1}/\Dob )&\rightarrow &(D6/\Dob )\, +\, (\Dsb /D2)\nonumber\\ &\rightarrow &(D6)\, + (\Dob) \, + \, (\Dsb /D2).\nonumber\end{eqnarray} \item When one starts the flow tree below the TH wall, the $\Dob$ binds to the $\Dsb$-D2 after the first split: this is referred to as the $\textsl{type A2}$ variant, and an example can be found on the right-hand side in figure \ref{typeA}. This split contributes in the background region below the TH wall, but it does not exist when starting above the wall. Schematically: \begin{eqnarray} (D4_{C_1}/\Dob )&\rightarrow &(D6)\, +\, (\Dsb /D2/\Dob )\nonumber\\ &\rightarrow &(D6)\, \, + \, (\Dsb /D2) + (\Dob).\nonumber \end{eqnarray}
\end{itemize}

Figure \ref{nottypeA1} illustrates that a type A1 split does not exist when the background is taken to be below the threshold wall.

%\EPSFIGURE[h]{nottypeA1.eps,height=8cm,trim=0 0 0 0}
%{The type A1 split cannot be found when the background is chosen below the anti-D0 threshold wall: After the first split into D6-anti-D0 and anti-D6 with flux dual to a degree one rational curve, the D6-D0bar branch crashes before either reaching a split or an attractor point.}
%\label{nottypeA1}

\begin{figure}[h]
\begin{tabular}{cc}
    \fbox{\includegraphics[width=0.42\textwidth,angle=0]{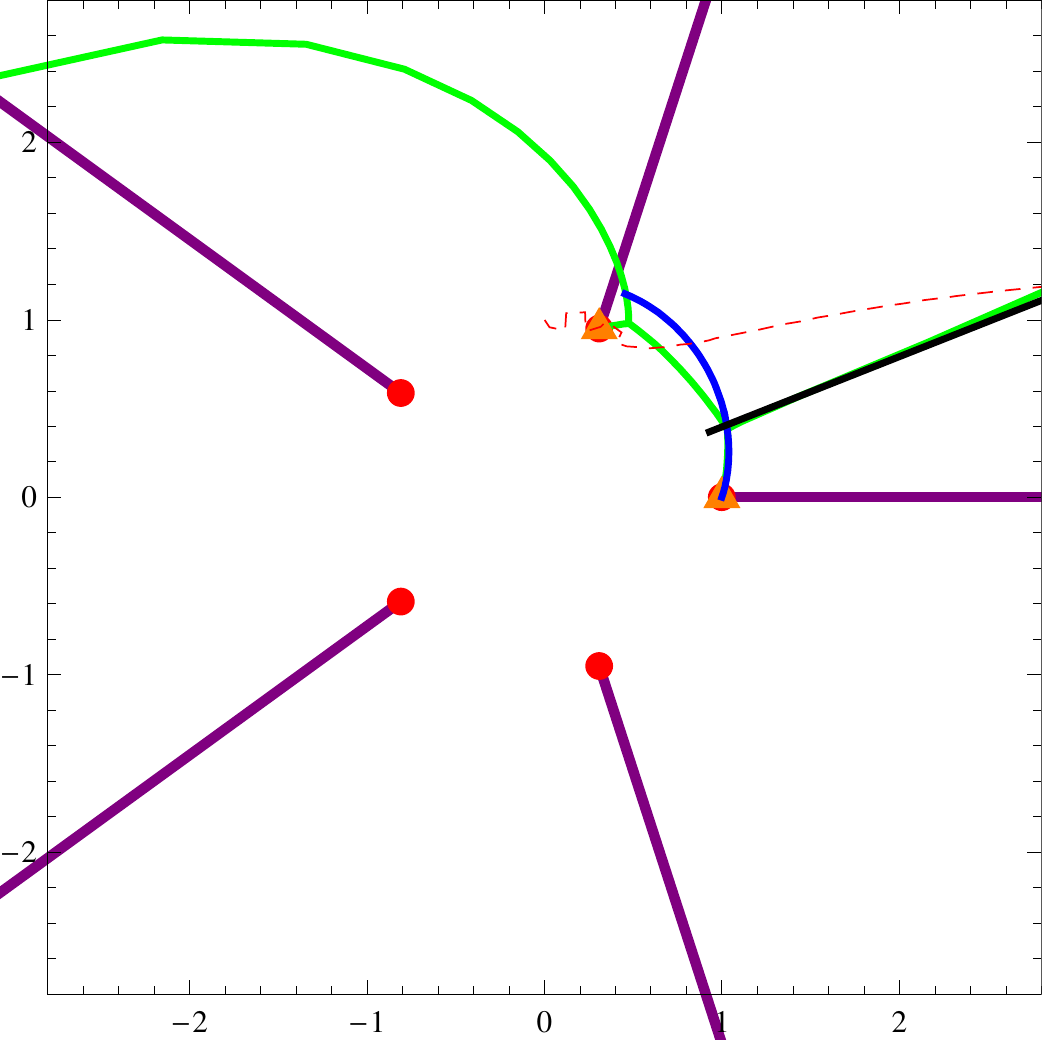}}&
    \fbox{\includegraphics[width=0.42\textwidth,angle=0]{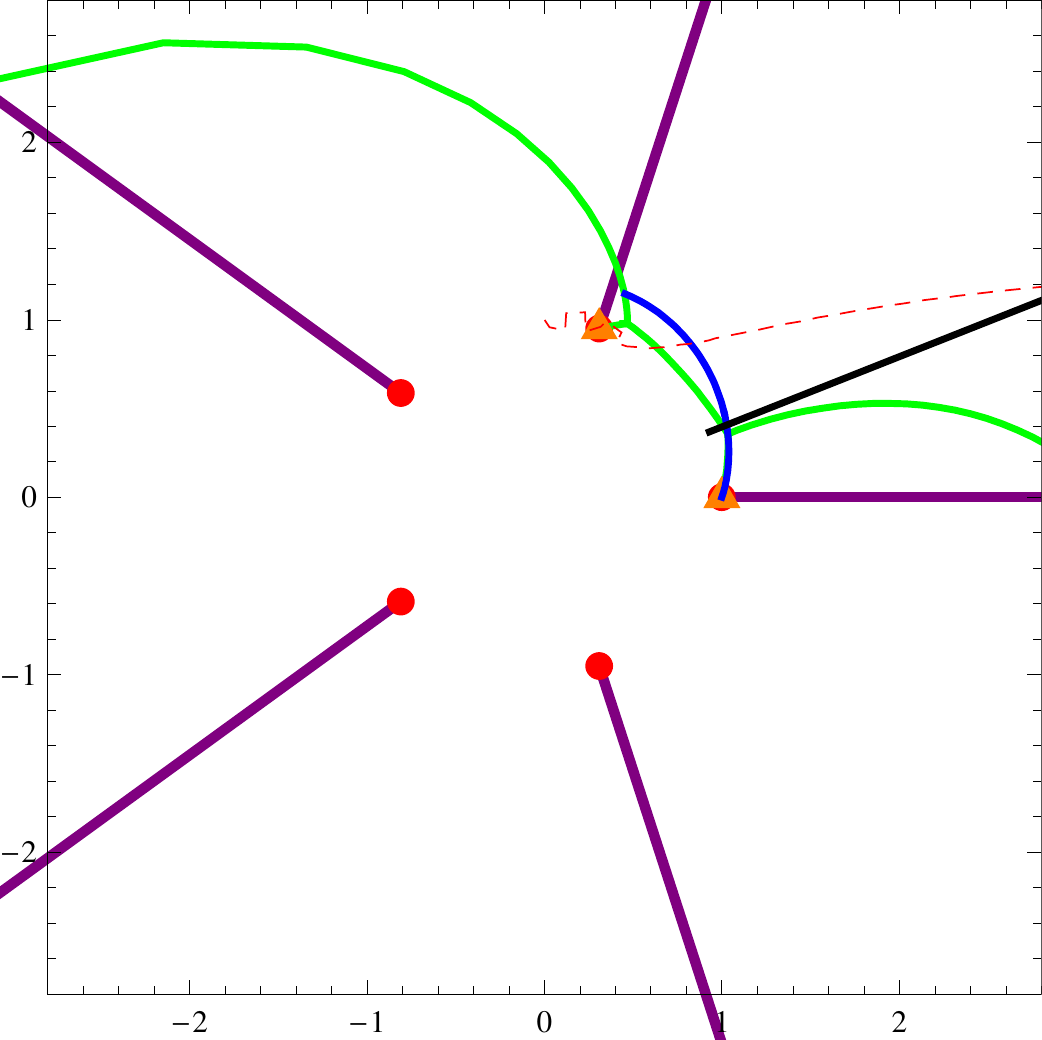}}
\end{tabular}
\caption{\textbf{Left: Type B split with background above TH wall.} A type B split with the background chosen above the threshold wall (and not too close to the conifold point). The charge splits into D6 and $\Dsb -D2-\Dob $. The D6-branch flows to a copy of the conifold point. The D2-D0 halo particle then splits off from the $\Dsb$ and flows off towards its attractor point outside of the fundamental wedge. The $\Dsb$ also flows towards a copy of the conifold point. \textbf{Right: Type B split with background below TH wall.} Example of a type B split flow tree with the background chosen below the threshold wall.}
\label{typeBfigs}
\end{figure}

The second basic type of split, which we denote a \textbf{type B} split also has three endpoints. However, this time, the D6 with one unit of flux and the $\Dsb$ do not carry any curves, and their flow branches each end on a copy of the conifold point. After the first split, the D2-D0 charge is on the $\Dsb$-side, and then leaves the $\Dsb$ at the second split point. Schematically: \begin{eqnarray} (D4_{C_1}/\Dob )&\rightarrow &(D6)\, +\, (\Dsb /D2/\Dob)\nonumber\\ &\rightarrow &(D6)\, + \, (\Dsb ) \, +(D2/\Dob ).\nonumber\end{eqnarray}The important message is that the existence of this type of split is not influenced by the threshold wall. \footnote{It does however require the $\Dsb$(core)-D2-D0 (halo) branch to reach the core-halo wall in order to be able to perform the second split (recall that the core-halo wall is plotted as a dotted red line in figure \ref{typeBfigs}), which is the case if one does not start too close to the copy of the conifold point lying on the upper branch cut boundary of the fundamental wedge. This latter fact would eventually lead to another topological sector for the final area code of this charge system, but we will ignore this subtlety, presently.}

Below the threshold wall the total index for our charge system is quite clear. It is calculated according to
\begin{equation}\label{correctind}
  \Omega=\Omega_{A2}+\Omega_{B}=(-1)^1\cdot 2\cdot N_{\textrm{DT}}(0,0)\cdot N_{\textrm{DT}}(1,2)=-1'138'500.
\end{equation}
Note that the DT-invariant $N_{\textrm{DT}}(1,2)$ counts all the $\Dsb$-D2-D0 states, so this is the appropriate enumeration for both the type A2 and the B flow trees. Had one not found the type B flow trees, one would have been overcounting the states, when using the DT invariant to enumerate the number of BPS states belonging to the $\Dsb$-branch.

Above the threshold wall, one has a split flow tree of type A1 contributing, which one can enumerate easily:
\begin{equation}\label{naiveind}
  \Omega_{A1}=(-1)^1\cdot 2\cdot N_{\textrm{DT}}(0,1)\cdot N_{\textrm{DT}}(1,1)=-1'150'000.
\end{equation}
However, one also has a type B flow tree above the wall, which explains why the two naively equivalent indices, \ref{correctind} and \ref{naiveind}, do not match. The apparent discrepancy lies in the fact, that, below the threshold wall, $N_{\textrm{DT}}(1,2)$ counts both the A2 and the B flow trees. Therefore the all flow trees starting below the threshold wall are taken care of by this index. 

However, when taking the background value of the modulus to start above the wall, the index with $N_{\textrm{DT}}(0,1)\cdot N_{\textrm{DT}}(1,1)$ only counts the A1 split. One still needs to account for the B split, which also exists in this sector of the moduli space. In principle, we would not be able to conclude how much the type B split contributes directly, but we do know that the total index cannot jump when crossing a wall of threshold stability. Therefore, we conclude that the type B split has to contribute $+11'500$ to the index above the wall, allowing us to also state $-1'138'500$ as the total index above the threshold wall. 

Note that these are at least valid statements for a large part of the background above the threshold wall. Namely, at least at a reasonable distance from the upper conifold point (in order for the $\Dsb$-D2-D0 branch `to be able to reach' the core-halo wall starting from the first split point). If the type B split disappears, one enters a third topological sector, but as we know that the index has to be same in that region, we will not examine it more closely.
\subsection{D4-brane, degree two rational curve and $\overline{D0}$: $\mathbf{\Delta q=2, \, \Delta q_0=-2}$ }\label{areacode22}
Here we discuss the area code of the system with $\hat{q}_0\approx -1.11$ in more detail. The total charge can be obtained by considering a (pure fluxed) D4 with additional flux dual to a degree two rational curve as well as one bound $\Dob$. This yields total charge
\begin{equation}
  \Gamma=H+\frac{9}{10}H^2-\frac{1}{60}H^3.
\end{equation}
Here we find three different basic types of flow trees. Two of these and their area code behave completely analogously to the system \ref{areacode12}, discussed previously. The only difference concerning these first two types of split flow trees, compared to the previous case study, is that one is dealing with a degree two instead of a degree one curve, and as a consequence of this difference in charge, the corresponding but analogous walls are slightly different. We will denote the possible flow trees in analogy.

\begin{figure}[h]
\begin{tabular}{cc}
    \fbox{\includegraphics[width=0.42\textwidth,angle=0]{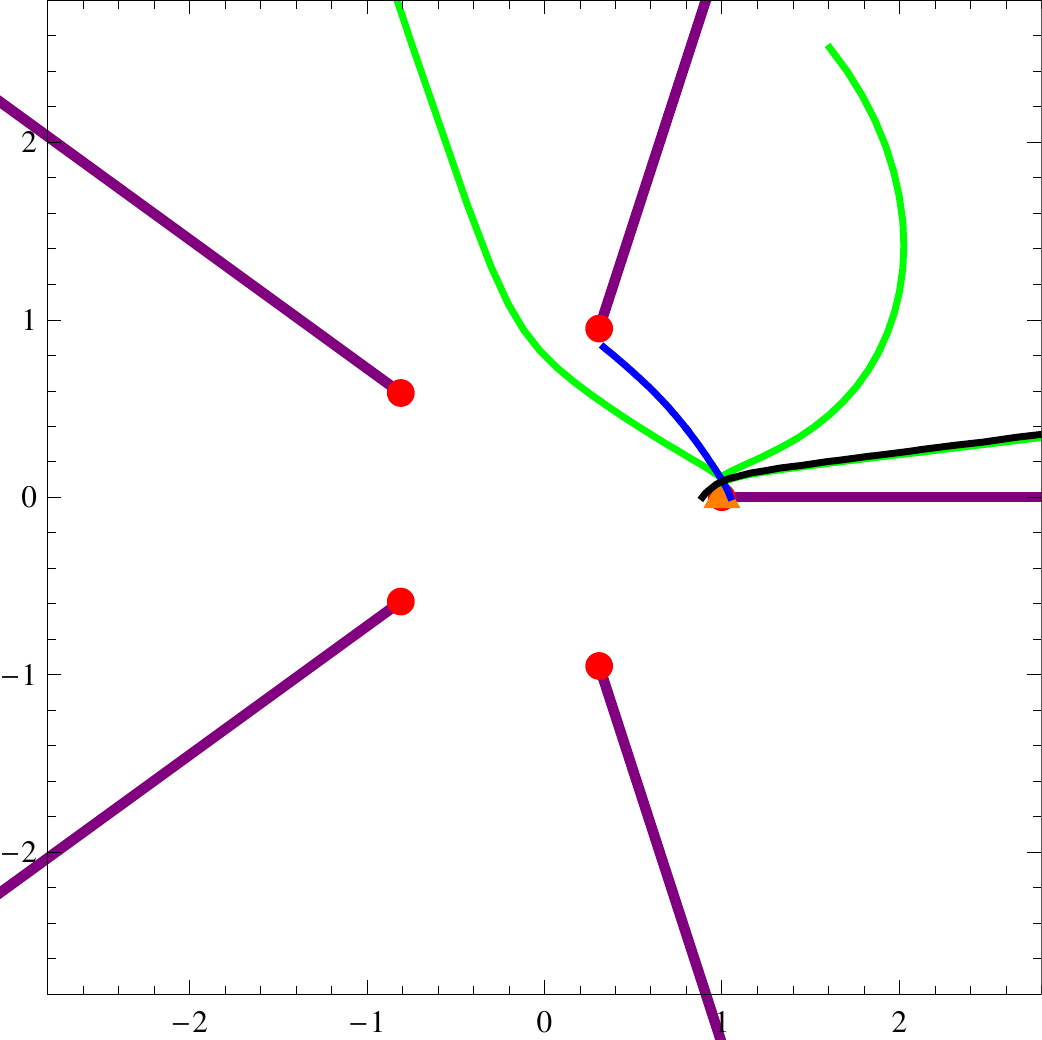}}&
    \fbox{\includegraphics[width=0.42\textwidth,angle=0]{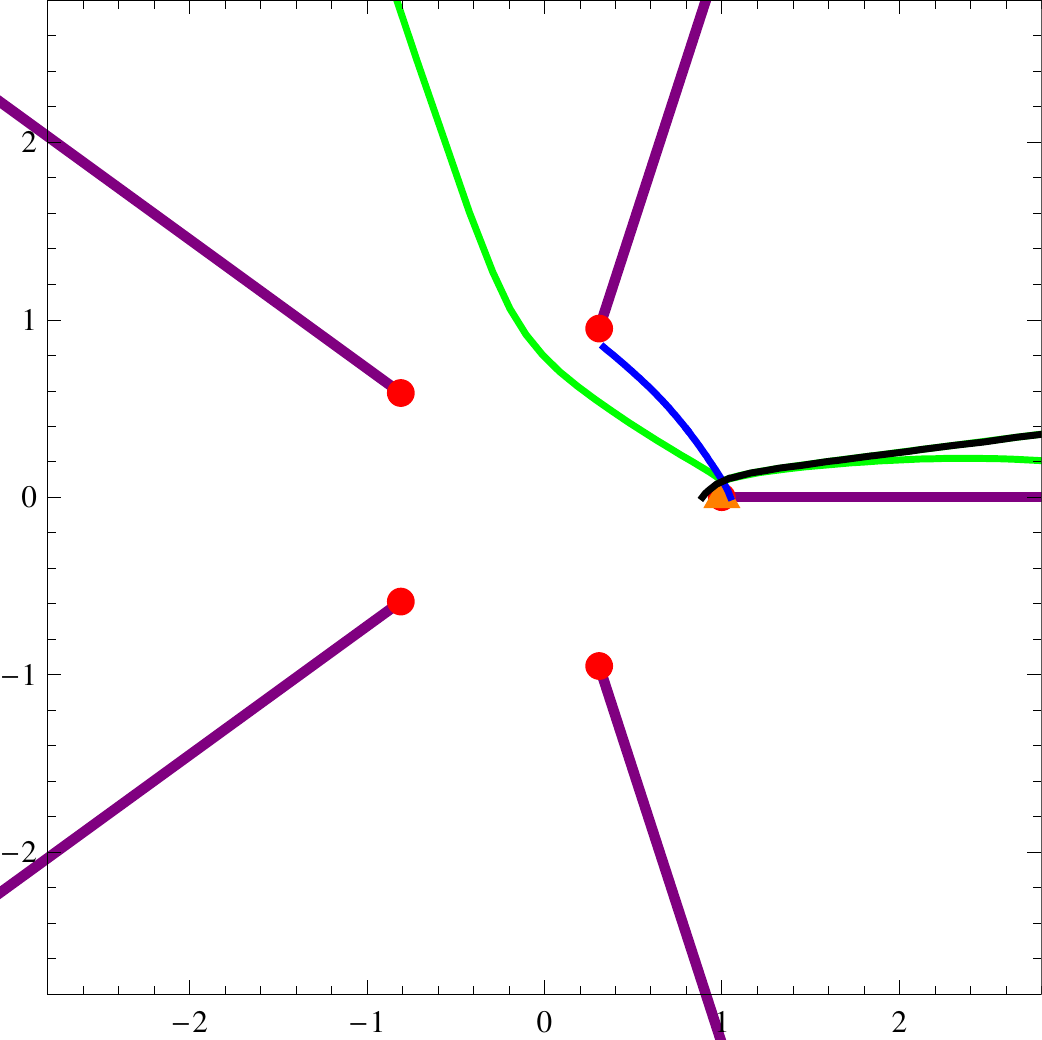}}
\end{tabular}
\caption{\textbf{Left: Type A1 split.} A split flow tree with the background above the threshold wall. The incoming branch reaches a wall of marginal stability and splits into D6-$\Dob$ and $\Dsb$-D2, where now the D2 is wrapped on a degree two rational curve. The $\Dsb$-D2-branch then flows off to an attractor point outside of the fundamental wedge, not shown in the figure. The D6 and the $\Dob$-brane split just slightly below the wall after which the $\Dob$ flows off to the LCS point, and the D6 (with one unit of worldvolume-flux) flows to a copy of the conifold point. 
\textbf{Right: Type A2 split.} A split flow tree with the background below the threshold wall. The incoming branch reaches a wall of marginal stability and splits into D6 and $\Dsb$-D2-$\Dob$, after which the D6 flows to a copy of the conifold point. The $\Dob$ splits off and heads towards the LCS point, while the $\Dsb$-D2 runs off towards an attractor point outside of the fundamental wedge, again not shown in the figure.}
\label{22Asplits}
\end{figure}

We can again list the \textbf{type A} flow trees, two variants between which a threshold wall interpolates.
\begin{itemize}
\item If one starts above the TH wall, the $\Dob$ remains on the D6-side after the first split (and then splits off to run towards the LCS point): this is denoted the \textsl{type A1} variant. The only difference to the case discussed in \ref{areacode12} is that the D2 is now placed on a degree two instead of a degree one curve. An example can be found on the left-hand side in figure \ref{22Asplits}. This split contributes in the background region above the TH wall, but it does not exist when starting below the wall: this is exemplified in figure \ref{nottypeA1}. Schematically: \begin{eqnarray} (D4_{C_1}/\Dob )&\rightarrow &(D6/\Dob )\, +\, (\Dsb /D2)\nonumber\\ &\rightarrow &(D6)\, + (\Dob) \, + \, (\Dsb /D2).\nonumber\end{eqnarray} \item When one starts the flow tree below the TH wall, the $\Dob$ binds to the $\Dsb$-D2 after the first split: this is referred to as the $\textsl{type A2}$ variant, and an example can be found on the right-hand side in figure \ref{22Asplits}. This split contributes in the background region below the TH wall, but it does not exist when starting above the wall. Schematically: \begin{eqnarray} (D4_{C_1}/\Dob )&\rightarrow &(D6)\, +\, (\Dsb /D2/\Dob )\nonumber\\ &\rightarrow &(D6)\, \, + \, (\Dsb /D2) + (\Dob).\nonumber \end{eqnarray}
\end{itemize}
The figure \ref{22Asplits} shows the two A-type flow trees.

Again, there is a \textbf{type B} flow tree, with the only difference in the halo charge. As this is straight forward, we will thus refrain from including another print and repeating a closer discussion.

\begin{figure}
\begin{center}
   \fbox{\includegraphics[width=0.42\textwidth,angle=0]{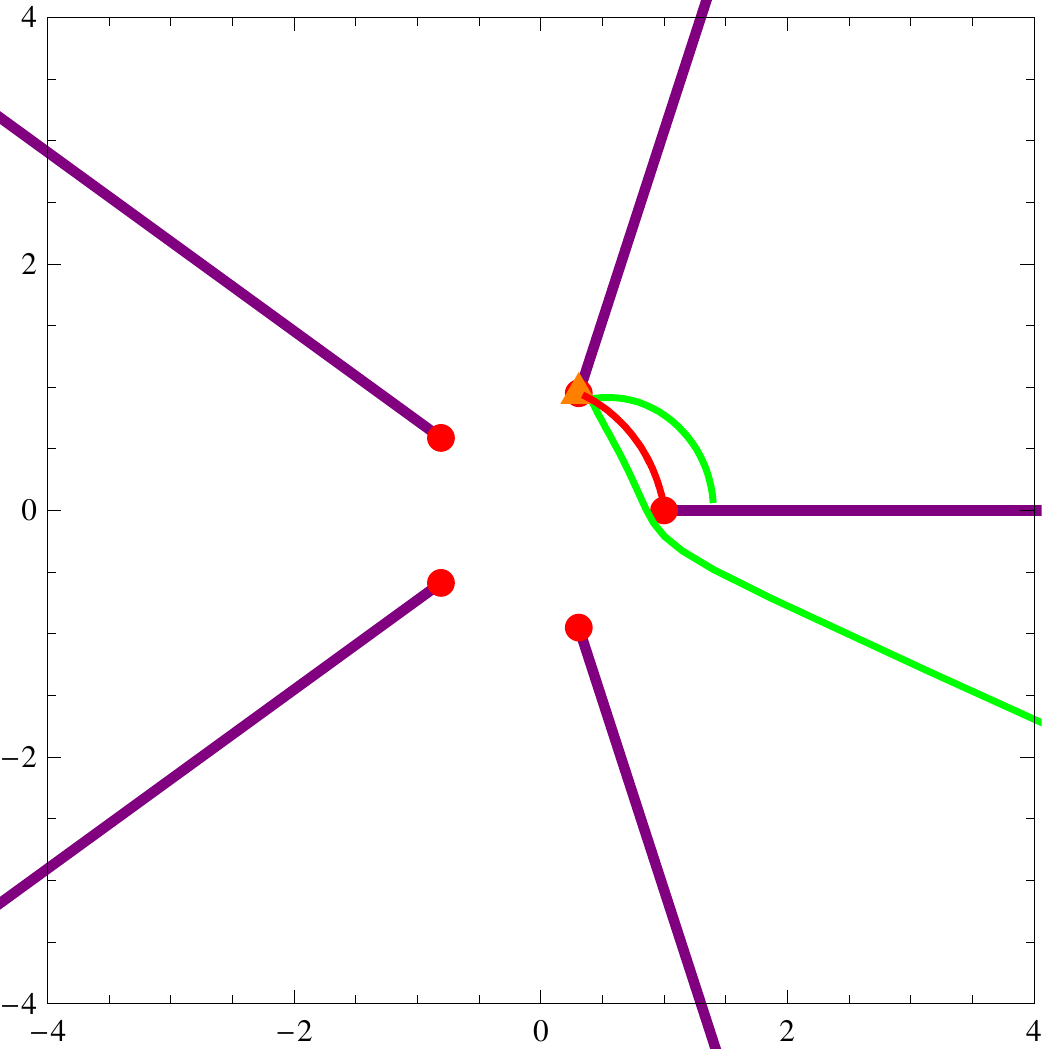}}
\end{center} 
\caption{\textbf{Type C1 split.} After the first split on the wall of marginal stability, the $\Dsb$ flows to a copy of the conifold point, and the D6 with a D2 on a degree three rational curve runs off to an attractor point outside of the fundamental wedge, not shown in this figure.}
\label{22typeC1b}
\end{figure}

However, one now has a third type of flow tree, which we will denote as \textbf{type C}. It is extremely convenient to analyze this charge configuration in a background where it is realized as a split flow with two end points. We will refer to this specific realization of the type C split, as \textsl{type C1}. The first center corresponds to a D6-brane with two units of worldvolume flux with a sheaf corresponding to a $\Dtb$ wrapped on a degree \textsl{three} rational curve. The second center corresponds to a $\Dsb$ with one unit of worldvolume-flux. The type C1 split flow tree requires the background to be `very close' to the wall of marginal stability between the two centers. An example of what is meant by `close' can be seen in figure \ref{22typeC1b}.

However, when one moves the background further `upwards in the figure', or `further away' from the wall of marginal stability, the first split point travels upwards `on the wall of marginal stability'. At some point, this first split point would seemingly cross the upper branch cut boundary of the fundamental wedge. This means that a topology change takes place. The index of BPS states is not allowed to jump, but flow trees can develop new branches. In the background region, where one finds the type C1 split flow tree, there is no single flow. However, by moving the background to the right, the incoming branch runs over the upper branch cut and then flows to an attractor point: a single flow enters the spectrum. This means that this topology change can be thought of as a trade off between the C1 split flow tree and a single flow (and possibly also new, different split flow trees). The contribution to the total index of BPS states cannot jump, when a topology change occurs. This example fits in nicely with the interpretations of \cite{Denef:2001xn}. Running this topology change in reverse, one could state: Starting in the background with the single flow, one crosses a branch cut when trying to `pull the flow through the conifold singularity', and one actually creates a new branch of the flow tree, namely the one ending on the nearby copy of the conifold point.

\begin{figure}[h!]
\begin{tabular}{cc}
    \fbox{\includegraphics[width=0.435\textwidth,angle=0]{22typeC1}}&
    \fbox{\includegraphics[width=0.45\textwidth,angle=0]{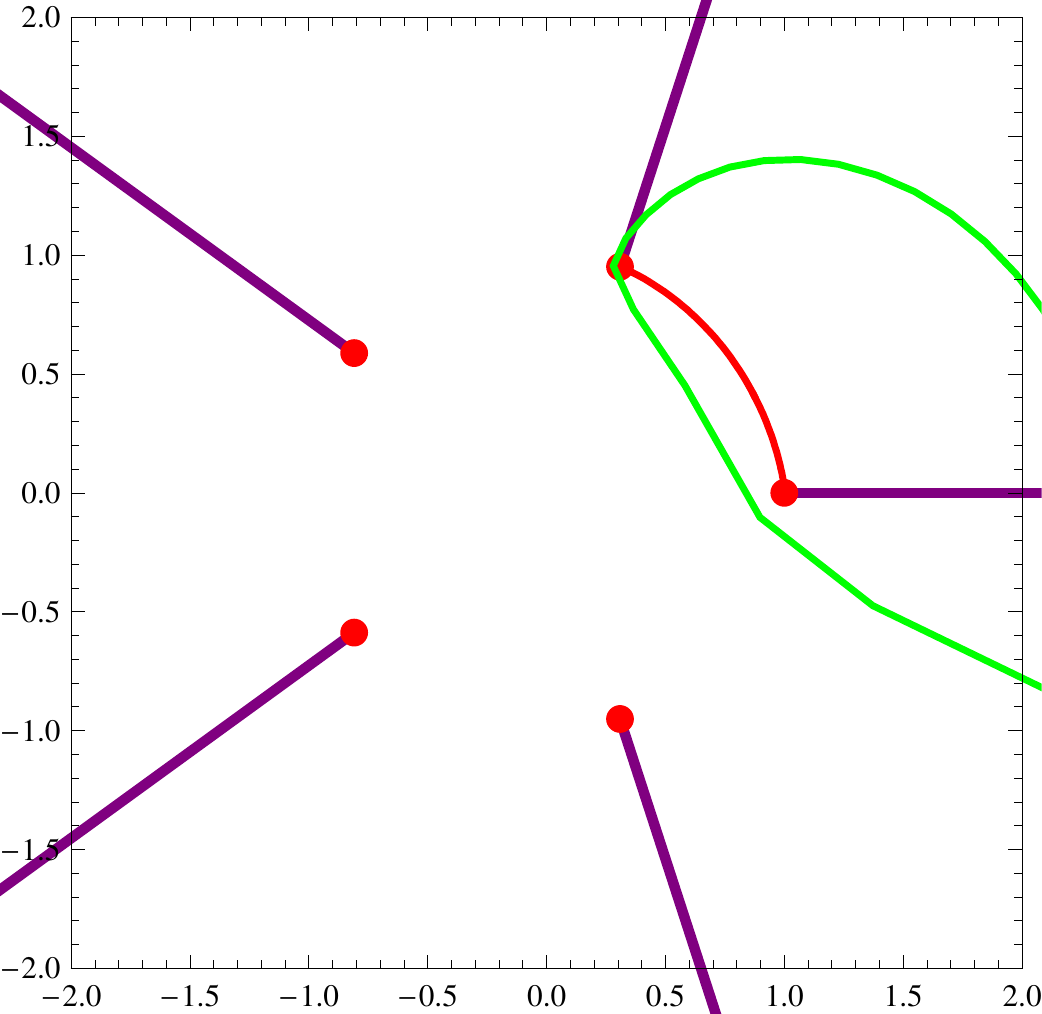}}
\end{tabular}
\caption{Topology change for the type C split. \textbf{Left: Type C1.} The charge supports a split flow. The first center corresponds to a D6 with two units of worldvolume flux and a $\Dtb$ wrapped on a degree three rational curve. The second center corresponds to a $\Dsb$ with one unit of worldvolume flux. As one moves the background modulus `upwards' or, alternatively, `further away' from the marginal stability wall, the first split point also moves upwards until hitting the branch cut boundary. \textbf{Right: Type C2, a single flow.} If one moves the background `too far' upwards or `too far' to the right, the incoming branch hits the branchcut before reaching the wall of marginal stability. Instead, this flow reaches an attractor point: a single flow enters the spectrum which is not supported in the background region where one finds the split flow on the left-hand side.}
\label{22Csplit}
\end{figure}

To conclude, taking the background very `close to the wall of marginal stability' between a D6 with a degree rational curve and the $\Dsb$, and also `close to the branch cut' forming the upper boundary of the fundamental wedge in w-plane, the total index belonging to our charge can be calculated as follows: 
\begin{equation}
  \Omega=\Omega_B+\Omega_{C1}=\Omega_B+(\Omega_C-\Omega_{C2})=124'762'875+317'206'375=441'969'250.
\end{equation}
This index has to remain invariant when varying the background modulus without crossing a wall of marginal stability, but the contributions might be quite different. According to the preceding remarks, one can see this as an interpolation between different topological sectors of moduli space. The reader can find this illustrated nicely in figure \ref{22Csplit}.

\end{appendix}

\end{document}